\documentclass[aps,pre,groupedaddress,twocolumn]{revtex4-2}

\usepackage{lineno}

\usepackage{amssymb,amsmath}

\usepackage{graphicx}
\usepackage{epsfig}
\usepackage{color}
\usepackage{bm}
\usepackage{tikz}
\usepackage{multirow}  

\usepackage{epsfig}
\usepackage{epstopdf}

\newcommand{\fig}[1]{\parbox{1.5cm}{\epsfig{file=#1.eps,width=1.5cm}}}
\newcommand{\figu}[1]{\parbox{1.5cm}{\epsfig{file=#1.eps,width=1.5cm}}}

\graphicspath{{fig/}}

\usepackage{amssymb}
\usepackage{pifont}%

\newcommand{\im}{{\rm i}}

\newcommand{\rd}{{\rm d}}
\newcommand{\tr}{{\rm tr}}

\definecolor{ForestGreen}{rgb}{0.13, 0.55, 0.13}

\begin{document}


\title{The influence of the Casimir effect on the binding potential for 3D wetting}



\author{Alessio~Squarcini}

\email{alessio.squarcini@uni.lu}
\affiliation{Complex Systems and Statistical Mechanics, Department of Physics and Materials Science, University of Luxembourg, 30 Avenue des Hauts-Fourneaux, L-4362 Esch-sur-Alzette, Luxembourg}

\author{Jos\'e M. Romero-Enrique}
\affiliation{
Departamento de F\'{\i}sica At\'omica, Molecular y
Nuclear, \'Area de F\'{\i}sica Te\'orica, Universidad de Sevilla,
Avenida de Reina Mercedes s/n, 41012 Seville, Spain}
\affiliation{
Instituto Carlos I de F\'{\i}sica Te\'orica y Computacional, Campus Universitario Fuentenueva,
Calle Dr. Severo Ochoa, 18071 Granada, Spain
}
\author{Andrew O. Parry}
\affiliation{
Department of Mathematics, Imperial College London, London SW7 2AZ, United Kingdom}


\date{\today}

\begin{abstract}
We provide comprehensive details of how a previously overlooked entropic, or low temperature Casimir contribution, $W_C$, to the total binding potential for 3D short-ranged wetting may be determined from a microscopic Landau-Ginzburg-Wilson Hamiltonian. The entropic contribution comes from the many microscopic configurations corresponding to a given interfacial one, which arise from bulk-like fluctuations about the mean-field (MF) constrained profile, and adds to the usual MF contribution $W_{MF}$. We determine the functional dependence of $W_C$ on the interface (and wall) shape using a boundary integral method which can be cast as a diagrammatic expansion with each diagram corresponding to successively higher-order exponentially decaying contributions. The decay of $W_C$ is qualitatively different for first-order and critical wetting with the change in form occurring at the MF tricritical point. Including the Casimir contribution to the binding potential preserves the global surface phase diagram but changes, radically, predictions for fluctuation effects at first-order and tricritical wetting, even when capillary-wave fluctuations are not considered.
\end{abstract}


\maketitle


\cleardoublepage

\section{Introduction}
Interfaces between bulk phases of matter exhibit an extraordinary wealth of physical properties both at, and away from, equilibrium. Pauli's playful remark that ``God created the bulk; surfaces were invented by the devil'' alludes to the difficulties encountered in attempts to model the rich phenomenology of surface physics. Of course, many advances have been made in the last century with progress often relying on adopting effective models or Hamiltonians which focus on the motion/fluctuations of a coarse-grained collective co-ordinate representing the local position of the interface. These effective models, which typically involve parameters such as the surface tension, are usually motivated by appeal to mesoscopic ideas together with restrictions based on symmetry principles and do not require a particularly detailed understanding of how they emerge from integrating out more microscopic degrees of freedom. The capillary-wave model of interface localization in a gravitational field \cite{BLS_1965,Weeks_1977}, the discrete-Gaussian model of the roughening transition \cite{Chui_Weeks} and the Kardar-Parisi-Zhang model of surface growth \cite{KPZ} demonstrate the power of such mesoscopic approaches.

However, the controversy surrounding the nature of the continuous wetting transition (referred to, for historical reasons, as critical wetting) in 3D systems with short-ranged forces stands out as a counter example where it has transpired that the precise details of the interfacial model, and understanding how it emerges out of a more microscopic framework is of crucial importance. The controversy here centres on the substantial discrepancy between the value of the correlation length critical exponent $\nu_\parallel$, as predicted by renormalization group (RG) analysis of an effective interfacial model, and the value extracted in careful simulation studies of critical wetting in the 3D Ising model. This discrepancy is important because 3D is a marginal dimension (an upper critical dimension) for short-ranged critical wetting and critical singularities are predicted to be non-universal, emerging from the sensitive interplay between fluctuations occurring on two different ranges of length scales: bulk fluctuations which occur on the microscopic scale of the bulk correlation length $\xi$, and interfacial fluctuations, resisted by the surface tension $\gamma$ (or, more generally the stiffness coefficient $\Sigma$ - see later), of the liquid-gas interfaces, occurring at much larger length scales which indeed become macroscopic on approaching the wetting transition. The predicted non-universality in 3D is therefore, a stringent test of our understanding of interfacial fluctuations and the critical wetting controversy has always hinted that some key physics is missing in the theoretical description.

In this paper, we shall show that previous effective models have, indeed, incorrectly modeled the contribution to the energy cost of interfacial configurations arising from bulk fluctuations. Specifically, the expression used for the binding potential in the interfacial model has missed an entropic contribution, equivalent to a (low temperature) thermal Casimir effect, which is entirely missing in mean-field (MF) treatments of wetting and interfacial models derived from them, and which points to a problem in the accepted picture of what constitutes thermal fluctuation effects at wetting transitions. This stems from the assumption that, since wetting transitions occur away from the bulk critical temperature, bulk-like fluctuations are unimportant, and that therefore, it is only the thermal wandering of the interface that leads to non-classical (non-MF) exponents. This, for example, is inherent in analyses of the Ginzburg criterion determining the upper critical dimension for wetting transitions. It is also explicit in attempts to systematically derive the interfacial Hamiltonian for short-ranged wetting  in which is identified as a constrained minimum of a more microscopic Landau-Ginzburg-Wilson (LGW) Hamiltonian equivalent to a saddle-point or MF treatment of the trace over microscopic degrees of freedom.

 However, this assumption contains a potential inconsistency since, having identified the interface model in this way, it is necessary to, ad hoc, replace the MF values for the tension/stiffness and bulk correlation length with their true equilibrium values, pertinent to the underlying microscopic model, and to, for example, correctly identify the appropriate value of the wetting parameter determining the non-universality. These quantities of course, take renormalized values precisely because of bulk-like fluctuations. This is an admission that bulk-fluctuations do matter, albeit that they can hopefully be allowed for by a sleight of hand replacement of the MF values of $\Sigma$ and $\xi$ with their true values. While this replacement is completely necessary - indeed, sum-rules demand that it is the equilibrium tension/stiffness that appears in the interfacial Hamiltonian - there is another aspect of bulk-like fluctuation effects that has not been considered and goes hand in hand with the renormalization of the surface tension and correlation length; namely, if bulk fluctuations renormalize the tension, then they also alter the form of, and add to, the wall - interface interaction potential, separately to considerations of interfacial fluctuations.

This can be seen in another way: in ignoring bulk-like fluctuations and identifying the interfacial Hamiltonian via a constrained minimization of the LGW model, one is also assuming that there is a unique one-to-one map between a given interfacial configuration and an underlying microscopic one. This is obviously not the case; there is a multiplicity of microscopic states, all arising from bulk-like fluctuations about the profile corresponding to the constrained minimum configuration, that map onto a given interfacial configuration. This feature has been recognized as being important in molecular descriptions of free interfaces in attempts to understand the rigidity or bending coefficient but its effect on the wall-interface interaction for wetting has not been considered \cite{Pedro_1, Pedro_2, Pedro_3,MCT}. This means that all binding potentials, whether functions or functionals, constructed previously have missed an entropic contribution, arising from the many-to-one map from microscopic configurations onto interfacial ones, which is generated when one integrates out the degrees of freedom. This entropic contribution is akin to a thermal Casimir effect - the fluctuation-induced force for a fluid confined between two substrates which arises from the restriction of bulk-fluctuations \cite{FdG, BCN, ES, HGDB}. This is present even when the average value of a microscopic field/order-parameter is zero; for example, in an Ising parallel plate geometry, at and above the critical temperature when there are no symmetry breaking surface fields. In this case, even though the equilibrium order-parameter profile across the geometry is zero, unperturbed from the bulk value, correlations are modified and there is still a finite-size contribution to the free-energy arising from fluctuations of the field which is not captured by MF treatments. The Casimir force is long ranged at the critical point, but it is always present, even away from $T_c$, where it decays exponentially on a scale set by the bulk correlation length. Within the description of short-ranged wetting we should therefore anticipate that there is an additional entropic or  low temperature Casimir contribution to the binding potential which competes with the established MF contribution, which has been overlooked in previous analyses.

In this paper, we determine this contribution using the recently developed non-local, diagrammatic, formalism for short-ranged wetting. Our central result is that the Casimir contribution to the binding potential functional, which includes the dependence on the shape of the interface and wall, can be represented by the diagram
    \begin{equation}
\beta W_{\rm C}[\ell,\psi] \approx \frac{1}{2}\;\; \fig{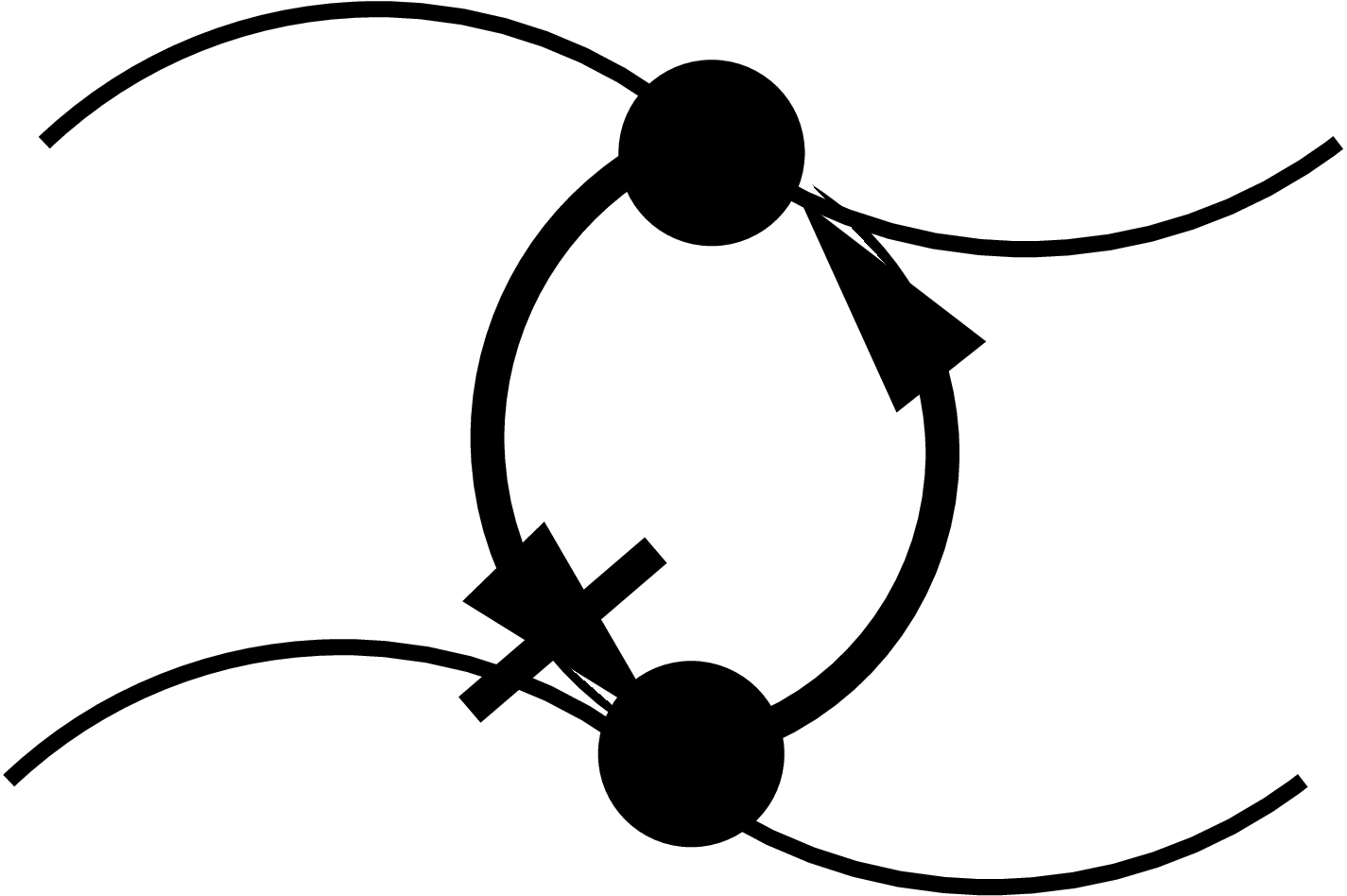}
\label{intro1}
\end{equation}
where $\beta=1/k_B T$, the upper and lower lines represent the interface and wall respectively, and the two connecting lines are Kernels arising from modified bulk and surface correlation functions. For planar interfacial and wall configurations, that is when the wetting film is of uniform thickness, $\ell$, this reduces to
\begin{equation}
\beta w_{C}(\ell) \approx \frac{ e^{-2\kappa\ell} }{ 32\pi\ell^{2} } \bigl[ 2(\kappa+g)\ell + 1 \bigr] \, ,
\label{intro2}
\end{equation}
determining the asymptotic decay of the Casimir contribution to the binding potential function. Here $\kappa=1/\xi$ while $g$ is the surface enhancement at the wall. The Casimir term must be added to the usual MF contributions to the binding potential in order to correctly determine critical singularities. We show how this forces a reappraisal of the accuracy of MF theory and previous RG predictions changing even the value of critical exponents at tricritical wetting and observables at first-order transitions. Readers uninterested in the technical details of the derivation of the Casimir contribution may skip to Section V where we discuss its impact on short-ranged wetting transitions. A preliminary account of some of our findings has appeared in \cite{SREP_2022,SREP2023}. 

\section{Wetting transitions and the critical wetting controversy}
To continue our introduction, we recall the basic phenomenology of wetting transitions and more details of the critical wetting controversy. The equilibrium contact angle $\theta$, of a macroscopic sessile liquid drop on an ideal planar substrate (wall), satisfies Young's equation of mechanical stability
\begin{equation}
\gamma_{wg}=\gamma_{wl}+\gamma\cos\theta\, ,
\end{equation}
where $\gamma_{wg}$ and $\gamma_{wl}$ are the surface tensions of the wall-gas and wall-liquid interfaces respectively. The modern theory of wetting began in 1977 when pioneering MF density functional studies by Cahn \cite{Cahn} and also Ebner and Saam \cite{EbnerSaam_1977} showed that it was possible that the contact angle could vanish as the temperature is increased to a wetting temperature $T_{w}$, marking the change from partial to complete wetting - for reviews see, for example \cite{gennes_wetting_1985, sullivantelodagama, dietrich_wetting_1988, BEIMR}. 

Clearly, the vanishing of the contact angle is equivalent to the vanishing of a singular contribution to the
 surface free-energy, 
$\gamma_{\rm sing}=\gamma_{wg}-\gamma_{wl}-\gamma$, which
 we can characterize by a surface critical exponent
\begin{equation}
\gamma_{\rm sing}\sim (T_{w}-T)^{2-\alpha_s}\, .
\end{equation}
Viewed in the grand-canonical ensemble, in which the wall is considered to be of infinite area and in contact with a bulk gas, at saturation chemical potential, the wetting transition corresponds to the divergence of the thickness $\langle \ell \rangle$ of an adsorbed layer of liquid which changes from being microscopic to macroscopic at $T_w$. The initial studies of both Cahn and Ebner and Saam \cite{Cahn,EbnerSaam_1977} indicated that the wetting transition was first-order corresponding to a discontinuous jump in the adsorption (with $\alpha_s=1$). However, subsequent mean-field studies by Sullivan \cite{Sullivan} and Nakanishi and Fisher \cite{NF}, both for systems with short-ranged forces, together with Abraham’s exact analysis 
of wetting in the 2D Ising model \cite{Abraham_1980} showed that critical wetting was also possible. In this case the growth
 of the wetting layer thickness, is continuous and characterized by an exponent
\begin{equation}
\langle \ell \rangle \sim (T_w-T)^{-\beta_s}\, .
\end{equation}
Associated with this
 is the continuous divergence of 
a parallel correlation length,
\begin{equation} 
\xi_\parallel \sim (T_w- T)^{-\nu_\parallel}\, ,
\end{equation}
arising from the 
build-up of capillary-wave-like fluctuations,
 due to the interfacial wandering of the liquid-gas interface as it unbinds for the 
wall. These interfacial fluctuations, cause the interfacial 
width, or perpendicular correlation length, $\xi_\perp$, 
to diverge - although this important feature is not captured by
 simple mean-field treatments of wetting. 3D 
is the marginal dimension for interfacial 
roughness (not to be confused with the 
marginal dimension for the wetting transition), 
for which we anticipate that $\xi_\perp$ 
diverges weakly, characterized by the
 famous capillary-wave relation \cite{BLS_1965}
\begin{equation}
\frac{\xi_\perp^2}{\xi^2}=\omega \ln (\xi_\parallel\Lambda)^2\, ,
\end{equation}
where $\Lambda$ is a suitable cut-off of order the inverse of a bulk correlation length. Here, in preparation for our discussion of short-ranged critical wetting, we have introduced the dimensionless wetting parameter
\begin{equation}
\omega = \frac{k_B T}{4\pi\Sigma\xi^2}\, ,
\end{equation}
where $\Sigma$ is the stiffness coefficient, which, 
for fluid interfaces, is identical to the surface 
tension $\gamma$.

The critical wetting exponents are not independent and, for example, satisfy the Rushbrooke-like relation
\begin{equation}
2-\alpha_s=2\nu_\parallel-2\beta_s\, .
\end{equation}
In 2D, where interfacial fluctuations
 are much stronger, and 
$\xi_\perp\sim \xi_\parallel^{1/2}$, the values of 
these exponents are well understood, particularly for systems with 
short-ranged forces, where mesoscopic treatments based on 
random-walk arguments \cite{fisher_walks_1984} and simple interfacial models are in  perfect agreement with Abraham's exact solution of the 
2D Ising model \cite{Abraham_1980} giving $\alpha_s=0$, $\beta_s=1$ and 
$\nu_{\parallel}=2$. In 3D, the fluctuation theory 
of wetting transitions relies, almost exclusively, 
on the use of interfacial models based on a collective co-ordinate 
$\ell({\bf{x}})$ representing the local position
 of the interface above the wall. These are 
generalizations of the Capillary-Wave model of interfacial broadening in a gravitational field \cite{Weeks_1977} and include the 
surface tension $\gamma$, or more generally the 
surface stiffness $\Sigma$ for lattice-based models
 (see, e.g. \cite{MPA_Fisher}), resisting fluctuations 
which increase the interface area, and a binding
 potential $w(\ell)$ to model the interaction with 
the wall which is a function of the film 
thickness. The standard, local interfacial
 Hamiltonian wetting then models the energy cost of long wavelength interfacial configurations as
\begin{equation}
H_I[\ell]=\int d{\bf{x}}\left(\frac{\Sigma}{2}(\nabla\ell)^2+w(\ell)\right)\, ,
\end{equation}
subject to an appropriate momentum cut-off 
$\Lambda \propto 1/\xi$. The binding potential is 
sensitive to the range of 
the intermolecular forces present and is 
constructed from more microscopic models, for 
example a Landau-Ginzburg-Wilson or related 
density functional model which is always treated 
in a mean-field approximation - an assumption 
which we shall return to and question shortly. For 
systems with dispersion forces this task is 
straightforward and amounts to simply integrating 
the intermolecular forces over the volume of the 
wetting layer which can be treated reliably as a 
structureless slab of liquid. In this case, 
interfacial fluctuation effects are not 
particularly important and the MF prediction 
$\beta_s=1$ for critical wetting has been 
confirmed convincingly in experimental studies for 
binary liquid mixtures \cite{Ragil}.

However, much greater care is required for systems
 with short-ranged forces, where the binding 
potential itself arises from bulk-like 
fluctuations in the wetting layer and decays 
exponentially, like the density profile itself, on 
the scale of the bulk correlation length of the 
adsorbed (liquid) phase $\xi$. In this case, for critical
 wetting, the
 binding potential was taken originally
 to be of the form
\begin{equation}
w(\ell)= ae^{-\kappa\ell}+be^{-2\kappa\ell}+\cdots\, ,
\end{equation}
together with a hard-wall condition that $\ell>0$. 
Here, $\kappa=1/\xi$, while the coefficient $a$ changes
 sign at the MF value of the critical
 wetting temperature and $b>0$ (see later). The above binding potential
 $w(\ell)$ encodes the underlying MF behavior
 corresponding to the critical exponents $\alpha_s=0$, $\beta_s=0(\ln)$
 and, most notably, $\nu_\parallel=1$. However, a Ginzburg 
criterion indicates that 3D is the upper critical 
dimension for short-ranged critical wetting 
suggesting that these exponents may be altered
 by interfacial fluctuations which are also marginal. 
This is indeed the case and RG calculations of the interfacial model, due to Brezin, Halperin and Leibler \cite{Brezin} and later Fisher and Huse \cite{FisherHuse}, predict very strong non-universal critical
 singularities controlled by the wetting parameter $\omega$, the value of which is temperature dependent. In particular, the correlation critical exponent, $\nu_\parallel$ is predicted to fall into one of three regimes: For $0<\omega<\frac{1}{2}$, an exact linear RG
 treatment of the interfacial model predicts that
\begin{equation}
\nu_\parallel=\frac{1}{1-\omega}\, ,
\end{equation}
which consistently recovers the MF result 
$\nu_\parallel=1$ on setting $\omega=0$, corresponding 
to an infinitely stiff interface. However, for 
$\frac{1}{2}<\omega<2$, the exponent behaves as
\begin{equation}
\nu_\parallel=\frac{1}{(\sqrt{2}-\sqrt{\omega})^2 }\, ,
\end{equation}
which departs very substantially from the 
MF prediction. 
For
$\omega>2$, the correlation length grows 
exponentially rather as a power-law.

Unfortunately, these very striking predictions are not supported by extremely careful Monte Carlo simulation studies of wetting in the 3D simple cubic Ising model due to Binder and co-workers \cite{Bryk_Binder_2013, Bryk_Terzyk_2013}. The Ising model is the simplest microscopic model that may be used to test the predictions of the effective Hamiltonian theory, for which, additionally, the temperature dependence of the wetting parameter is known quite accurately (although its value was initially overestimated). For the simple cubic Ising model, the wetting parameter takes the value $\omega\approx 0.5$ near the roughening temperature $T_R\approx 0.54T_c$, and increases rapidly remaining near constant and approaching a universal value $\omega\approx 0.8$
 near the bulk critical temperature $T_c$ \cite{EHP_1992, FW_1992}. This suggests that very substantial departures from MF prediction $\nu_\parallel=1$ should be observable for critical wetting. On the positive side, the simulation studies, for different surface fields and surface coupling parameters, confirm the expected form of the surface phase diagram, as predicted by Nakanishi and Fisher \cite{NF}. This shows lines of critical wetting and first-order wetting which meet at a tricritical point. However, a detailed study of thermodynamic observables, and in particular the growth of the surface susceptibility, for critical wetting transitions occurring at different temperatures do not reveal the predicted strong non-universal critical singularities. Originally the observed singularities appeared to be much closer to the predictions of MF theory although small deviations from classical behavior were observable. The interpretation of this discrepancy has now been sharpened and subsequent Ising model simulations, performed on larger systems and using more sophisticated finite-size scaling analyses, have significantly improved estimates of both the wetting temperature and the critical exponents \cite{Bryk_Binder_2013}. For example, for the critical wetting transition occurring at $T_w=0.9 T_c$, for which $\omega\approx 0.8$ and RG theory predicts $\nu_\parallel\approx 3.7$, the simulations measure that $\nu_\parallel\approx 2$. While for critical wetting occurring at the lower temperature $T_w=0.63 T_c$ for which $\omega\approx 0.7$ and RG theory predicts $\nu_\parallel\approx 3.0$, the simulations measure that $\nu_\parallel\approx 1.9$. These indicate that non-classical exponents are present but not in any way to the degree predicted.
 
\section{Mean-field contribution to the binding functional}
In this Section, we revisit how the binding potential may be calculated from first-principles by using the non-local formalism and evaluated for specific interface and wall configurations. We divide our presentation into several sub-sections. These provide the necessary background material and summarize the main results and technical challenges beginning with the familiar LGW model and MF theory of short-ranged wetting.

\subsection{The LGW Hamiltonian}
The starting point for our analysis is the standard LGW Hamiltonian for adsorption at a wall based on a scalar magnetization-like order parameter
\begin{equation}
H_{\rm LGW}[m] = \int_{\mathcal{V}} d\mathbf{r} \left(\frac{1}{2} (\boldsymbol\nabla m)^2 +\phi(m)\right) + \int_{S_1} d\mathbf{s}  \phi_1 (m(\mathbf{s})) \, .
\label{HLGW}
\end{equation}
Here $\phi(m)$ is a double well potential modeling bulk phase coexistence below a critical temperature $T_c$. We shall assume an Ising symmetry and denote the spontaneous magnetization $m_0$ and inverse of the bulk correlation length $\kappa = 1/\xi$. The explicit calculations presented here are done for the reliable double parabola potential, which in zero bulk field reads
\begin{equation}
\phi(m)=\frac{\kappa^2}{2}(|m|-m)^2 \, .
\label{DP}
\end{equation}
We consider a semi-infinite geometry where the bounding wall $\mathcal{S}_1$ has a (smooth) shape described -- within Monge's parametrization --  by a height function $\psi(\bf{x})$ with local area element $d{\bf{s}}=\sqrt{1+(\nabla \psi)^2}d\bf{x}$ - although in most of our applications we will consider the wall planar. We denote the volume of the region above the wall $\mathcal{V}$. The surface potential couples to the local surface magnetization, which we devote $m(\bf{s})$, via
\begin{equation}
\phi_1(m) = - \frac{g}{2}m^2-h_{1}m \, .
\end{equation}
where $g$ is the surface enhancement and $ h_{1}$ is the surface field. Alternatively, this can be written
\begin{equation}
\phi_{1}(m) = - \frac{g}{2}(m_{s}-m)^2 \, ,
\end{equation}
where $m_{s} = - gh_{1}$ is the preferred magnetization at the wall. We suppose that $m_{s}$ is positive and that the bulk magnetization is $-m_{0}$ (corresponding to bulk field $h={0}^{-}$) so that the wetting film is of net positive magnetization. Using the LGW Hamiltonian, and in particular within the above DP approximation, our central task here is to evaluate, or best approximate, the partition function
\begin{equation}
Z=\int Dm e^{-\beta H_{\rm LGW}[m]} \, ,
\end{equation}
corresponding to the usual functional integral over all magnetization configurations weighted by the usual Boltzmann factor.

\subsection{MF theory and the Nakanishi-Fisher phase diagram}
\begin{figure}[tbp!]
\includegraphics[width=8.6cm]{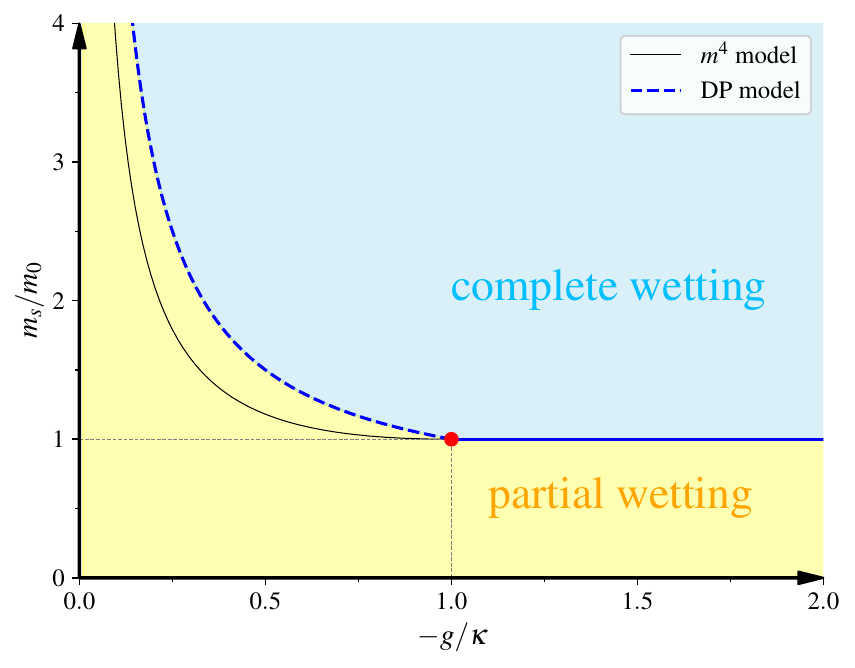}
\caption{Mean-field wetting phase diagram of the LGW model for a flat wall, as a function of the enhancement $g$ and the surface preferred magnetization $m_s$. The continuous blue line corresponds to critical wetting transition line, and the dashed blue line to the first-order wetting transition within the DP model. The red bullet locates the conditions for tricritical wetting. For comparison, the branch for the full $m^4$ model is shown with a thin black line.}
\label{NF_phase_diagram}
\end{figure}

In its simplest formulation, MF theory corresponds to neglecting all configurations except the one that minimizes the Hamiltonian and therefore gives the greatest Boltzmann weight in the partition function. Hence, the MF magnetization profile is found from the solution of the unconstrained minimization
\begin{equation}
\frac{\delta H_{\rm LGW}[m]}{\delta m}=0 \, .
\end{equation}
For a planar wall ($\psi=0$), situated in the $z=0$ plane, the equilibrium MF profile, $m(z)$, then follows from the Euler-Lagrange equation
\begin{equation}
m^{\prime\prime}(z)=\phi^{\prime}(m)
\end{equation}
which has a well-known first-integral
\begin{equation}
m^{\prime}(z)=-\sqrt{\phi(m)-\phi(-m_0)}
\end{equation}
the latter is solved with the boundary condition $-gm^{\prime}(0)=h_1$. This has an elegant graphical interpretation for the profile $m(z)$, valid for quite general potentials $\phi(m)$, first described by Cahn \cite{Cahn} and later used by Nakanishi and Fisher \cite{NF} to determine the global surface phase diagram - see Fig. \ref{NF_phase_diagram}. For negative surface enhancement this shows a line of critical wetting transitions occurring along $ h_1= - g m_0$, which is equivalent to $ m_{s} = m_{0}$, when $ -g > \kappa$ and a line of first-order wetting transitions when $ -g < \kappa$. A tricritical wetting transition occurs exactly when $g=-\kappa$ and $ m_{s} = m_{0}$. 
For completion we mention that the complete wetting transition refers to divergence of $\langle \ell \rangle$ and $\xi_\parallel$ as $h \to 0$ for $T>T_w$, although this transition, which is much easier to understand, will not be our concern here.

The logarithmic divergence of the film thickness at critical and tricritical wetting follows directly from the MF profile $m(z)$. Similarly, the solution of the Ornstein-Zernike equation for the magnetization correlation function, under the assumption that there are only Gaussian fluctuations about the MF equilibrium profile, determines directly that the parallel correlation diverges with exponent $\nu_{\parallel} =1$, for critical wetting, and with exponent $\nu_{\parallel}=3/4$ for tricritical wetting. A Ginzburg criterion confirms that these MF predictions are invalidated in dimension $d<3$ by the Gaussian soft modes associated with the translations and fluctuations of the interface which has always been interpreted to mean that 3D is the upper critical dimension. As we shall show, this conclusion remains true for critical wetting (and for complete wetting) but is incorrect for tricritical wetting since other fluctuations also invalidate the MF predictions, even for $d>3$, when the interfacial modes are irrelevant.

\subsection{The Constrained Trace definition of $H_I[\ell]$}
The MF results were used by a number of authors to construct the binding potential $w(\ell)$ appearing in an interfacial model, concluding that it was a sum of exponentially decaying terms and identifying the coefficients which are related to the MF phase boundaries. For example, Lipowsky, Kroll and Zia \cite{LKZ_1983} determined the interfacial model $H_I[\ell]$ directly from the LGW Hamiltonian by evaluating it for profiles which are local translations of the MF solution $m(z)$ controlled by the interface position $\ell(\bf{x})$. This, of course, already assumes that non-classical critical behavior at wetting transitions only emerges from the translations and long-wavelength fluctuations of the interface. This assumption is made even more explicit in the later study of Fisher and Jin \cite{FJ_1991}, who sought to systematically derive the interfacial model from the LGW Hamiltonian. To this end, we suppose that the partition function is evaluated in two steps: first, a constrained trace over all magnetization configurations that correspond to a given interfacial one and then the trace over all interfacial configurations. That is,
\begin{equation}
Z=\int D\ell \int Dm^{\prime} e^{-\beta H_{\rm LGW}[m]} \, ,
\end{equation}
where the prime on the measure denotes the constrained class of microscopic profiles. This separation into constrained microscopic and interfacial configurations obviously preserves the partition function and clarifies that the interfacial Hamiltonian should be identified via the partial partition function
\begin{equation}
\label{03062022_1604}
e^{-\beta H_I[\ell]-\beta \gamma_{wl}\mathcal{A}_\psi}=\int Dm^{\prime} e^{-\beta H_{\rm LGW}[m]} \, ,
\end{equation}
where $\gamma_{wl}$ is the free energy per unit area of a non-interfacial configuration of a bulk $+m_0$, i.e. liquid state, in contact with the substrate $S_1$ of area $\mathcal{A}_\psi$.
The interface position must also be defined in a suitable way, for example, using a simple crossing criterion \cite{FJ_1991}, which is the one we adopt here. This asserts that the magnetization vanishes for position vectors $\bf{s}_{\ell}=(x,\ell(\bf{x}))$ along the interface $\mathcal{S}_2$. That is, we impose a constraint on all configurations in the partial partition function that
\begin{equation}
m({\bf{s}}_{\ell})=0 \, .
\end{equation}
We denote the region above the interface by $\mathcal{V}_{-}$, and the region between wall and interface by $\mathcal{V}_{+}$, corresponding to the wetting layer. Regardless of the definition of the interface we anticipate that, for isotropic fluid interfaces, the interfacial Hamiltonian identified in this way takes the form \begin{equation}
H_I[\ell]=\gamma\mathcal{A}_\ell + W[\ell,\psi]
\end{equation}
where the first term is the equilibrium surface tension times the interfacial area describing the free interface (ignoring higher-order corrections related to the curvature) and $W[\ell,\psi]$ is the binding potential functional describing the interaction with the wall. The evaluation of this term the focus of this paper. For near planar interfaces it is legitimate to approximate the Hamiltonian for the free interface, by $\frac{\gamma}{2} \int d \textbf{x} (\nabla \ell)^2$. For anisotropic systems, for example those based on lattice models, it is also necessary to replace the tension with the stiffness coefficient $\Sigma$. For a planar wall ($\psi=0$) of area $L_{\parallel}^{2}$ and a wetting layer of uniform thickness $\ell$ the binding potential functional reduces to the usual binding potential function
\begin{equation}
w(\ell) = \frac{W[\ell,\psi]}{L_\parallel^2}
\end{equation}
which appears in local interfacial Hamiltonians and whose derivative is related to the older concept of a disjoining pressure \cite{Derjaguin}.

\subsection{The Saddle point approximation: local and non-local models}
\begin{figure}[tbp!]
\includegraphics[width=8.6cm]{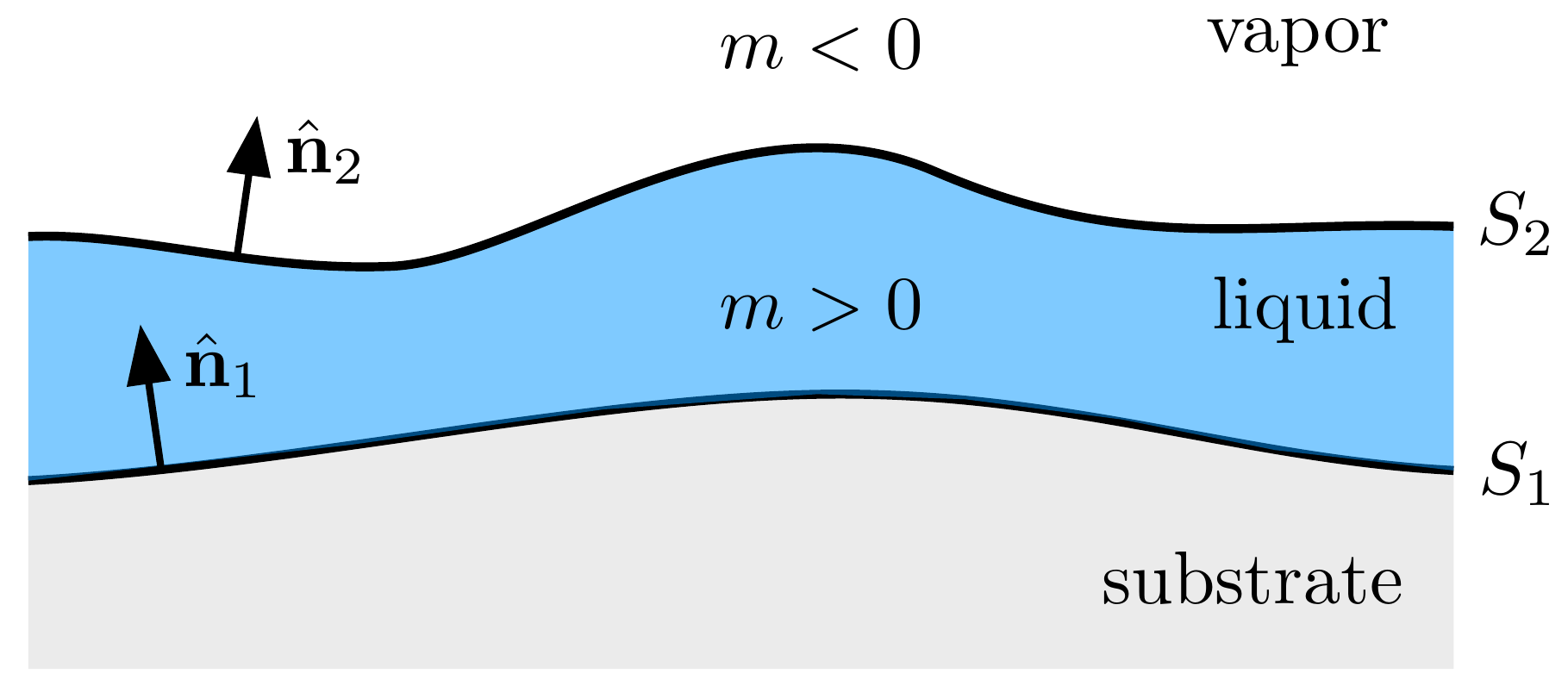}
\caption{Schematic illustration of a nonplanar interfacial configuration (top line) for a constrained wetting film of liquid at a nonplanar wall (bottom line). Conventions for the surface normals are shown.}
\label{fig_geometry}
\end{figure}

To evaluate the constrained trace, it is now customary to ignore all bulk fluctuations and make a saddle point or MF-like approximation which identifies
\begin{equation}
H_I[\ell] = H_{\rm{LGW}}[m_\Xi]-\gamma_{wl}\mathcal{A}_\psi
\end{equation}
where $m_{\Xi}(\textbf{r})$ is the \emph{unique} profile that minimizes the LGW Hamiltonian subject to the crossing criterion. This is the explicit one-to-one map from microscopic to interfacial states which clearly ignores any entropic contribution to $H_{I}$ arising from all the other microscopic configurations that also respect the crossing criterion. Within the DP model, and for a general non-planar wall, the constrained profile, which we will only need to specify within the region $\mathcal{V}_{+}$ of the wetting layer, follows from solution of the Helmholtz equation
\begin{equation}
\label{03062022_1408}
\nabla^{2} m_\Xi =\kappa^{2}( m_{\Xi} - m_{0})
\end{equation}
together with the boundary conditions
\begin{eqnarray}
\label{03062022_1409a}
&& m_{\Xi}(\textbf{r}) = 0 \, , \qquad \textbf{r} \in S_2 \\
\label{03062022_1409b}
&& \mathbf{n}_{1} \cdot \boldsymbol\nabla m_{\Xi} = -g (m_{\Xi}-m_{1})  \, , \qquad \textbf{r} \in S_1 \, ,
\end{eqnarray}
where $\mathbf{n}_1$ is the normal at the wall, as shown in Fig.~\ref{fig_geometry}. In the region occupied by the vapor instead (\ref{03062022_1408}) is replaced by
\begin{equation}
\label{03062022_1410}
\nabla^{2} m_\Xi =\kappa^{2}( m_{\Xi} + m_{0}) \, ,
\end{equation}
which is subject to the crossing criterion, Eq. (\ref{03062022_1409a}), and the asymptotic requirement $m_{\Xi} \rightarrow -m_{0}$ as $z \rightarrow + \infty$.

The constrained profile is a functional of the interface and wall shape. For a planar-wall and a wetting layer of uniform thickness $\ell$, the constrained minimization identifies the binding potential function as \cite{PRBRE_2006}
\begin{equation}
\label{ }
w_{\rm MF}(\ell) = \frac{H_{\rm {LGW}}[m_{\pi}]}{L_\parallel^2}-\gamma-\gamma_{wl} \, ,
\end{equation}
where $L_\parallel^2=\mathcal{A}_\psi=\mathcal{A}_\ell$, is the planar area. Here $m_{\pi}(z;\ell)$ is the planar constrained profile satisfying
\begin{equation}
\label{ }
\partial^{2}_{z} m_{\pi} = \phi^{\prime}(m)
\end{equation}
together with the boundary condition $\partial_{z} m_{\pi}(0;\ell) = \phi^{\prime}(m_ \pi(0;\ell))$ and crossing criterion $m_\pi(\ell;\ell)=0$. We have put a subscript MF on the binding potential to emphasize that this is the contribution determined by the saddle point approximation. Within the DP model this can determined exactly as
\begin{equation}
w_{\rm MF}(\ell) =a e^{-\kappa\ell} + b e^{-2\kappa\ell} + \cdots
\end{equation}
with coefficients
\begin{equation}
a = - \gamma \frac{2 gt}{g-\kappa} \, ,\hspace{1cm} b= \gamma \frac{g+\kappa}{g-\kappa} \, ,
\end{equation}
where $\gamma=\kappa^2 m_0$ is the DP result for the surface tension and $t=(m_{0}-m_{s})/m_{0}$ \cite{PRBRE_2006}. Constructed in this way the binding potential obviously encodes the MF wetting behavior with its minimum occurring at the equilibrium film thickness while its curvature at this point determines $\xi_{\parallel} =\sqrt{\gamma/w^{\prime\prime}_{\rm{MF}}(\ell)}$. The coefficient $b$ is positive for critical wetting in which case $t$ is the scaling field, which vanishes at the phase boundary. For tricritical wetting, $b=0$, while for first-order wetting $b<0$ and it is necessary to include the next order term in the expansion, which in general, is of order $e^{-3\kappa\ell}$. In their derivation of the effective Hamiltonian for wetting at planar wall, Fisher, Jin and Parry \cite{FJP_1994} note that the ansatz $m_{\Xi} \approx m_\pi(z;\ell(\bf{x}))$, corresponding to a local shift of the planar constrained profile, approximately solves the Helmholtz equations for small interfacial gradients. This leads to the identification
\begin{equation}
H_I[\ell]=H_{\rm LGW}[m_\pi(z;\ell(\mathbf{x}))]-\gamma_{wl}L_\parallel^2
\end{equation}
which is very similar to the original approach of Lipowksy Kroll and Zia \cite{LKZ_1983} but who didn't emphasize the aspects arising from the constraints. For near planar interfacial configurations this leads to the local interfacial model
\begin{equation}
\label{03062022_1421}
H_I[\ell] = \int d{\bf{x}} \biggl[ \frac{\Sigma(\ell)}{2} (\nabla \ell)^{2} + w_{MF}(\ell) \biggr]
\end{equation}
where
\begin{equation}
\Sigma(\ell) = \Sigma+a e^{-\kappa\ell}-2b \kappa\ell e^{-2\kappa\ell}+\cdots
\end{equation}
is an explicit position dependence appearing in the stiffness coefficient and we have ignored the constant contribution, $\gamma L_\parallel^2$, to $H_I[\ell]$. When the position dependence in $\Sigma(\ell)$ is neglected, the local Hamiltonian reduces to the original model used in the RG analysis of Brezin, Halperin and Leibler \cite{BHL_1983} and Fisher and Huse \cite{FH_1985} leading to the predictions of strong non-universality for 3D critical wetting.

Despite its systematic derivation, the effective Hamiltonian (\ref{03062022_1421}) is fundamentally unsatisfactory. When fluctuations are included, using a RG treatment, the coupling between the flows of the binding potential and position dependent stiffness leads, inevitably, to an instability which reverses the order of the wetting transitions in the Nakanishi-Fisher phase diagram \cite{FJ_1992,JF_1993_prb}. This prediction is certainly incorrect. But it is also unsatisfactory for other reasons. For example, it gives no indication as to the physical origin of the different exponential terms in $W(\ell)$ and $\Delta\Sigma(\ell) = \Sigma(\ell) - \Sigma$ and how the coefficients are related, hinting that there is something more unifying. Moreover, the model cannot be applied to wetting in relatively simple geometries - for example, in an acute wedge, since one cannot consistently model the wall-interface binding near the wedge vertex and far from it \cite{PREL_2004}.

These problems are overcome within the non-local description which can be derived from the Helmholtz equations using a rigorous boundary integral technique \cite{RESPG_2018}. The proof of this is outlined below. First we note that within the DP approximation a simple application of Green's theorem identifies the constrained Hamiltonian $H_{LGW}[m_{\Xi}]$ as surface integrals of the normal derivatives of the constrained magnetization evaluated over the interface, $S_2$, and wall, $S_1$, surfaces \cite{PRBRE_2006, PRBRE_2007, PRBRE_2008_prl, PREBR_2008, BPRRE_2009, PRBRE_2008}. These integrals can be expressed in terms of a rescaled bulk Green function, which is the solution of the equation $(-\nabla_{\mathbf{s}}^2+\kappa^2)K(\textbf{s},\textbf{s}^{\prime})=2\kappa\delta(\mathbf{s}-\mathbf{s}^\prime)$, i.e.
\begin{equation}
\label{03062022_1616-0}
K(\textbf{s},\textbf{s}^{\prime}) = 2 \left(\frac{\kappa}{2\pi}\right)^{\frac{d}{2}} \frac{K_{\frac{d-2}{2}}\left(\kappa|\textbf{s}-\textbf{s}^{\prime}|\right)}{|\textbf{s}-\textbf{s}^{\prime}|^{\frac{d-2}{2}}}\, ,
\end{equation}
where $d$ is the spatial dimensionality and $K_\nu(x)$ is the modified Bessel function of second kind. For 3D it reduces to
\begin{equation}
\label{03062022_1616}
K(\textbf{s},\textbf{s}^{\prime}) = \frac{\kappa}{2\pi} \frac{\textrm{e}^{-\kappa|\textbf{s}-\textbf{s}^{\prime}|}}{|\textbf{s}-\textbf{s}^{\prime}|}
\end{equation}
where $\textbf{s}$ and $\textbf{s}^{\prime}$ will only be evaluated at either the interface or the wall. We introduce the convenient short-hand notation
\begin{equation}
\label{20102022_2058}
K(\mathbf{s},\mathbf{s}')\equiv \fig{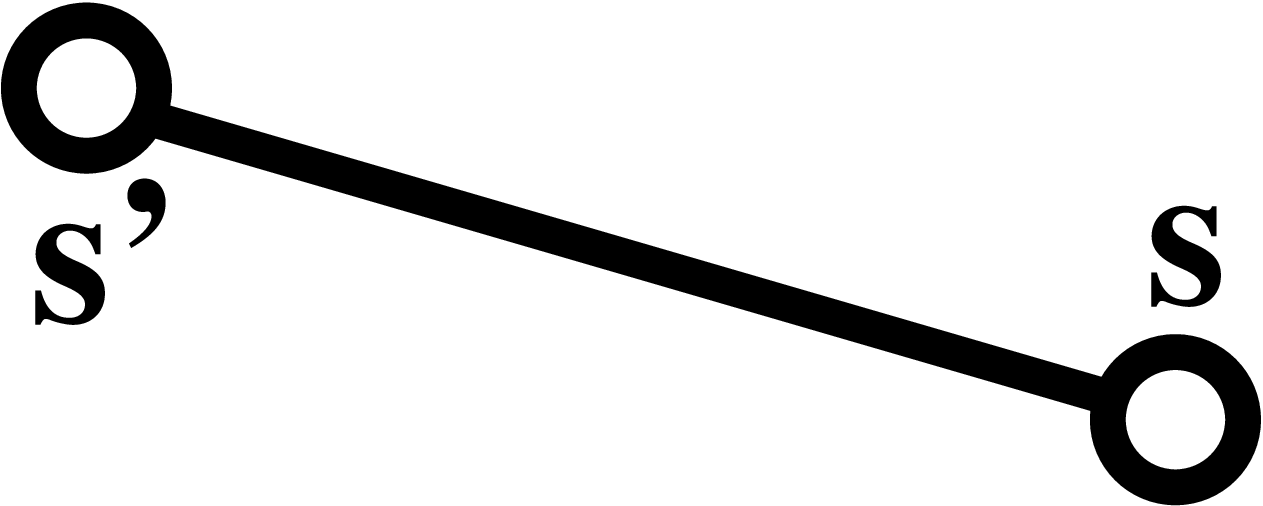} \, .
\end{equation}
The binding potential $W_{\rm {MF}}[\ell,\psi]$ can be written as an exact perturbation series of boundary integral convolutions involving diagrams (\ref{20102022_2058}) that connect the interface and the wall, and related diagrams associated to two other kernels acting either on $S_1$ or $S_2$:
\begin{equation}
\begin{aligned}
\label{20102022_2059}
U(\textbf{s},\textbf{s}^{\prime}) & = K(\textbf{s},\textbf{s}^{\prime}) - \delta(\textbf{s},\textbf{s}^{\prime}) \\
& \equiv \quad \fig{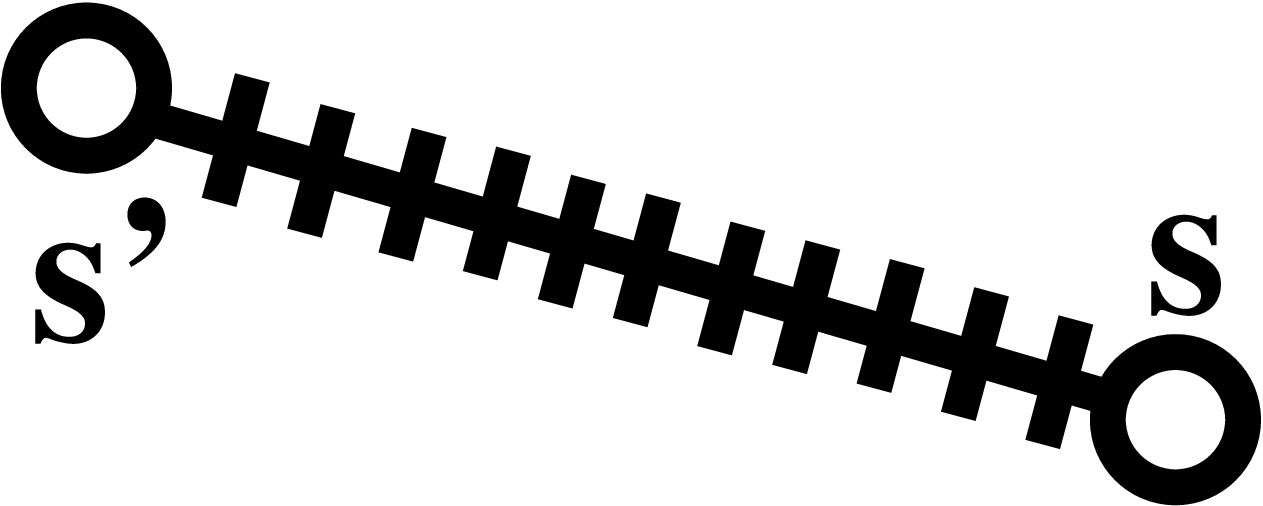}
\end{aligned}
\end{equation}
and
\begin{equation}
\label{defdiagrams}
\frac{1}{\kappa}\mathbf{n}_{1}(\mathbf{s})\cdot\boldsymbol\nabla_{\mathbf{s}} K(\mathbf{s},\mathbf{s}') 
\equiv \quad \fig{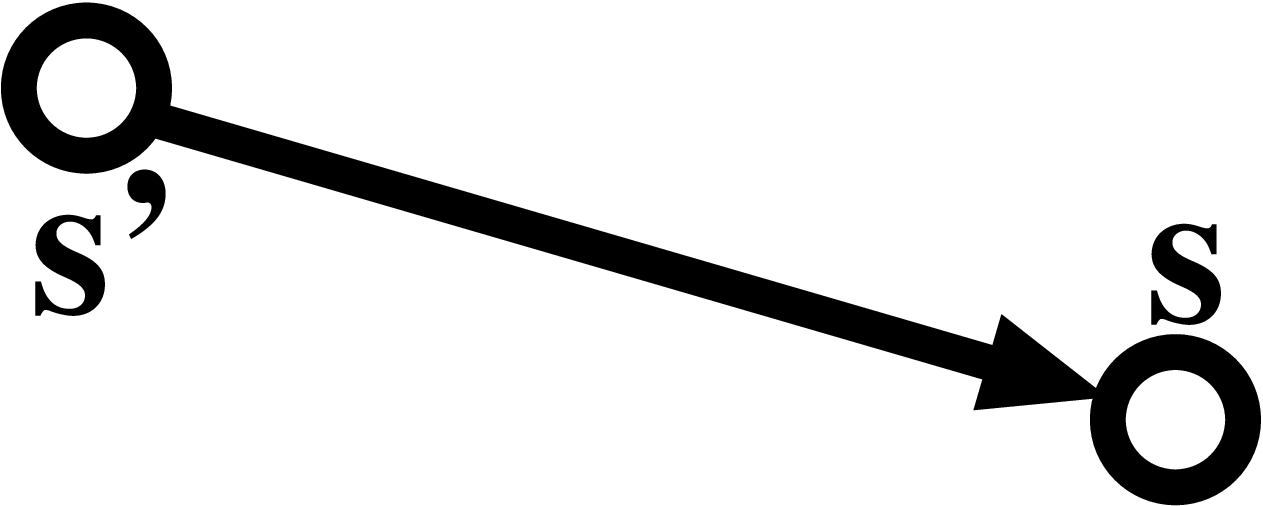}\, .
\end{equation}
These latter diagrams fully  incorporate the functional dependence on the shape of the interface and the wall, including curvature corrections \cite{RESPG_2018}. However, if one focuses only on near planar and parallel interfacial and wall configurations the leading two terms in the binding potential functional can be written just as
\begin{equation}
W_{\rm MF}[\ell,\psi] = a \Omega_{1}^{1} + b \Omega_{1}^{2}
\label{WMFfunctional}
\end{equation}
where
\begin{equation}
\Omega_{1}^{1} = \fig{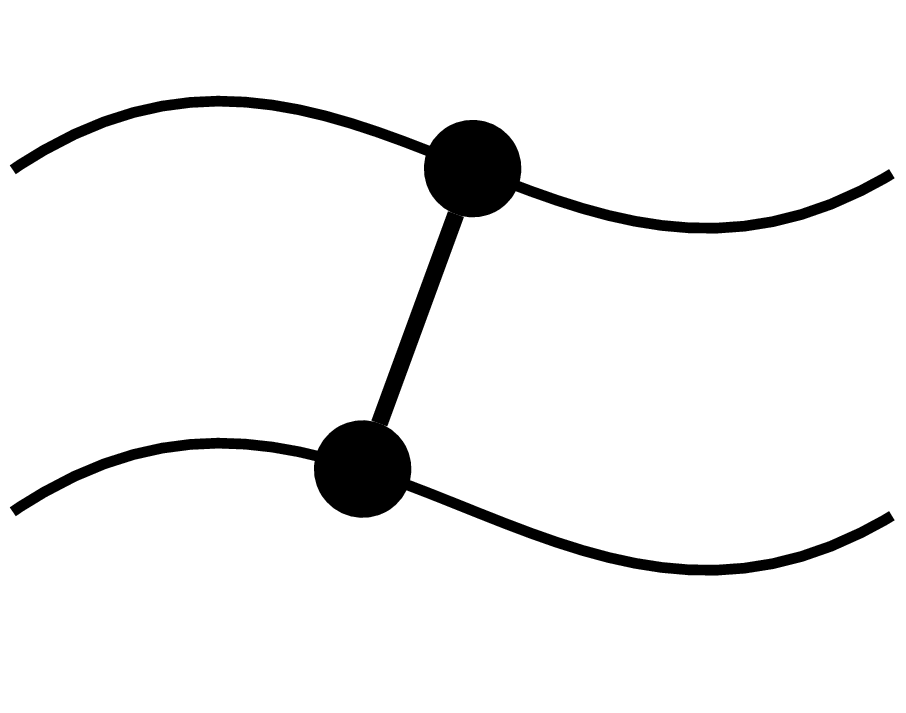}
\end{equation}
and 
\begin{equation}
\Omega_{1}^{2} = \fig{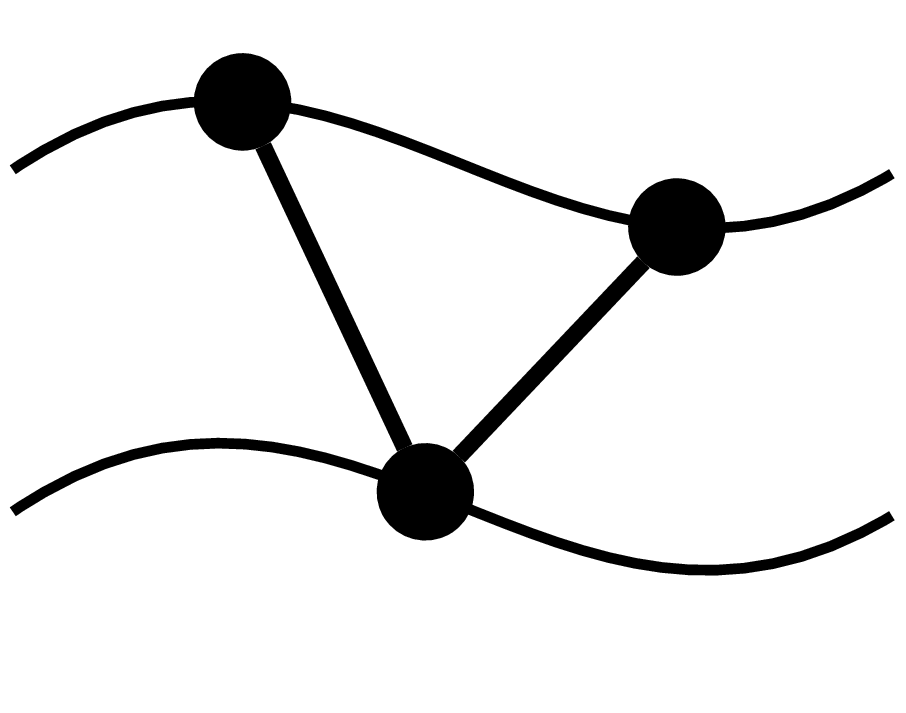}
\end{equation}
In these diagrams the upper and lower wavy lines represent the interface and wall, respectively, while the black dot denotes integration over the surface with the appropriate area element. For wetting on a planar wall, these diagrams simplify considerably. The leading term is local
\begin{equation}
\begin{aligned}
\label{omega11}
\Omega_{1}^{1} & = \fig{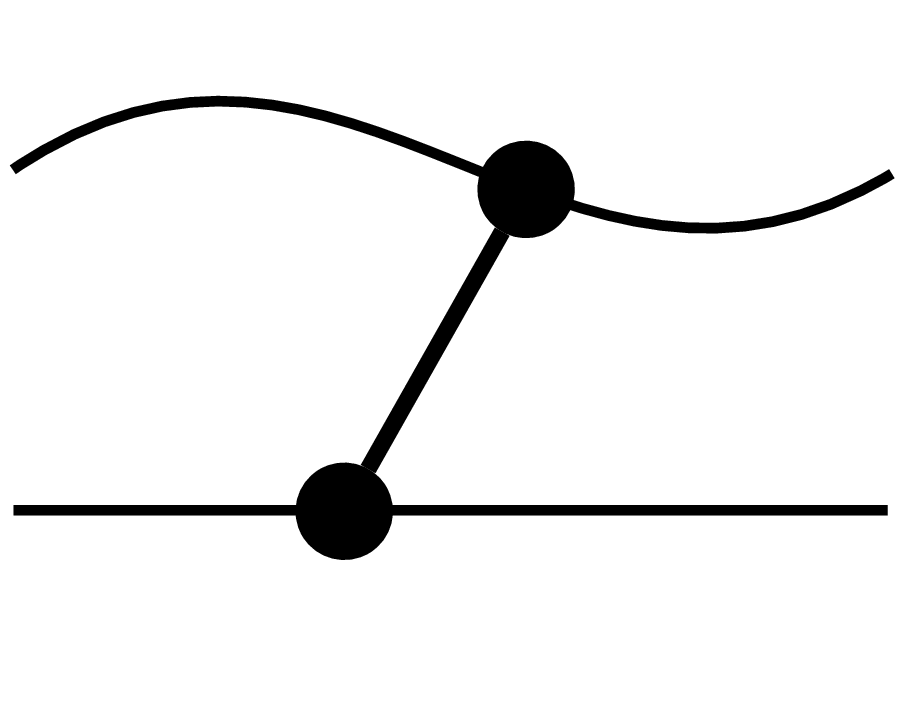} \\
& = \int d\textbf{s} \, e^{-\kappa\ell}
\end{aligned}
\end{equation}
where $d\textbf{s} = \sqrt{1+(\nabla \ell)^2} d\textbf{x}$ and explains why the leading order exponentials in $W$ and $\Delta \Sigma$ are the same. However, the term representing the repulsion, for critical wetting remains non-local reducing to the two-body interfacial interaction 
\begin{equation}
\begin{aligned}
\label{omega12}
\Omega_{1}^{2} & = \fig{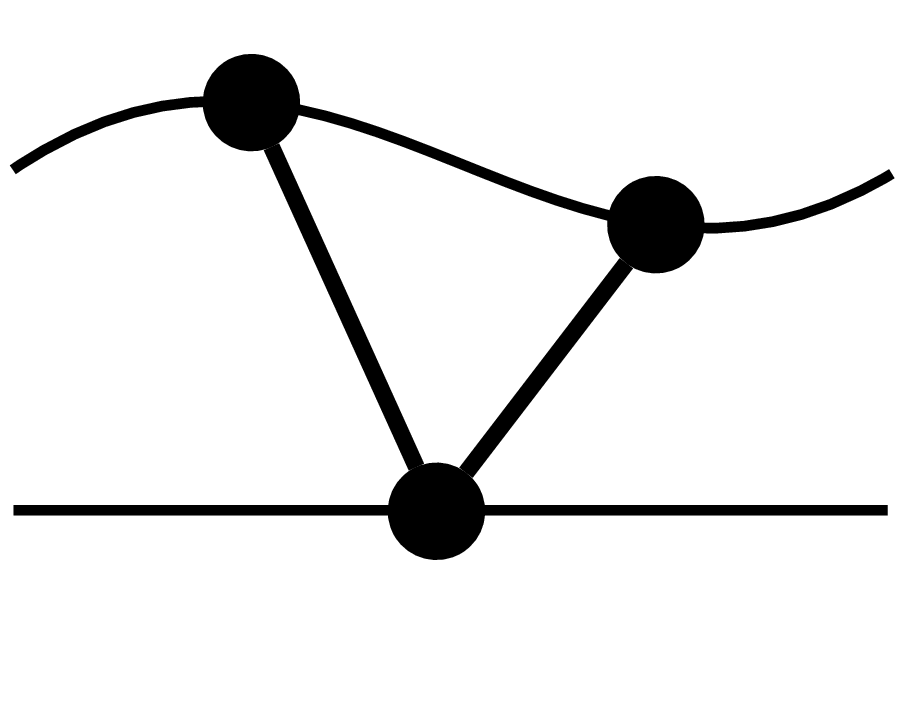} \\
& = \iint d\textbf{s}_{1} d\textbf{s}_{2} \, S(\textbf{x}_{12}; \bar{\ell})
\end{aligned}
\end{equation}
where $\bar{\ell} = [\ell(\textbf{x}_{1}) + \ell(\textbf{x}_{2})]/2$ and
\begin{equation}
S(\textbf{x}_{12}; \bar{\ell}) = \frac{\kappa^2}{2 \pi} \int_{2 \kappa \ell}^{\infty} d\tau \frac{e^{-\sqrt{\tau^{2} + \kappa^{2}\ell^{2}}}}{\sqrt{\tau^{2} + \kappa^{2}\ell^{2}}} \, .
\end{equation}
For small interfacial gradients this recovers the Fisher-Jin Hamiltonian \cite{FJ_1991, FJ_1992, JF_1993_prb, JF_1993} but clarifies that the expansion is only valid for Fourier modes with wavevectors $q^2 \ll \kappa/ \ell$. RG and simulation studies show that the non-local model does not exhibit any stiffness instability which is clear since for $b>0$ the two-body interaction is repulsive at all wavelengths \cite{PREL_2004,PREBR_2008}.

A further improvement can be done if now we consider the substrate and the interface are not approximately parallel, but we can still neglect curvature effects \cite{RESPG_2018}. Now,  an infinite set of decorated diagrams do contribute to both $\Omega_1^1$ and $\Omega_1^2$. However, it is possible to resum them in order to express the diagrammatic expansion Eq.~(\ref{WMFfunctional}) in a compact way. For example, 
we can rewrite $\Omega_1^1$ and $\Omega_1^2$ for the $g\to -\infty$ case as
\begin{equation}
\Omega_1^1=
-\frac{1}{2}\fig{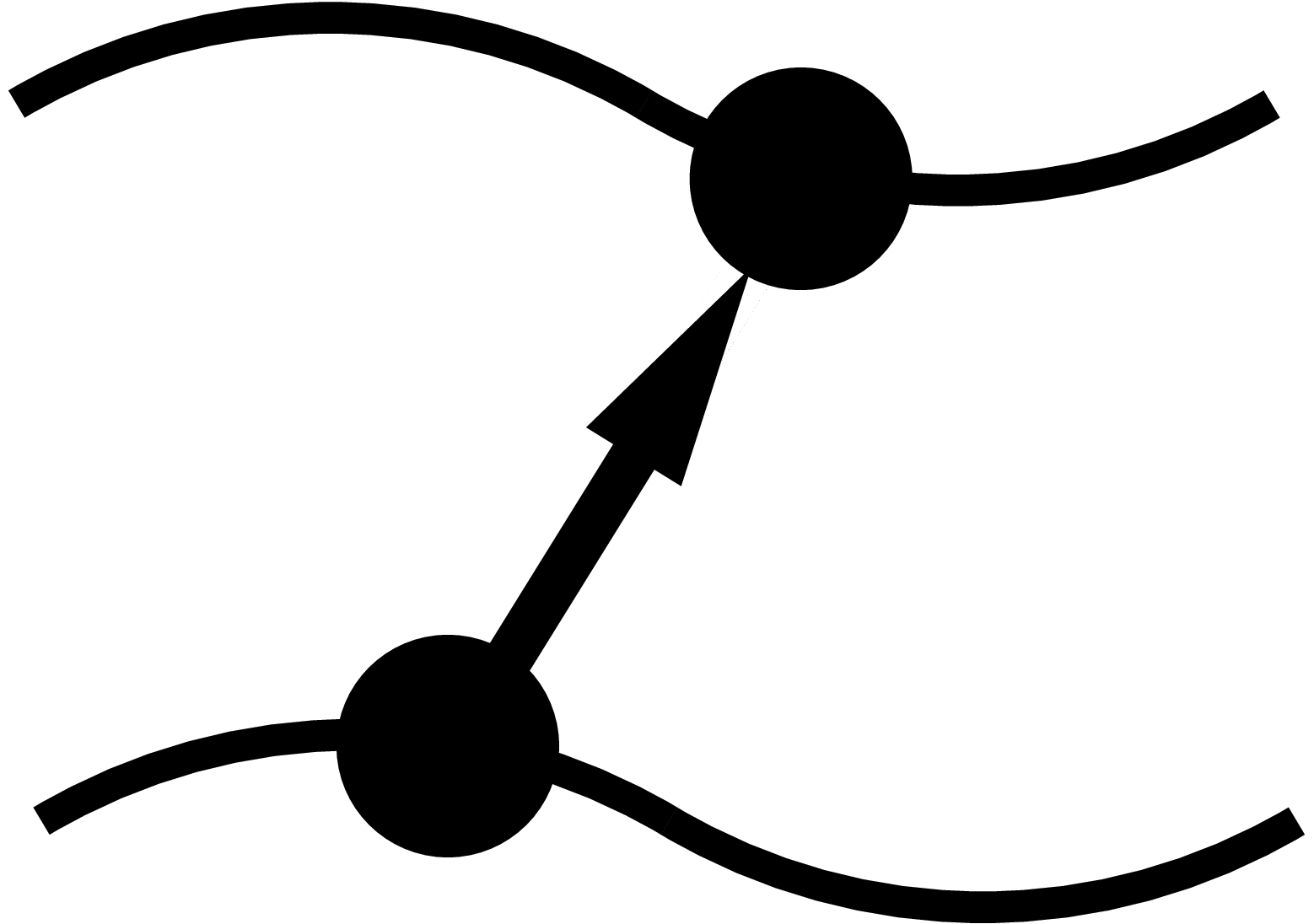}+\frac{1}{2}\fig{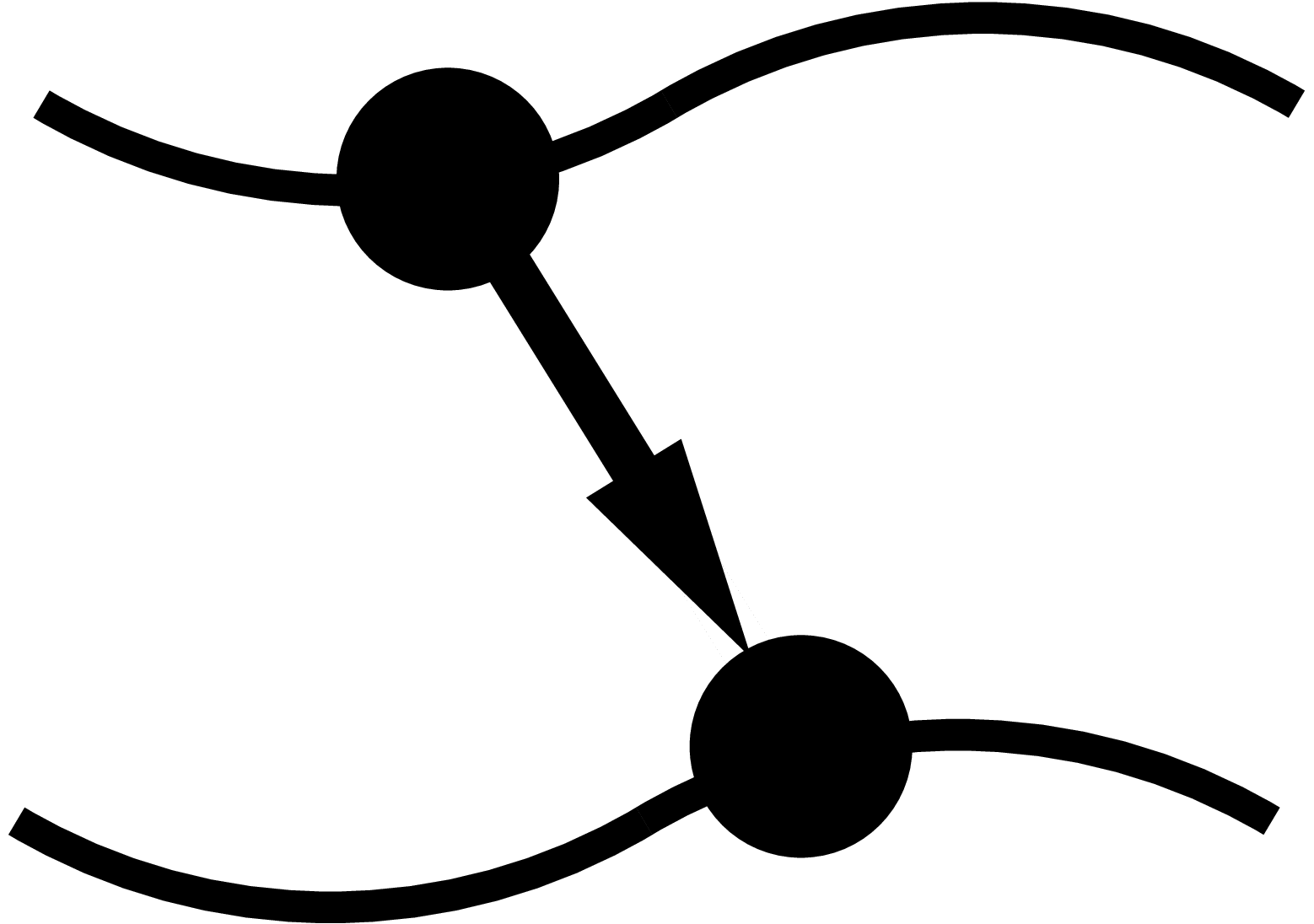}
\end{equation}
and 
\begin{equation}
\Omega_{1}^{2} = -\fig{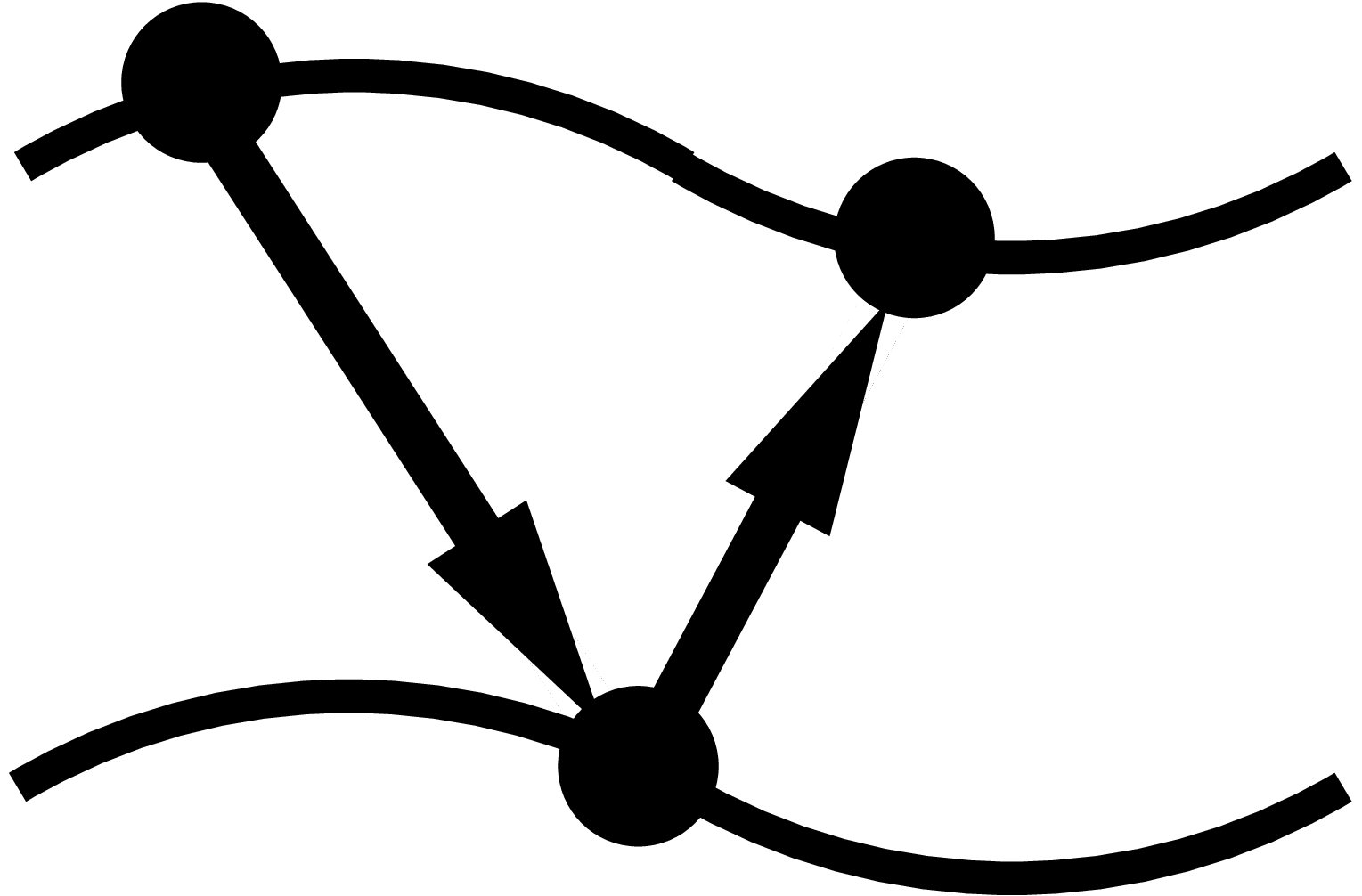} \ .
\end{equation}
Analogous expressions can be obtained for general $g$. 

Finally, the constrained magnetization in the wetting layer can also be determined as a perturbative expansion \cite{RESPG_2018}. Formally, it can be expressed as
\begin{eqnarray}
\delta m_\Xi(\mathbf{r})&\equiv& m_\Xi(\mathbf{r})-m_0
\nonumber\\&=&\delta m_1 \int_{S_1} \rd \mathbf{s}\int_{S_1} \rd \mathbf{s}^\prime O_{11}(\mathbf{s},\mathbf{s}^\prime )K(\mathbf{s}^\prime,\mathbf{r})\nonumber\\
&+&\delta m_1 \int_{S_1} \rd \mathbf{s}\int_{S_2} \rd \mathbf{s}^\prime O_{12}(\mathbf{s},\mathbf{s}^\prime )K(\mathbf{s}^\prime,\mathbf{r})\nonumber\\
&-& m_0 \int_{S_2} \rd \mathbf{s}\int_{S_1} \rd \mathbf{s}^\prime O_{21}(\mathbf{s},\mathbf{s}^\prime )K(\mathbf{s}^\prime,\mathbf{r})\nonumber\\
&-& m_0 \int_{S_2} \rd \mathbf{s}\int_{S_2} \rd \mathbf{s}^\prime O_{22}(\mathbf{s},\mathbf{s}^\prime )K(\mathbf{s}^\prime,\mathbf{r})\, ,
\label{constrainedm}
\end{eqnarray}
where $\delta m_1=m_1-m_0$ and $O_{ij}(\mathbf{s},\mathbf{s}^\prime)$, with $\mathbf{s}\in S_i$ and $\mathbf{s}^\prime\in S_j$, is a boundary operator that can be expanded as a series of boundary integral convolutions of diagrams (\ref{20102022_2058}), (\ref{20102022_2059}) and (\ref{defdiagrams}) (see Eq. (113) in Ref. \cite{RESPG_2018}).

\section{Determination of the Casimir contribution to the binding potential functional}
\label{sec_diagrammatic}
In this Section we show how the calculation of the Casimir interaction in wetting layers can be efficiently tackled using diagrammatic techniques. In Sec.~\ref{sec_5_1} we prepare the ground for the following developments by showing how to split the MF contribution from the fluctuations on top of it. In Sec.~\ref{sec_5_2} we identify the ``one-loop'' Casimir free energy and show its explicit calculation which we carry out by means of an adaptation of the Li-Kardar formalism \cite{LK_1991} for the case of Dirichlet boundary conditions, which is extended to the surface field case in Sec.~\ref{sec_5_5}. In Sec.~\ref{sec_5_4} we reobtain the expression of the Casimir interaction by means of boundary integral equations \cite{RESPG_2018}, which naturally leads to the diagrammatic expansion described in Sec.~\ref{sec_5_6}.

\subsection{General formulation}
\label{sec_5_1}
An advantage of the crossing criterion definition of the interface location is that there is no contribution to the binding potential functional $W[\ell, \psi]$ from the trace over microscopic degrees of freedom within $\mathcal{V}_{+}$ since this region is shielded from the wall. Consequently, we need only focus on the constrained trace (\ref{03062022_1604}) within the wetting layer, corresponding to the region $\mathcal{V}_{-}$. For the field theory quadratic in the field this can be done exactly. Let us define the field
\begin{equation}
\phi(\textbf{r})
\equiv m(\textbf{r}) - m_{\Xi}(\textbf{r})
\end{equation}
to denote the fluctuations about the constrained minimum solution. The fluctuations of this field determine entropic/Casimir contribution to the binding potential and correspond to all the microscopic configurations that map onto a given interfacial one. The crossing criterion [Eq.~(\ref{03062022_1409a})] demands that this field vanishes at the interface so that we must impose
\begin{equation}
\phi(\textbf{r}) = 0 \, , \quad \textbf{r} \in S_2
\label{boundary}
\end{equation}
in the constrained trace. Moreover, since the Hamiltonian is quadratic, we can re-write the LGW Hamiltonian, within the wetting layer, exactly as
\begin{equation}
H_{\rm LGW}[m] = H_{\rm LGW}[m_{\Xi}] + \Delta H_{\rm LGW}[\phi]
\end{equation}
where
\begin{equation}
\begin{aligned}
\label{deltahlgw}
\Delta H_{\rm LGW}[\phi] & = \int d\mathbf{r} \left(\frac{1}{2} (\boldsymbol\nabla \phi)^2 + \frac{\kappa^{2}}{2} \phi^{2} \right) \\
&  - \frac{g}{2} \int_{\mathcal{S}_{1}} d\mathbf{s} \, \phi^{2}
\end{aligned}
\end{equation}
is a Gaussian Hamiltonian containing only a surface enhancement coupling at the wall. 
The absence of a coupling to the surface field is some indication that the Casimir contribution to the binding potential will be simpler than the MF contribution. It is worth noting that this expression can be obtained from the boundary-independent Hamiltonian
\begin{equation}
\frac{1}{2}\int d\mathbf{r} \phi(-\nabla^2 +\kappa^2)\phi \, ,
\label{hamiltonianfluct}
\end{equation}
when evaluated using fields $\phi(\mathbf{r})$ constrained by the conditions Eq. (\ref{boundary}) and
\begin{equation}
\phi(\textbf{s})+\frac{1}{g}\mathbf{n}(\mathbf{s})\cdot\nabla \phi(\mathbf{s}) = 0 \, , \quad \textbf{s} \in S_1 \, .
\label{boundary2}
\end{equation}
This condition reduces to the Dirichlet boundary condition in the limit $g\to -\infty$.
Taking into account that $\phi \nabla^2 \phi =\nabla \cdot (\phi \nabla \phi)-(\nabla \phi)^2$ and applying the divergence theorem, (\ref{hamiltonianfluct}) reads
\begin{eqnarray}
\int d\mathbf{r} \left(\frac{1}{2} (\boldsymbol\nabla \phi)^2 
+ \frac{\kappa^{2}}{2}\right)
-
\frac{1}{2}\int_{S_2} d\mathbf{s} \phi(\mathbf{s}) \mathbf{n}(\mathbf{s})\cdot \boldsymbol\nabla \phi(\mathbf{s})\nonumber\\
+\frac{1}{2}\int_{S_1}  d\mathbf{s} \phi(\mathbf{s}) \mathbf{n}(\mathbf{s})\cdot \boldsymbol\nabla \phi(\mathbf{s}) \, ,
\label{hamiltonianfluct2}
\end{eqnarray}
which reduces to (\ref{deltahlgw}) after considering the constraints (\ref{boundary}) and (\ref{boundary2}).

It follows that the constrained trace (\ref{03062022_1604}) can be divided exactly into a MF contribution, which has already been determined, together with a constrained fluctuation sum over the bulk fluctuations about it
\begin{eqnarray}
e^{-\beta H_{\rm I}[\ell]-\beta\gamma_{wl}\mathcal{A}_\psi}&=& e^{-\beta H_{\rm LGW}[m_{\Xi}]} \nonumber\\
&\times&\int ({\mathcal D}\phi)^{\prime} \, e^{-\beta \Delta H_{\rm LGW}[\phi]} \, ,
\end{eqnarray}
where the order parameter $\phi$ in the path integral is constrained by conditions (\ref{boundary}) and (\ref{boundary2}).
Strictly speaking, the constrained trace on $\phi$ for a given interfacial profile includes magnetization profiles in which domains with negative magnetization are nucleated within the wetting layer. This is expected to occur in a neighborhood of the liquid-vapour interface, but it should be more unlikely as we move away from that region. Thus, we assume that the contribution of these sign-changing magnetization profiles is to renormalize the surface tension $\gamma_{lv}$ and that this effect is independent of the distance between the substrate and the liquid-vapour interface is the wetting-layer width is large enough. Under this hypothesis, the binding potential separates additively into a MF contribution and Casimir contribution
\begin{equation}
W[\ell,\psi] = W_{\rm MF}[\ell,\psi] + W_{\rm C}[\ell,\psi] \, .
\end{equation}
The calculation of $W_{\rm C}[\ell,\psi]$ is the subject of the following Section. The contribution of the fluctuations to the wetting-layer free energy can be obtained as the usual one-loop contribution:
\begin{equation}
\label{oneloop0}
\beta F_{fl}=\frac{1}{2} \textrm{tr} \ln \textsf{K}_W \, ,
\end{equation}
where $\textsf{K}_W$ is an integral operator within the wetting layer ${\mathcal V}_+$ with the constrained rescaled Green function $K_W(\mathbf{r},\mathbf{r}^\prime)\equiv 2\kappa[\langle \phi(\mathbf{r})\phi(\mathbf{r}^\prime)\rangle -\langle \phi(\mathbf{r})\rangle\langle\phi(\mathbf{r}^\prime)\rangle ]$ ($\mathbf{r},\mathbf{r}^\prime \in {\mathcal V}_+$) as its kernel. This function satisfies
\begin{equation}
\label{ }
\left( - \nabla^{2} + \kappa^{2} \right) K_W(\textbf{r},\textbf{r}^{\prime}) = 2\kappa \delta(\textbf{r}-\textbf{r}^{\prime})\, ,
\end{equation}
where $\delta(\mathbf{r})$ is Dirac's delta function, with boundary conditions
\begin{eqnarray}
Y_{\mathbf{s}}K_W(\mathbf{s},\mathbf{r})& = & 0 \, , \qquad \textbf{s} \in S_1 \, ,\\
K_W(\textbf{s},\textbf{r})& = & 0 \, , \qquad \textbf{s} \in S_2 \, ,
\end{eqnarray}
where $Y_{\mathbf{s}}\equiv \left( 1 + g^{-1}\mathbf{n}_{1} \cdot \boldsymbol{\nabla}_{\textbf{s}} \right)$.
We can write $K_W(\mathbf{r},\mathbf{r}^\prime)$ as $K(\mathbf{r},\mathbf{r}^\prime)+\Delta K_W(\mathbf{r},\mathbf{r}^\prime)$, where $K(\mathbf{r},\mathbf{r}^\prime)$ is the rescaled bulk Green function (\ref{03062022_1616}), and $\Delta K_W$ satisfies
\begin{equation}
\label{ }
\left( - \nabla^{2} + \kappa^{2} \right) \Delta K_W(\textbf{r},\textbf{r}^{\prime}) = 0\, ,
\end{equation}
with boundary conditions
\begin{eqnarray}
Y_{\mathbf{s}} \Delta K_W(\textbf{s},\textbf{r}) & = & -Y_{\mathbf{s}} K(\textbf{s},\textbf{r}) \, , \qquad \textbf{s} \in S_1 \, ,\\
\Delta K_W(\textbf{s},\textbf{r})& = & -K(\textbf{s},\textbf{r}) \, , \qquad \textbf{s} \in S_2 \, .
\end{eqnarray}
We can obtain a formal expression for $\Delta K_W$ in a similar way as the mean-field constrained magnetization, by substituting $\delta m_1$ by $-Y_sK(\mathbf{s},\mathbf{r}^\prime)$ and $m_0$ by $K(\mathbf{s},\mathbf{r}^\prime)$ in Eq. (\ref{constrainedm}), i.e.
\begin{eqnarray}
&&\Delta K_W(\mathbf{r},\mathbf{r}^\prime)= 
\nonumber\\&-& \int_{S_1} \rd \mathbf{s}\int_{S_1} \rd \mathbf{s}^\prime Y_{\mathbf{s}}K(\mathbf{s},\mathbf{r}^\prime)O_{11}(\mathbf{s},\mathbf{s}^\prime )K(\mathbf{s}^\prime,\mathbf{r})\nonumber\\
&-& \int_{S_1} \rd \mathbf{s}\int_{S_2} \rd \mathbf{s}^\prime Y_{\mathbf{s}}K(\mathbf{s},\mathbf{r}^\prime) O_{12}(\mathbf{s},\mathbf{s}^\prime )K(\mathbf{s}^\prime,\mathbf{r})\nonumber\\
&-& \int_{S_2} \rd \mathbf{s}\int_{S_1} \rd \mathbf{s}^\prime K(\mathbf{s},\mathbf{r}^\prime) O_{21}(\mathbf{s},\mathbf{s}^\prime )K(\mathbf{s}^\prime,\mathbf{r})\nonumber\\
&-& \int_{S_2} \rd \mathbf{s}\int_{S_2} \rd \mathbf{s}^\prime K(\mathbf{s},\mathbf{r}^\prime)O_{22}(\mathbf{s},\mathbf{s}^\prime )K(\mathbf{s}^\prime,\mathbf{r})\, .
\label{deltakw}
\end{eqnarray}
As $K_W\approx K$ away from the surfaces, we can define the surface contribution to the free energy associated with the fluctuations as
\begin{eqnarray}
\frac{1}{2} \textrm{tr} \ln \textsf{K}_W &-& \frac{1}{2} \textrm{tr} \ln \textsf{K} =  \frac{1}{2} \textrm{tr} \ln [\textsf{K}_W  \textsf{K}^{-1}]
\\=\frac{1}{2} \textrm{tr} \ln [\textsf{I}+\Delta\textsf{K}_W  \textsf{K}^{-1}]&=& \frac{1}{2} \sum_{n=1}^\infty \frac{(-1)^{n+1}}{n}\textrm{tr}\left(\Delta\textsf{K}_W  \textsf{K}^{-1}\right)^n
\nonumber
\end{eqnarray}
where $\mathsf{K}$ is the integral operator with the rescaled bulk Green function $K(\mathbf{r},\mathbf{r}^\prime)$ as kernel, $\mathsf{K}^{-1}$ its inverse and $\mathsf{I}$ the identity operator. The integral operator $\Delta \mathsf{K}_W\mathsf{K}^{-1}$ acts on the space of functions with support on $S_1\cup S_2$, with kernel
\begin{eqnarray}\nonumber
- \int_{S_1} \rd \mathbf{s}^\prime Y_{\mathbf{s}^\prime}K(\mathbf{s}^\prime,\mathbf{r})O_{1i}(\mathbf{s}^\prime,\mathbf{s} )
- \int_{S_2} \rd \mathbf{s}^\prime K(\mathbf{s}^\prime,\mathbf{r}) O_{2i}(\mathbf{s}^\prime,\mathbf{s}) \\
\label{deltakw11}
\end{eqnarray}
if $\mathbf{s}\in S_i$. Note that the trace of any power of this operator only involves convolutions of its kernel evaluated on $\mathbf{r}\in S_1\cup S_2$. Now, we define the boundary integral operator $\Delta \mathsf{K}_S$ as the action of $\Delta \mathsf{K}_W\mathsf{K}^{-1}$ when imposing that $\mathbf{r}\in S_1\cup S_2$ in its kernel, and $\mathsf{K}_S=\mathsf{I}_S+\Delta \mathsf{K}_S$, where $\mathsf{I}_S$ is the identity operator in the space of functions with support on $S_1\cup S_2$. Therefore, 
\begin{equation}
    \frac{1}{2} \textrm{tr} \ln [\textsf{K}_W  \textsf{K}^{-1}]=\frac{1}{2}\textrm{tr}_S \ln \mathsf{K}_S\, ,
\end{equation}
where the subscript $\textrm{tr}_S$ in the right-hand side indicates that the trace is performed only on the wetting layer boundary $S_1\cup S_2$. In order to obtain the Casimir contribution to the binding potential, the surface contributions of the fluctuations associated to uncoupled interface and wall must be subtracted. This mimicks the situation when the interface and wall are infinitely far apart and corresponds to the fluctuation contributions to $\gamma_{lv}$ and $\gamma_{wl}$. This can be obtained similarly as $(1/2)\textrm{tr}_S \mathsf{K}_S^\infty$, 
where the boundary integral operator $\mathsf{K}_S^\infty$ is defined analogously to $\mathsf{K}_S$, substituting in Eq. (\ref{deltakw11}) $O_{ij}$ by $O_{ij}^\infty$, which are the contributions to the boundary integral operators $O_{ij}$ that do not involve $K$-diagrams connecting the interface and the wall (so $O_{12}^\infty=O_{21}^\infty=0$). Thus
\begin{eqnarray}
\beta W_{\rm C}[\ell,\psi] &=& \frac{1}{2} \textrm{tr}_S \ln \textsf{K}_S - \frac{1}{2}\textrm{tr}_S \ln \textsf{K}_S^\infty\nonumber\\&=&\frac{1}{2} \textrm{tr}_S \ln [\textsf{K}_S (\textsf{K}_S^\infty)^{-1}] \, .
\label{oneloop}
\end{eqnarray}
Although we can use the boundary integral techniques used in Ref. \cite{RESPG_2018} to determine the Casimir contribution to the binding potential, a direct approach is too convoluted and, instead, we will determine it by a different procedure in the next subsections.

\subsection{Field-theoretic derivation: Dirichlet boundary condition on the wall}
\label{sec_5_2}
We are now in the position to illustrate how to isolate the ``one-loop'' contribution to the binding potential, providing a proof of Eq.~(\ref{oneloop}). The method we will employ is closely related to those used in the literature for the analysis of the Casimir interaction energy \cite{BRW_1985, LK_1991,  Bordag_2006}. More specifically, the formalism we will use is based on statistical-mechanical ideas, Gaussian field theories, and follows as an adaptation of a method devised by Li and Kardar \cite{LK_1991}.

Suppose we want to compute the partition function $Z^\prime$ associated with constrained order parameter fluctuations 
\begin{equation}
\label{wl_01}
Z^\prime = \int \left( \mathcal{D}\phi \right)^{\prime} \, \textrm{e}^{- \frac{\beta}{2}\int_{\mathbb{R}^{3}} d\mathbf{r} \phi (-\nabla^2 +\kappa^2)\phi}
\end{equation}
where $\left( \mathcal{D}\phi \right)^{\prime}$ denotes the path integral measure with respect to fields $\phi(\textbf{r})$ that satisfy the Dirichlet boundary conditions $\phi(\textbf{s} \in S_{1}) = \phi(\textbf{s} \in S_{2})=0$ at the surfaces $S_{1}$ and $S_{2}$. These surfaces correspond to the solid-liquid interface ($S_{1}$) and to the liquid-vapor interface ($S_{2}$). For the sake of notational convenience, we write $S = S_{1} \cup S_{2}$. The sum over paths is restricted to satisfy the Dirichlet boundary conditions can be conveniently implemented by supplying the unrestricted measure $\mathcal{D}\phi$ with the Dirac delta function, which can be represented through the Fourier representation 
\begin{equation}
\label{wl_02}
\prod_{ \textbf{s} \in S } \delta( \phi(\textbf{s}) ) = \int \mathcal{D}b \, \exp\left( \im \int_{S}\rd\textbf{s} \, b(\textbf{s}) \phi(\textbf{s}) \right) \, ,
\end{equation}
in terms of the auxiliary field $b(\textbf{s})$ and up to an overall constant which can be reabsorbed into the path integral measure. The partition function $Z^\prime$ is thus written as
\begin{eqnarray} \nonumber
\label{wl_03}
Z^\prime & = & \int \mathcal{D}b \int \mathcal{D}\phi \, \exp\Biggl[ - \int_{\mathbb{R}^{3}}\rd\textbf{r} \Bigl[ \frac{1}{2} \phi (-\nabla^2+\kappa^2)\phi) \Bigr] \\
& + & \im \int_{S}\rd\textbf{s} \, b(\textbf{s}) \phi(\textbf{s}) \Biggr] \, .
\end{eqnarray}
In order to make progresses, we rewrite (\ref{wl_03}) in a way that is suitable for calculations. To this end, we cast the argument in the exponential in a quadratic form in the fields. This can be achieved by writing the argument in the form
\begin{equation}
\label{wl_04}
- \int_{\mathbb{R}^{3}}\rd\textbf{r} \Bigl[ \frac{1}{2} \phi(-\nabla^2+\kappa^2)\phi - \phi(\textbf{r}) B(\textbf{r}) \Bigr] \, ,
\end{equation}
where
\begin{equation}
\label{wl_05}
B(\textbf{r}) = \im  \int_{S}\rd\textbf{s} \, b(\textbf{s}) \delta( \textbf{r} - \textbf{s} ) \, .
\end{equation}
Then, (\ref{wl_04}) is diagonalized by rearranging to read
\begin{equation}
\label{wl_06}
- \int_{\mathbb{R}^{3}}\rd\textbf{r} \Bigl[ \frac{1}{2} (\phi-\phi_{0})(-\nabla^2+\kappa^2)(\phi-\phi_{0}) + C \Bigr] \, ,
\end{equation}
for some $\phi_{0}$ and $C$ that can be fixed as shown in the following. We equate (\ref{wl_04}) and (\ref{wl_05}), and getting rid of vanishing surface terms, we obtain
\begin{eqnarray}
\label{wl_07} \nonumber
&& \int_{\mathbb{R}^{3}}\rd\textbf{r} \Bigl[ \phi(\textbf{r})\left( \textsf{G}_{0}^{-1}\phi_{0}(\textbf{r}) - B(\textbf{r}) \right) \\
&& - \phi_{0}(\textbf{r}) \textsf{G}_{0}^{-1}\phi_{0}(\textbf{r}) - C \Bigr] = 0 \, .
\end{eqnarray}
where $\textsf{G}_{0}^{-1} = -\nabla^{2} + \kappa^{2}$. Since this has to be solved for any $\phi(\textbf{r})$, it  follows that
\begin{equation}
\label{wl_08}
\textsf{G}_{0}^{-1} \phi_{0}(\textbf{r}) = B(\textbf{r}) \, .
\end{equation}
This, in turn, implies that $\phi_{0}$ is given by
\begin{eqnarray} \nonumber
\label{wl_09}
\phi_{0}(\textbf{r}) & = & 
\int_{\mathbb{R}^{3}}\rd\textbf{r}^{\prime} \, G_{0}(\textbf{r}, \textbf{r}^{\prime}) \textsf{G}_{0}^{-1} \phi_{0}(\textbf{r}^{\prime}) \\
& = & \im \int_{S}\rd\textbf{s} \, b(\textbf{s}) G_{0}(\textbf{r},\textbf{s}) \, ,
\end{eqnarray}
where $G_0(\textbf{r},\textbf{r}^\prime)=K(\textbf{r},\textbf{r}^\prime)/2\kappa$ is the free Green function associated to the differential operator $\textsf{G}_{0}^{-1}$, so $\textsf{G}_0$ is the integral operator with kernel $G_0$.
Then, by plugging (\ref{wl_09}) into (\ref{wl_07}), we find that $C$ is given by
\begin{equation}
\label{wl_10}
C = \frac{1}{2} \int_{S}\rd\textbf{s} \int_{S}\rd\textbf{s}^{\prime} \, b(\textbf{s}) G_{0}(\textbf{s},\textbf{s}^{\prime}) \delta( \textbf{r} - \textbf{s}^{\prime} ) b(\textbf{s}^{\prime}) \, .
\end{equation}

Collecting the above results, the partition function becomes
\begin{eqnarray}
\label{wl_11}
Z^\prime & = & Z_{0} \int \mathcal{D}b \, \textrm{e}^{ - \frac{1}{2} \int_{S}\rd\textbf{s} \int_{S}\rd\textbf{s}^{\prime} \, b(\textbf{s}) G_{0}(\textbf{s},\textbf{s}^{\prime}) b(\textbf{s}^{\prime}) } \, ,
\end{eqnarray}
where $Z_{0}$ is the path integral with respect to $\phi$ in (\ref{wl_03}), which can be written in terms of the shifted field $\psi=\phi-\phi_{0}$
\begin{eqnarray} \nonumber
\label{wl_12}
Z_{0} & = & \int \mathcal{D}\psi \exp\Biggl[ - \int_{\mathbb{R}^{3}}\rd\textbf{r} \Bigl( \frac{1}{2} \psi (-\nabla^2 +\kappa^2)\psi) \Bigr)\Biggr] \, .
\end{eqnarray}
Since the resulting Euclidean action -- or, Hamiltonian -- in (\ref{wl_11}) is quadratic, the partition function $Z^\prime$ is given by
\begin{eqnarray} \nonumber
\label{wl_13}
Z^\prime & \propto & \left( \det \textsf{K}_S \right)^{-1/2} \, ,
\end{eqnarray}
where $\textsf{K}_S$ is the boundary integral operator with kernel $K(\textbf{s},\textbf{s}^{\prime}) \equiv 2\kappa G_{0}(\textbf{s},\textbf{s}^{\prime})$, with $\textbf{s},\textbf{s}^{\prime} \in S$. It thus follows that information coming from the boundaries are codified by the operator $\textsf{K}_S$.

The total free energy $F = - k_{\rm B}T \ln Z^\prime$ reads
\begin{eqnarray} \nonumber
\label{wl_14}
F & = & \frac{1}{2} k_{\rm B}T \ln \left( \det \textsf{K}_S \right) \\
& = & \frac{1}{2} k_{\rm B}T \tr_{S} \left( \ln \textsf{K}_S \right) \, ,
\end{eqnarray}
where $\tr_{S}(\dots)$ denotes the trace over a complete set of functions with support on the union of the boundaries, $S=S_{1} \cup S_{2}$. This quantity in reality is infinite. Consequently we have to subtract, from the total free energy, the contribution stemming from the two infinitely separated surfaces $S_{1}$ and $S_{2}$. The contribution associated with isolated (meaning uncorrelated) surfaces is given by the boundary integral operator $\textsf{K}_S^\infty$ with kernel
\begin{equation}
\label{wl_15}
K_S^{\infty}(\textbf{s},\textbf{s}^{\prime}) = 
\begin{cases}
2\kappa G_{0}(\textbf{s},\textbf{s}^{\prime})      & \textbf{s}, \textbf{s}^{\prime} \in S_{1} \,\, \textrm{or} \,\, \textbf{s}, \textbf{s}^{\prime} \in S_{2} \\
0       & \textbf{s} \in S_{1}, \textbf{s}^{\prime} \in S_{2} \\
0	& \textbf{s} \in S_{2}, \textbf{s}^{\prime} \in S_{1} \, .
\end{cases}
\end{equation}
Then, the associated free energy is
\begin{equation}
\label{wl_16}
F_{\infty} = \frac{1}{2} k_{\rm B}T \tr_{S} \left( \ln \textsf{K}_S^{\infty} \right) \, .
\end{equation}
The Casimir contribution to the free energy is defined by
\begin{eqnarray} \nonumber
\label{wl_17}
W_C & = & F - F_{\infty} \\ \nonumber
& = & \frac{1}{2} k_{\rm B}T \bigl[ \tr_{S} \left( \ln \textsf{K}_S \right) - \tr_{S} \left( \ln \textsf{K}_S^{\infty} \right) \bigr] \\
& = & \frac{1}{2} k_{\rm B}T \tr_{S} \bigl[ \ln \left( (\textsf{K}_S^{\infty})^{-1}\textsf{K}_S \right) \bigr] \, ;
\end{eqnarray}
which proves the result (\ref{oneloop}). It is now clear that the residual free energy due to the subtraction of the free energy associated with infinitely separated surfaces is actually \emph{finite}. The analysis that follows aims at expressing (\ref{wl_17}) in more explicit fashion.

Both the operators $\textsf{K}_S$ and $\textsf{K}_S^\infty$ can be expressed as block matrices in which matrix elements correspond to the boundary integral operator when the arguments of the associated kernels are evaluated either on different parsing of surfaces. More explicitly, we write
\begin{equation}
\label{wl_18}
\textsf{K}_S = 
\left(\begin{array}{cc}
\textsf{K}_{11} & \textsf{K}_{12} \\
\textsf{K}_{21} & \textsf{K}_{22}
\end{array}\right) \, ,
\end{equation}
where the explicit expression of the kernels of the boundary integral operators read $K_{ij}(\textbf{s},\textbf{s}^{\prime}) = 2\kappa G_{0}(\textbf{s},\textbf{s}^{\prime})$ with
\begin{equation}
\label{wl_19}
K_{ij}(\textbf{s},\textbf{s}^{\prime}) \leftrightarrow \binom{\textbf{s}}{\textbf{s}^{\prime}} \in \binom{S_{i}}{S_{j}} \, \quad i,j=1,2 \, .
\end{equation}
The off-diagonal entries, which satisfy $\textsf{K}_{21}=\textsf{K}_{12}^{\dag}$, are absent in the operator
\begin{equation}
\label{wl_20}
\textsf{K}_S^{\infty} = 
\left(\begin{array}{cc}
\textsf{K}_{11} & 0 \\
0 & \textsf{K}_{22}
\end{array}\right) \, ,
\end{equation}
because in this case the two surfaces are completely uncorrelated.

It is now simple to check that the matrix which appear in (\ref{wl_17}) has the following structure
\begin{eqnarray}
\label{wl_21} \nonumber
(\textsf{K}_S^{\infty})^{-1} \textsf{K}_S & = &
\left(\begin{array}{cc}
\textsf{I}_1 & \textsf{K}_{11}^{-1}\textsf{K}_{12} \\
\textsf{K}_{22}^{-1}\textsf{K}_{21} & \textsf{I}_2
\end{array}\right) \\
& = & \textsf{I} + \textsf{M}
\end{eqnarray}
where $\textsf{I}_1$, $\textsf{I}_1$ and $\textsf{I}$ are the identity operators defined on the space of functions with support on $S_1$, $S_2$ and $S$, respectively, and $\textsf{M}$ is the traceless matrix
\begin{equation}
\label{wl_22}
\textsf{M} = 
\left(\begin{array}{cc}
0 & \textsf{K}_{11}^{-1}\textsf{K}_{12} \\
\textsf{K}_{22}^{-1}\textsf{K}_{21} & 0
\end{array}\right) \, .
\end{equation}
Returning to (\ref{wl_17}), the Casimir interaction energy becomes
\begin{eqnarray} \nonumber
\label{wl_23}
W_C & = &  \frac{1}{2} k_{\rm B}T \tr_{S} \bigl[ \ln \left(  \textsf{I} + \textsf{M} \right) \bigr] \\
& = &  \frac{1}{2} k_{\rm B}T \sum_{n=1}^{\infty} \frac{(-1)^{n+1}}{n} \tr_{S}\left( \textsf{M}^{n} \right) \, .
\end{eqnarray}
It is simple to check by inspection, and then by induction on $n$, that
\begin{equation}
\label{wl_24}
\tr_{S}\left( \textsf{M}^{2n+1} \right) = 0 \, ,
\end{equation}
and
\begin{eqnarray}
\label{wl_25}
\tr_{S}\left( \textsf{M}^{2n} \right) &=& 2 \tr_{S_1} \Bigl[ \left( \textsf{K}_{11}^{-1} \textsf{K}_{12} \textsf{K}_{22}^{-1} \textsf{K}_{21} \right)^{n} \Bigr]\\
&=& 2\tr_{S_2} \Bigl[ \left(\textsf{K}_{21}\textsf{K}_{11}^{-1} \textsf{K}_{12} \textsf{K}_{22}^{-1}  \right)^{n} \Bigr] \, ,
\end{eqnarray}
where $\tr_{S_i}$ is the trace over a complete set of functions with support on $S_i$. The above properties mean the infinite series in (\ref{wl_23}) can be resummed and the Casimir interaction energy can be expressed as the trace-log of a certain operator; thus,
\begin{eqnarray}
\nonumber
W_C & = & \frac{1}{2} k_{\rm B}T \tr_{S_2} \ln\left( \textsf{I}_2 - \textsf{K}_{21}\textsf{K}_{11}^{-1} \textsf{K}_{12} \textsf{K}_{22}^{-1}  \right) \\
& = & \frac{1}{2} k_{\rm B}T \tr_{S_2} \ln\left( \textsf{I}_2 - \textsf{N} \right) \, ,
\label{wl_26} 
\end{eqnarray}
where $\textsf{N}=\textsf{K}_{21}\textsf{K}_{11}^{-1} \textsf{K}_{12} \textsf{K}_{22}^{-1}$.

For general wall shapes and interfacial configurations, the formal expression Eq. (\ref{wl_26}),  cannot be used directly to evaluate the Casimir contribution to the binding potential. However, it is possible to obtain  expressions for $W_C$ in special cases. For example, taking into account the representation of the kernel $K(\mathbf{r},\mathbf{r}')$ in Cartesian, cylindrical and spherical coordinates shown in Appendix \ref{repKcoordinates}, it is possible to obtain diagonal representations of the boundary integral operators $\mathsf{K}_{ij}$ and $\mathsf{K}^{-1}_{ii}$ ($i=1,2$) in the slab geometry and in cylindrical and spherical shells. For a slab geometry with two parallel walls separated by a distance $\ell$, we use as the set of complete orthonormal functions on each surface $S_1$ and $S_2$ as $|\mathbf{q}\rangle_i\equiv \exp(-\textrm{i}\mathbf{q}\cdot \mathbf{x})/\sqrt{L_\parallel^{(d-1)}}$ with $\mathbf{x}_i\in S_i$ ($i=1,2$). Thus, the eigenvalues of $\mathsf{K}_{ij}$ and $\mathsf{K}^{-1}_{ii}$ are given by
\begin{eqnarray}
    _1\langle \mathbf{q}|\mathsf{K}_{12}|\mathbf{q}\rangle_2&=& _2\!\langle \mathbf{q}|\mathsf{K}_{21}|\mathbf{q}\rangle_1=\frac{\kappa}{\kappa_q}e^{-\kappa_q \ell}\label{eigen1}\\
    _1\langle \mathbf{q}|\mathsf{K}_{11}^{-1}|\mathbf{q}\rangle_1&=& _2\!\langle \mathbf{q}|\mathsf{K}_{22}^{-1}|\mathbf{q}\rangle_2=\frac{\kappa_q}{\kappa}\, ,\label{eigen2}
\end{eqnarray}
with $\kappa_q=\sqrt{\kappa^2+q^2}$, so $\mathsf{N}$ and is also diagonal operators in this representation, with eigenvalues
\begin{equation}
_2\langle\mathbf{q}|\mathsf{N}|\mathbf{q}\rangle_2=(_1\!\langle \mathbf{q}|\mathsf{K}_{12}|\mathbf{q}\rangle_2)^2(_1\!\langle \mathbf{q}|\mathsf{K}_{11}^{-1}|\mathbf{q}\rangle_1)^2=e^{-2\kappa_q\ell}\label{eigen3} \, .
\end{equation}
Consequently, $\ln(\mathsf{I}_2-\mathsf{N})$ is also diagonal with eigenvalues $\ln(1-\exp(-2\kappa_q \ell))$. Thus, the Casimir binding potential contribution can be expressed as
\begin{equation}
W_C=\frac{1}{2} k_{\rm B}T \sum_{\mathbf{q}} \ln\left( 1 - \textrm{e}^{-2\kappa_{q}\ell} \right) \, ,
\label{slit_dirichlet0}
\end{equation}
which, as $L_\parallel\to \infty$, reduces for $d>1$ to
\begin{equation}
\label{slit_dirichlet}
W_C = \frac{1}{2} k_{\rm B}T L_{\parallel}^{d-1} \int \frac{\rd\textbf{q}}{(2\pi)^{d-1}} \, \ln\left( 1 - \textrm{e}^{-2\kappa_{q}\ell} \right) \, ,
\end{equation}
in agreement with previous reported results \cite{SREP_2022,SREP2023} (see also Appendix \ref{sec_cylinders}).
For the sake of completeness we also provide the corresponding result in dimension $d=1$
\begin{equation}
\label{slitgeneral_22}
W_C = \frac{1}{2} k_{\rm B}T \ln\left( 1 - \textrm{e}^{-2\kappa\ell} \right) \, .
\end{equation}

Now we turn to the case of a cylindrical shell, where $S_1$ and $S_2$ are coaxial cylinders of radius $R$ and $R+\ell$, respectively. Now, the set of complete orthonormal on functions on $S_1$ and $S_2$ are $|n,q_z\rangle_1\equiv \exp(-\textrm{i}(n\varphi+q_z z))/\sqrt{2\pi R L_z}$ and $|n,q_z\rangle_2\equiv \exp(-\textrm{i}(n\varphi+q_z z))/\sqrt{2\pi (R+\ell) L_z}$, where the point on each surface is parametrized by its cylindrical coordinates $\rho(=R\textrm{ or }R+\ell),\varphi$ and $z$. Again $\mathsf{K}_{ij}$ and $\mathsf{K}_{ii}^{-1}$ are diagonal, with eigenvalues
\begin{eqnarray}
    _1\langle n,q_z|\mathsf{K}_{12}|n q_z\rangle_2&=& _2\!\langle n,q_z|\mathsf{K}_{21}|n,q_z\rangle_1=2\kappa \sqrt{R(R+\ell)}\nonumber\\
    &\times&K_{|n|}(\kappa_{q_z}(R+\ell))I_{|n|}(\kappa_{q_z}R)\label{eigen1_2}\\
    _1\langle n,q_z|\mathsf{K}_{11}^{-1}|n,q_z\rangle_1&=& 2\kappa R K_{|n|}(\kappa_{q_z}R)I_{|n|}(\kappa_{q_z}R)\label{eigen1_2}\label{eigen2_2}\\
    _2\!\langle n,q_z|\mathsf{K}_{22}^{-1}|n,q_z\rangle_2&=& 2\kappa (R+\ell) K_{|n|}(\kappa_{q_z}(R+\ell))\nonumber\\&\times&I_{|n|}(\kappa_{q_z}(R+\ell))
    \label{eigen2_2_2}\, .
\end{eqnarray}
Thus, the eigenvalues of $\mathsf{N}$ are
\begin{eqnarray}
_2\langle n,q_z|\mathsf{N}|n,q_z\rangle_2&=&(_1\!\langle n,q_z|\mathsf{K}_{12}|,q_z\rangle_2)^2 _1\!\langle n,q_z|\mathsf{K}_{11}^{-1}|n,q_z\rangle_1\nonumber\\
&\times&_2\!\langle n,q_z|\mathsf{K}_{22}^{-1}|n,q_z\rangle_2
\nonumber\\&=& \frac{K_{|n|}(\kappa_{q_z}(R+\ell))I_{|n|}(\kappa_{q_z}R)}{K_{|n|}(\kappa_{q_z}R)I_{|n|}(\kappa_{q_z}(R+\ell))}\label{eigen3_2} \, ,
\end{eqnarray}
and, consequently,
\begin{align}
\beta W_C & = \frac{1}{2}  \sum_{n=-\infty}^{\infty}\sum_{q_z} \ln \biggl[ 1 - \frac{K_{|n|}(\kappa_{q_z}(R+\ell))I_{|n|}(\kappa_{q_z}R)}{K_{|n|}(\kappa_{q_z}R)I_{|n|}(\kappa_{q_z}(R+\ell))} \biggr] \nonumber\\
& = \frac{1}{2} \sum_{q_z}\Bigg[\ln\left( 1 - \frac{K_{0}(\kappa_{q_z}(R+\ell))I_{0}(\kappa_{q_z}R)}{K_{0}(\kappa_{q_z}R)I_{0}(\kappa_{q_z}(R+\ell))} \right) \nonumber\\
& +  2\sum_{n=1}^\infty \ln\left( 1 - \frac{K_{n}(\kappa_{q_z}(R+\ell))I_{n}(\kappa_{q_z}R)}{K_{n}(\kappa_{q_z}R)I_{n}(\kappa_{q_z}(R+\ell))} \right) \Bigg]\, ,
\label{cylinder_general}
\end{align}
which for $L_z\to \infty$ reads
\begin{align}
\beta W_C & = \frac{L_z}{4\pi} \int_{-\infty}^\infty dq_z \Bigg[ \ln\left( 1 - \frac{K_{0}(\kappa_{q_z}(R+\ell))I_{0}(\kappa_{q_z}R)}{K_{0}(\kappa_{q_z}R)I_{0}(\kappa_{q_z}(R+\ell))} \right) \nonumber\\
& + 2\sum_{n=1}^\infty \ln\left( 1 - \frac{K_{n}(\kappa_{q_z}(R+\ell))I_{n}(\kappa_{q_z}R)}{K_{n}(\kappa_{q_z}R)I_{n}(\kappa_{q_z}(R+\ell))} \right) \Bigg]\, .
\label{cylinder_general2}
\end{align}
This expression is in complete agreement with the direct derivation of $W_C$ for cylindrical shells by using zeta regularization described in the Appendix \ref{sec_cylinders}.

Finally we will consider the case of a spherical shell, where $S_1$ and $S_2$ are concentrical spheres of radius $R$ and $R+\ell$, respectively. Now, the set of complete orthonormal on functions on $S_1$ and $S_2$ are $|l,m\rangle_1\equiv Y_l^m(\theta,\varphi)/R$ and $|l,m\rangle_2\equiv Y_l^m(\theta,\varphi)/(R+\ell)$, where the point on each surface is parametrized by its spherical coordinates $r(=R\textrm{ or }R+\ell),\theta$ and $\varphi$ and $Y_l^m$ are the spherical harmonics. The eigenvalues of $\mathsf{K}_{ij}$ and $\mathsf{K}_{ii}^{-1}$ are 
\begin{eqnarray}
    _1\langle l,m|\mathsf{K}_{12}|l,m\rangle_2&=& _2\!\langle l,m|\mathsf{K}_{21}|l,m\rangle_1=2\kappa \sqrt{R(R+\ell)}\nonumber\\
    &\times&K_{l+\frac{1}{2}}(\kappa (R+\ell))I_{l+\frac{1}{2}}(\kappa R)\label{eigen1_3}\\
    _1\langle l,m|\mathsf{K}_{11}^{-1}|l,m\rangle_1&=& 2\kappa R K_{l+\frac{1}{2}}(\kappa R)I_{l+\frac{1}{2}}(\kappa R)\label{eigen1_2}\label{eigen2_3}\\
    _2\!\langle l,m|\mathsf{K}_{22}^{-1}|l,m\rangle_2&=& 2\kappa (R+\ell) K_{l+\frac{1}{2}}(\kappa (R+\ell))\nonumber\\&\times&I_{l+\frac{1}{2}}(\kappa (R+\ell))
    \label{eigen2_3_2}\, .
\end{eqnarray}
Thus, $\mathsf{N}$ is diagonal with eigenvalues
\begin{eqnarray}
_2\langle l,m|\mathsf{N}|l,m\rangle_2&=&(_1\!\langle l,m|\mathsf{K}_{12}|,l,m\rangle_2)^2 _1\!\langle l,m|\mathsf{K}_{11}^{-1}|l,m\rangle_1\nonumber\\
&\times&_2\!\langle l,m|\mathsf{K}_{22}^{-1}|l,m\rangle_2
\nonumber\\&=& \frac{K_{l+\frac{1}{2}}(\kappa (R+\ell))I_{l+\frac{1}{2}}(\kappa R)}{K_{l+\frac{1}{2}}(\kappa R)I_{l+\frac{1}{2}}(\kappa(R+\ell))}\label{eigen3_3} \, ,
\end{eqnarray}
and, consequently,
\begin{align}
\beta W_C & = \frac{1}{2} \sum_{l=0}^{\infty}\sum_{m=-l}^l \ln \Biggl[ 1 - \frac{K_{l+\frac{1}{2}}(\kappa(R+\ell))I_{l+\frac{1}{2}}(\kappa R)}{K_{l+\frac{1}{2}}(\kappa R)I_{l+\frac{1}{2}}(\kappa(R+\ell))} \Biggr] \nonumber\\
& = \frac{1}{2} \sum_{l=0}^{\infty}(2l+1)\ln \Biggl[ 1 - \frac{K_{l+\frac{1}{2}}(\kappa(R+\ell))I_{l+\frac{1}{2}}(\kappa R)}{K_{l+\frac{1}{2}}(\kappa R)I_{l+\frac{1}{2}}(\kappa(R+\ell))} \Biggr] ,
\label{sphere_general}
\end{align}
in complete agreement with the zeta-regularization derivation of $W_C$ described in Appendix \ref{sec_cylinders}.

\subsection{Field-theoretic derivation: surface fields at the wall}
\label{sec_5_5}
We now consider the Casimir interaction for a substrate which is characterized by the surface potential $-(g/2)\int_{S_{1}}\rd\textbf{s}_{1} \, (\phi(\textbf{s}_{1}))^2$, with enhancement parameter $g<0$ so that it provides an energy penalty when the fluctuating field $\phi$ is non-zero on $S_1$. As was described in Sec. \ref{sec_5_1}, the effect of the surface field can be addressed by computing the partition function of a free field subjected to Robin boundary conditions on $S_1$. On the other hand, Dirichlet boundary conditions are applied on $S_2$. Following the procedure described in Sec. \ref{sec_5_2}, the constrained partition function (\ref{wl_01}) can be related to the unrestricted expression by the introduction of the constraints as products of Dirac delta functions
\begin{equation}
\prod_{\mathbf{s}\in S_1}\delta\left(\phi(\mathbf{s})+\frac{1}{g}\mathbf{n}\cdot\boldsymbol{\nabla}\phi(\mathbf{s})\right)\prod_{\mathbf{s}\in S_2}\delta(\phi(\mathbf{s})) \, ,
\end{equation}
which admits a Fourier representation akin to Eq. (\ref{wl_02})
\begin{eqnarray}
\int \mathcal{D}b \, \exp\Bigg( \im \int_{S_1}\rd\textbf{s} \, b(\textbf{s})\left[\phi(\textbf{s})+\frac{1}{g}\mathbf{n}\cdot\boldsymbol{\nabla}\phi(\mathbf{s})\right]\nonumber\\+\im\int_{S_2}\rd\textbf{s} \, b(\textbf{s}) \phi(\textbf{s}) \Bigg) \, .
\end{eqnarray}
Thus, the action of the partition function $Z^\prime$ can be written in the form (\ref{wl_04}), with $B(\mathbf{r})$ now defined as
\begin{eqnarray}
B(\textbf{r}) &=& \im  \int_{S_1}\rd\textbf{s}b(\mathbf{s})\left(\delta(\mathbf{r}-\mathbf{s})-\frac{1}{g}\mathbf{n}(\mathbf{s})\cdot\boldsymbol{\nabla}_{\mathbf{r}}\delta(\mathbf{r}-\mathbf{s})\right)\nonumber\\
&+&\im\int_{S_2}\rd\textbf{s} \, b(\textbf{s}) \delta( \textbf{r} - \textbf{s} ) \, .
\end{eqnarray}
Diagonalization of (\ref{wl_04}), i.e. Eq. (\ref{wl_06}), can be done with
\begin{eqnarray} \nonumber
\phi_{0}(\textbf{r}) & = & \im \int_{S_1} d\mathbf{s} b(\mathbf{s})Y_{\mathbf{s}}G_0(\mathbf{r},\mathbf{s})
 \\
& + & \im \int_{S_2}\rd\textbf{s} \, b(\textbf{s}) G_{0}(\textbf{r},\textbf{s}) \, ,
\end{eqnarray}
where we have defined $Y_{\mathbf{s}}=1+(1/g)\mathbf{n}(\mathbf{s})\cdot \nabla_{\mathbf{s}}\equiv 1+(1
/g)\partial_n$, with the normal derivative $\partial_n$ taken before putting the argument onto the surface, and
$C$ given by
\begin{eqnarray}
C &=& \frac{1}{2} \int_{S_1}\rd\textbf{s} \int_{S_1}\rd\textbf{s}^{\prime} \, b(\textbf{s}) Y_{\mathbf{s}}Y_{\mathbf{s}'}G_{0}(\textbf{s},\textbf{s}^{\prime}) \delta( \textbf{r} - \textbf{s}^{\prime} ) b(\textbf{s}^{\prime}) \nonumber\\
&+&\frac{1}{2} \int_{S_1}\rd\textbf{s} \int_{S_2}\rd\textbf{s}^{\prime} \, b(\textbf{s}) Y_{\mathbf{s}}G_{0}(\textbf{s},\textbf{s}^{\prime}) \delta( \textbf{r} - \textbf{s}^{\prime} ) b(\textbf{s}^{\prime})\nonumber\\
&+&\frac{1}{2} \int_{S_2}\rd\textbf{s} \int_{S_1}\rd\textbf{s}^{\prime} \, b(\textbf{s}) Y_{\mathbf{s}'}G_{0}(\textbf{s},\textbf{s}^{\prime}) \delta( \textbf{r} - \textbf{s}^{\prime} ) b(\textbf{s}^{\prime}) \nonumber\\
&+&\frac{1}{2} \int_{S_2}\rd\textbf{s} \int_{S_2}\rd\textbf{s}^{\prime} \, b(\textbf{s}) G_{0}(\textbf{s},\textbf{s}^{\prime}) \delta( \textbf{r} - \textbf{s}^{\prime} ) b(\textbf{s}^{\prime}) \, .
\end{eqnarray}
Therefore, the partition function $Z^\prime$ can be recast as 
\begin{align}
Z^\prime & = Z_{0} \int \mathcal{D}b \, \exp\Biggl[ - \frac{1}{2} \int_{S_1}\rd\textbf{s} \int_{S_1}\rd\textbf{s}^{\prime} b(\textbf{s}) Y_{\mathbf{s}}Y_{\mathbf{s}'} \nonumber \\
&  G_{0}(\textbf{s},\textbf{s}^{\prime}) b(\textbf{s}^{\prime}) \nonumber\\
& - \frac{1}{2} \int_{S_1}\rd\textbf{s} \int_{S_2}\rd\textbf{s}^{\prime} \, b(\textbf{s}) Y_{\mathbf{s}} G_{0}(\textbf{s},\textbf{s}^{\prime}) b(\textbf{s}^{\prime})\nonumber\\
& - \frac{1}{2} \int_{S_2}\rd\textbf{s} \int_{S_1}\rd\textbf{s}^{\prime} \, b(\textbf{s}) Y_{\mathbf{s}'}G_{0}(\textbf{s},\textbf{s}^{\prime}) b(\textbf{s}^{\prime})\nonumber\\
& - \frac{1}{2} \int_{S_2}\rd\textbf{s} \int_{S_2}\rd\textbf{s}^{\prime} \, b(\textbf{s}) G_{0}(\textbf{s},\textbf{s}^{\prime}) b(\textbf{s}^{\prime})
\Biggr]\, .
\end{align}

With these results, the Casimir interaction energy is again given by Eq. (\ref{wl_17}), where the operators $\mathsf{K}_S$ and $\mathsf{K}_S^\infty$ can still be written as block matrices, i.e. Eqs. (\ref{wl_18}) and (\ref{wl_20}). However, now the boundary integral operators $\mathsf{K}_{11}$,
$\mathsf{K}_{12}$, $\mathsf{K}_{21}$ and $\mathsf{K}_{22}$ have as kernels, $Y_{\mathbf{s}}J(\textbf{s},\textbf{s}^{\prime})$, $Y_{\mathbf{s}}K(\textbf{s},\textbf{s}^{\prime})$, $J(\textbf{s},\textbf{s}^{\prime})$ and $K(\textbf{s},\textbf{s}^{\prime})$, respectively, where $K(\mathbf{s},\mathbf{s}^\prime)\equiv2\kappa G_0(\mathbf{s},\mathbf{s}^\prime)$ and $Y_{\mathbf{s}^\prime}K(\textbf{s},\textbf{s}^{\prime})\equiv J(\textbf{s},\textbf{s}^{\prime})=K(\textbf{s},\textbf{s}^{\prime})+(1/g)\mathbf{n}(\mathbf{s}^\prime)\cdot \boldsymbol{\nabla}_{\mathbf{s}^\prime}K(\mathbf{s},\mathbf{s}^\prime)$. Note that integration in the boundary integral operator is done on the argument $\mathbf{s}^\prime$. Thus, and in analogy to the Dirichlet case, the Casimir interaction energy can be recast as Eq. (\ref{wl_26}). As in the Dirichlet case, explicit expressions for the Casimir interaction energy can be obtained for special geometries. For example, in the slab geometry we can use the same planar wave representation as in the Dirichlet case which  diagonalize the operators $\mathsf{K}_S$ and $\mathsf{K}_S^\infty$. Now, the diagonal terms of $\mathsf{K}_{ij}$ are
\begin{eqnarray}\nonumber
\label{alt_06}
 &&_1\langle\mathbf{q}|{\textsf{K}}_{11}|\mathbf{q}\rangle_1 \nonumber\\&=& \left( 1 + g^{-1} \partial_{n} \right) \left( 1 + g^{-1} \partial_{n^{\prime}} \right) \frac{\kappa}{\kappa_{q}} \textrm{e}^{-\kappa_{q}|z-z^\prime|} \bigg\vert_{z, z^\prime \rightarrow 0} \\ \nonumber
& = & \left( 1 + g^{-1} \partial_{z_{>}} \right) \left( 1 + g^{-1} \partial_{z_{<}} \right) \frac{\kappa}{\kappa_{q}} \textrm{e}^{-\kappa_{q}(z_{>}-z_{<})} \bigg\vert_{z_{<}, z_{>} \rightarrow 0} \\
& = & \frac{\kappa}{\kappa_{q}}  \left( 1 - \frac{\kappa_{q}}{g} \right) \left( 1 + \frac{\kappa_{q}}{g} \right) \, ,
\end{eqnarray}
while $_2\langle\mathbf{q}|{\textsf{K}}_{22}|\mathbf{q}\rangle_2=\kappa/\kappa_{q}$ and the off-diagonal term reads
\begin{eqnarray}\nonumber
\label{alt_07}
&&_1\langle\mathbf{q}|{\textsf{K}}_{12}|\mathbf{q}\rangle_2 =  _2\!\langle\mathbf{q}|{\textsf{K}}_{21}|\mathbf{q}\rangle_1\nonumber\\&=&
\left( 1 + g^{-1} \partial_{z_{<}} \right) \frac{\kappa}{\kappa_{q}} \textrm{e}^{-\kappa_{q}(z_{>}-z_{<})} \bigg\vert_{z_{<} \rightarrow 0,z_>\rightarrow \ell}\nonumber\\
&=&
\frac{\kappa}{\kappa_{q}}  \left( 1 + \frac{\kappa_{q}}{g} \right) \textrm{e}^{-\kappa_{q}\ell} \, .
\end{eqnarray}
Therefore, the eigenvalues of the operator $\textsf{N}$ yields
\begin{equation}
\label{alt_08}
_2\langle\mathbf{q}|{\textsf{N}}|\mathbf{q}\rangle_2 = \frac{g+\kappa_{q}}{g-\kappa_{q}} \textrm{e}^{-2\kappa_{q}\ell} \, .
\end{equation}
From this result we find the Casimir interaction energy for the slit geometry with boundary fields; for $d>1$ we obtain
\begin{equation}
\label{slitgeneral_21}
F_{\rm Casimir} = \frac{k_{\rm B}T}{2}  L_{\parallel}^{d-1}\int \frac{\rd\textbf{q}}{(2\pi)^{d-1}} \, \ln\left( 1 - \frac{g+\kappa_{q}}{g-\kappa_{q}} \textrm{e}^{-2\kappa_{q}\ell} \right) \, ,
\end{equation}
in agreement with previously reported results \cite{SREP_2022,SREP2023}, while for $d=1$
\begin{equation}
F_{\rm Casimir}^{(d=1)} = \frac{1}{2} k_{\rm B} T \ln\left( 1 - \frac{g+\kappa}{g-\kappa} \textrm{e}^{-2\kappa\ell} \right) \, .
\end{equation}
As expected, by performing the limit $g \rightarrow - \infty$ in the above expressions we retrieve (\ref{slit_dirichlet}) and (\ref{slitgeneral_22}) for Dirichlet boundary conditions.

A further simplification can be made. Let's define the operator $\mathsf{Y}_1$ such as, for any function $|\phi \rangle_1$ with support in $S_1$, $\mathsf{Y}_1|\phi \rangle_1 =\phi(\mathbf{s})+(1/g)\mathbf{n}(\mathbf{s})\cdot \boldsymbol{\nabla}\phi(\mathbf{s})$. With this definition, $\mathsf{K}_{12}=\mathsf{Y}_1\bar{\mathsf{K}}_{12}$, where $\bar{\mathsf{K}}_{12}$ is the integral operator with kernel $K(\mathbf{s},\mathbf{s}^\prime)$, with $\mathbf{s}\in S_1$ and $\mathbf{s}^\prime\in S_2$. On the other hand, $\mathsf{K}_{11}=\mathsf{Y}_1\bar{\mathsf{K}}_{11}$, where $\bar{\mathsf{K}}_{11}$ is the integral operator with kernel $J(\mathbf{s},\mathbf{s}^\prime)$, with $\mathbf{s},\mathbf{s}^\prime \in S_1$. Therefore, the operator $\mathsf{N}$ can be recast as
\begin{eqnarray}
\mathsf{N}&=&\textsf{K}_{21}\textsf{K}_{11}^{-1} \textsf{K}_{12} \textsf{K}_{22}^{-1}=\mathsf{K}_{21}(\mathsf{Y}_1 \bar{\textsf{K}}_{11})^{-1}\mathsf{Y}_1\bar{\textsf{K}}_{12}\textsf{K}_{22}^{-1}
\nonumber\\&=&
\mathsf{K}_{21} \bar{\textsf{K}}_{11}^{-1} \mathsf{Y}_1^{-1} \mathsf{Y}_1\bar{\textsf{K}}_{12}\textsf{K}_{22}^{-1}=\mathsf{K}_{21} \bar{\textsf{K}}_{11}^{-1}\bar{\textsf{K}}_{12}\textsf{K}_{22}^{-1}\, .
\label{defN2}
\end{eqnarray}
In this expression, the integral operator $\bar{\mathsf{K}}_{11}$, acting on a function $|\phi\rangle_1$ with support on $S_1$, can be recast as
\begin{eqnarray}
&&\bar{\mathsf{K}}_{11}|\phi\rangle_1=\int_{S_1} \rd \mathbf{s}^\prime\phi(\mathbf{s}^\prime)\Biggl(K(\mathbf{s},\mathbf{s}^\prime)\nonumber\\&+&
\frac{1}{g}\frac{d K(\mathbf{s},\mathbf{s}^\prime+\epsilon\mathbf{n}(\mathbf{s}^\prime))}{d\epsilon}\Bigg |_{\epsilon=0^+}\Biggr)\nonumber\\&=&\int_{S_1} 
\rd \mathbf{s}^\prime\phi(\mathbf{s}^\prime)\Biggl(K(\mathbf{s},\mathbf{s}^\prime)+\frac{1}{g}
\mathbf{n}(\mathbf{s}^\prime)\cdot
\boldsymbol{\nabla}_{\mathbf{s}^\prime}K(\mathbf{s},\mathbf{s}^\prime)\nonumber\\
&-&\frac{\kappa}{g}\delta(\mathbf{s}-\mathbf{s}^\prime)\Biggr )\, ,
\end{eqnarray}
where the gradient is evaluated on $\mathbf{s}^\prime \in S_1$ and the integral must be understood in the sense of Cauchy's principal value.

\subsection{Boundary-integral method derivation}
\label{sec_5_4}
The Casimir contribution can be also obtained by the direct evaluation of the correlation function of the liquid phase in contact with the wall, $K_1(\mathbf{s},\mathbf{s}^\prime)$, where $\mathbf{r}$ and $\mathbf{r}^\prime$ are above $S_1$. This procedure was considered for a different physical system in Ref.~\cite{Bordag_2006}. The function $K_1$ satisfies
\begin{equation}
\label{sec5_3-1}
\left( - \nabla^{2}_{\mathbf{r}} + \kappa^{2} \right) K_1(\textbf{r},\textbf{r}^{\prime}) = 2\kappa \delta(\mathbf{r}-\mathbf{r}^\prime) \, ,
\end{equation}
subject to the boundary conditions $Y_{\mathbf{s}}K_1(\mathbf{s},\mathbf{r}^\prime)\equiv K_1\mathbf{s},\mathbf{r}^\prime)+(1/g)\mathbf{n}(\mathbf{s})\cdot\boldsymbol{\nabla}_{\mathbf{s}}K_1(\mathbf{s},\mathbf{r}^\prime)=0$ for $\mathbf{s}\in S_1$. We can obtain the Casimir interaction energy by the procedure outlined in Sec. \ref{sec_5_2}, by substituting the wetting layer boundary $S$ by the interface $S_2$ and the bulk correlation function $K(\mathbf{s},\mathbf{s}^\prime)$ by $K_1(\mathbf{s},\mathbf{s}^\prime)$. We note that $K_1^\infty(\mathbf{s},\mathbf{s}^\prime)=K(\mathbf{s},\mathbf{s}^\prime)$. Thus,
\begin{equation}
\label{sec5_3-1_2}
W_C=\frac{k_\textrm{B}T}{2}\tr_{S_2}[\ln \mathsf{L}_1 \mathsf{K}_{22}^{-1}]\, ,
\end{equation}
where $\mathsf{K}_{22}$ is defined as the integral operator acting on the functions with support on $S_2$ with kernel $K(\mathbf{s},\mathbf{s}^\prime)$, and $\mathsf{L}_1$ is the integral operator with kernel $K_1(\mathbf{s},\mathbf{s}^\prime)$, $\mathbf{s},\mathbf{s}^\prime\in S_2$.

As for the determination of the MF contribution, the technical challenge here is to evaluate $K_1$ systematically and, again, this can be done using a boundary integral technique by expanding about the bulk solution (\ref{03062022_1616}). To this end we define
\begin{equation}
\Delta K_{1}(\textbf{r},\textbf{r}^{\prime}) = K_{1}(\textbf{r},\textbf{r}^{\prime}) - K(\textbf{r},\textbf{r}^{\prime})
\end{equation}
which satisfies the Helmholtz equation
\begin{equation}
\label{03062022_1630}
\left( - \nabla_{\mathbf{r}}^{2} + \kappa^{2} \right) \Delta K_{1}(\textbf{r},\textbf{r}^{\prime}) = 0 \, ,
\end{equation}
together with the boundary conditions
\begin{equation}
\label{bc1}
Y_{\mathbf{s}} \Delta K_{1}(\textbf{s},\textbf{r}^\prime) = - Y_{\mathbf{s}}K(\textbf{s},\textbf{r}^\prime) \, , \qquad \textbf{s} \in S_1 \, .
\end{equation}
This function can be determined using the boundary integral techniques described in Ref.~\cite{RESPG_2018}. Its single-layer representation is
\begin{equation}
    \label{sec5_3-2}
    \Delta K_1(\mathbf{r},\mathbf{r}^\prime)=\int_{S_1} \rd \mathbf{s} b(\mathbf{r}^\prime,\mathbf{s})K(\mathbf{s},\mathbf{r})\, ,
\end{equation}
where the auxiliar field $b(\mathbf{r}^\prime,\mathbf{s})$, under application of the boundary condition (\ref{bc1}), satisfies the boundary integral equation
\begin{eqnarray}
    \label{sec5_3-3}
    &-&J(\mathbf{s},\mathbf{r}^\prime)\equiv -Y_{\mathbf{s}}K(\mathbf{s},\mathbf{r}^\prime)=
    \int_{S_1} \rd \mathbf{s}^\prime b(\mathbf{r}^\prime,\mathbf{s}^\prime) Y_{\mathbf{s}}K(\mathbf{s}^\prime,\mathbf{s})\nonumber\\
    &=&\int_{S_1} \rd \mathbf{s}^\prime b(\mathbf{r}^\prime,\mathbf{s}^\prime)\Biggl(K(\mathbf{s},\mathbf{s}') + \frac{1}{g}\mathbf{n}_1(\mathbf{s})\mathbf{\cdot} \boldsymbol\nabla_{\mathbf{s}}K(\mathbf{s},\mathbf{s}') \nonumber\\&-& \frac{\kappa}{g}\delta(\mathbf{s}-\mathbf{s}')\Biggr)\, 
\end{eqnarray}
for $\textbf{s} \in S_1$. Thus
\begin{equation}
\label{gen_10}
b(\mathbf{r}^\prime,\mathbf{s})= - \int_{S_{1}}\rd\textbf{s}^\prime  \, J(\textbf{s},\textbf{r}^\prime) X(\textbf{s},\textbf{s}^{\prime}) \, ,
\end{equation}
where $X(\textbf{s},\textbf{s}^{\prime})$ is the kernel of the inverse of the boundary integral operator with kernel $Y_{\mathbf{s}}K(\mathbf{s}^\prime,\mathbf{s})$. Therefore, $\Delta K_1(\mathbf{r},\mathbf{r}^\prime)$ reads
\begin{equation}
    \label{sec5_3-3}
    \Delta K_1(\mathbf{r},\mathbf{r}^\prime)=-\int_{S_1} \rd \mathbf{s} \int_{S_{1}}\rd\textbf{s}^\prime \, J(\textbf{s},\textbf{r}^\prime) X(\textbf{s},\textbf{s}^{\prime}) K(\mathbf{s},\mathbf{r})\, 
\end{equation}
and, consequently, the boundary integral operator with kernel $\Delta K_1$, $\Delta \mathsf{L}_1$, can be written in terms of the boundary integral operators defined in the previous subsection as
\begin{equation}
    \Delta \mathsf{L}_1 = \mathsf{K}_{21} \bar{\mathsf{K}}_{11}^{-1} \bar{\mathsf{K}}_{12}\, .
\end{equation}
Taking into account that $\mathsf{L}_1=\mathsf{K}_{22}+\Delta\mathsf{L}_1$, we obtain that $\mathsf{L}_1 \mathsf{K}_{22}^{-1}=\mathsf{I}_2+\Delta \mathsf{L}_1 \mathsf{K}_{22}^{-1} = 
\mathsf{I}_2-\mathsf{K}_{21} \bar{\mathsf{K}}_{11}^{-1} \bar{\mathsf{K}}_{12} \mathsf{K}_{22}^{-1}=\mathsf{I}_2-\mathsf{N}$, where the boundary integral operator $\mathsf{N}$ is given by Eq. (\ref{defN2}). Thus, Eq. (\ref{sec5_3-1_2}) leads to the expression (\ref{wl_26}) for the Casimir contribution to the binding potential.

\subsection{Diagrammatic expansion}
\label{sec_5_6}
Our next goal is to obtain a diagrammatic expansion for the Casimir interaction akin to the mean-field non-local binding potential. For this purpose, the formal expression (\ref{wl_26}) of the Casimir contribution can be written as an expansion
\begin{equation}
\label{20102022_2057}
\beta W_{\rm C}[\ell,\psi] = \frac{1}{2} \sum_{n=1}^{\infty} (\Omega_{n}^{n})_{\rm C}\, ,
\end{equation}
where
\begin{equation}
\label{omega_n_n}
(\Omega_C)_{n}^{n} = - \tr_{S_2} \frac{\mathsf{N}^n}{n} \, .
\end{equation}
This notation is introduced in analogy to the different terms of the mean field non-local binding potential \cite{PRBRE_2006}, and their curvature corrections \cite{RESPG_2018}. Now we notice that the kernel of the operator $\mathsf{N}$ involves two $K$-kernels connecting the wall and the interface. Since both $K$ and $\textbf{n} \cdot \nabla K$ decay exponentially with the distance between $S_{1}$ and $S_{2}$, when their arguments are on different surfaces, this means that, for each $\textbf{n}$, we can anticipate that $(\Omega_{n}^{n})_{\rm C} \sim e^{-2 \kappa n d(\ell,\psi)}$ where $d(\ell,\psi)$ is the minimum distance between the wall and interface. Here we focus on the leading terms in (\ref{20102022_2057}) corresponding to large separations between the interfaces and large radii of curvature as pertinent to discussions of wetting transitions.

Let's start with the case of Dirichlet boundary conditions on the wall, i.e. $g\to -\infty$. We now recast our expression diagrammatically using the same dictionary introduced earlier [Eqs.~(\ref{20102022_2058})-(\ref{defdiagrams})]. The diagrammatic representation of the kernels of all the operators $\mathsf{K}_{ij}$ is already known \cite{RESPG_2018}:
\begin{equation}
\label{wl_27}
K_{11}^{-1} =  \figu{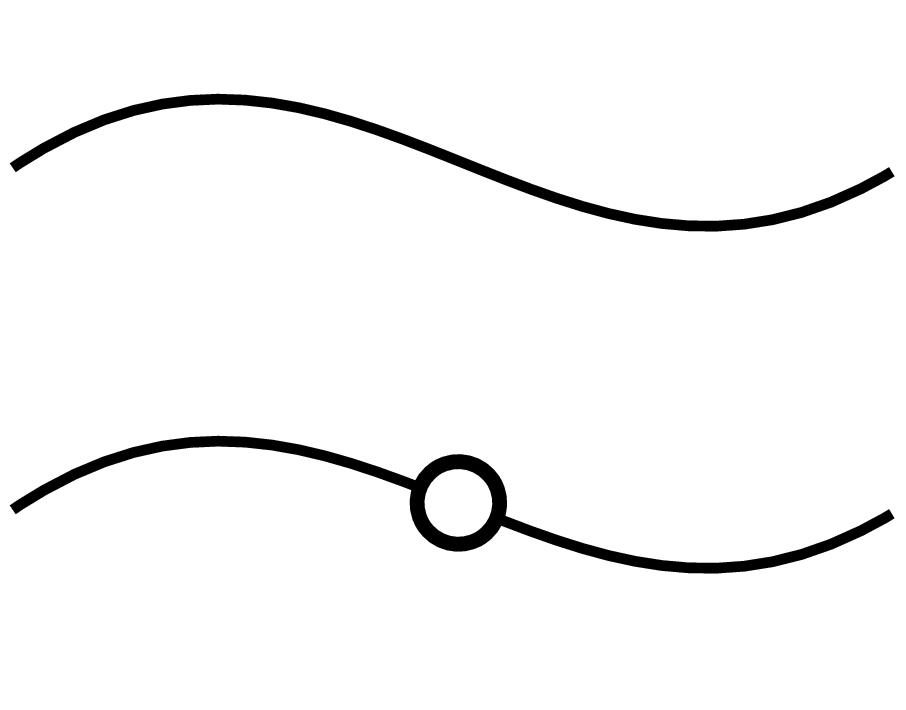} -  \figu{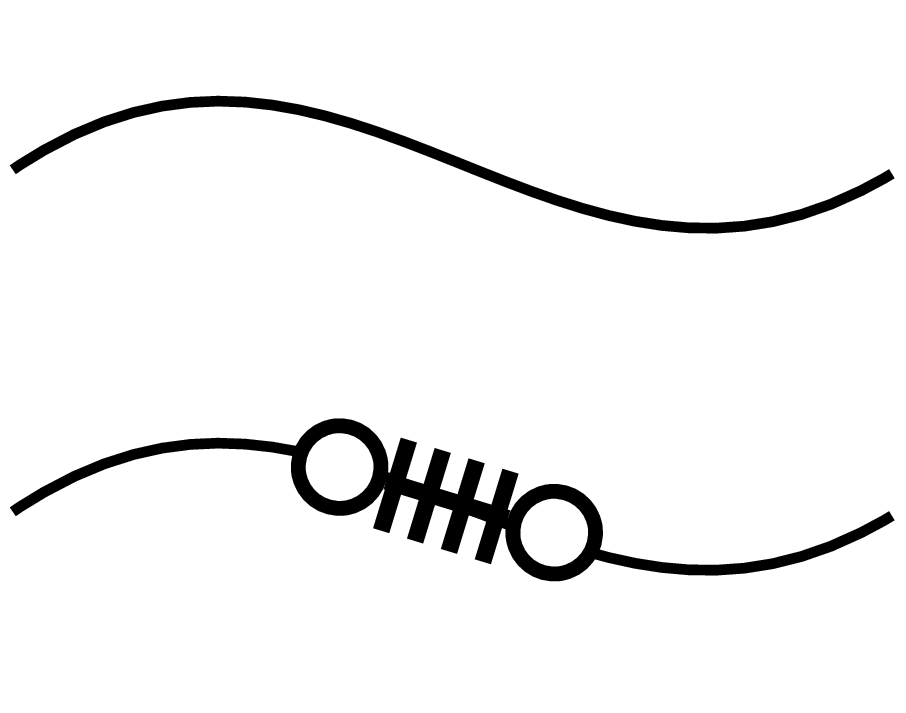} + \figu{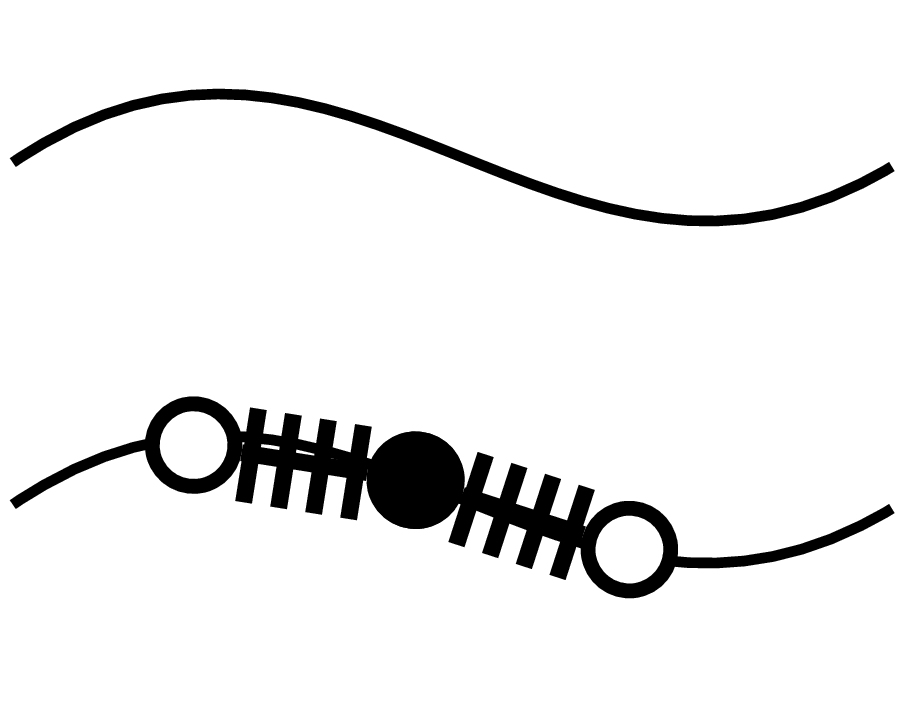} + \dots \, ,\\
\end{equation}
\begin{equation}
\label{wl_28}
K_{22}^{-1} =  \figu{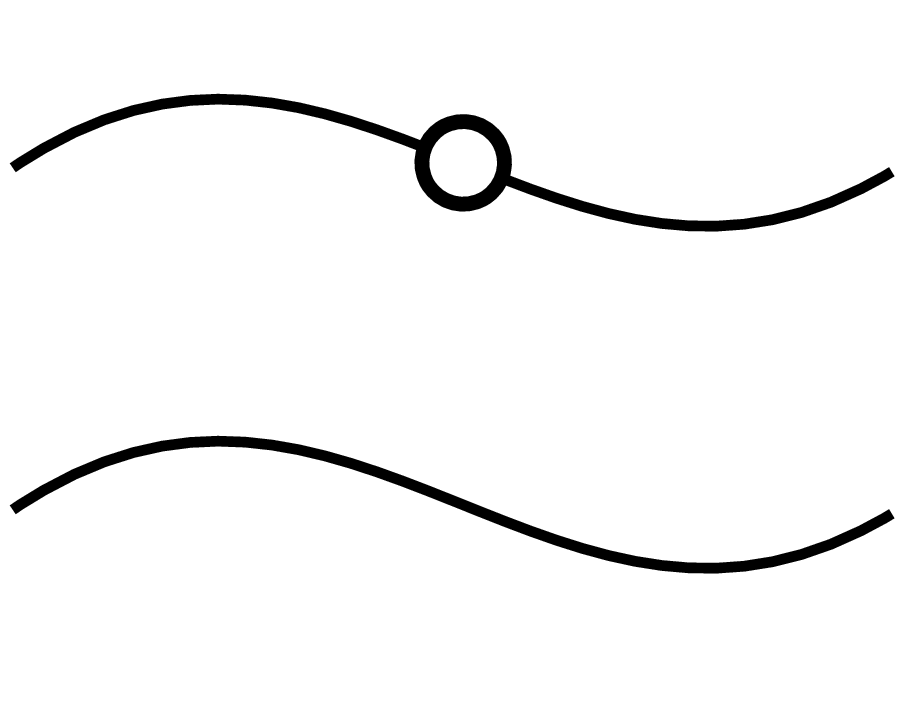} -  \figu{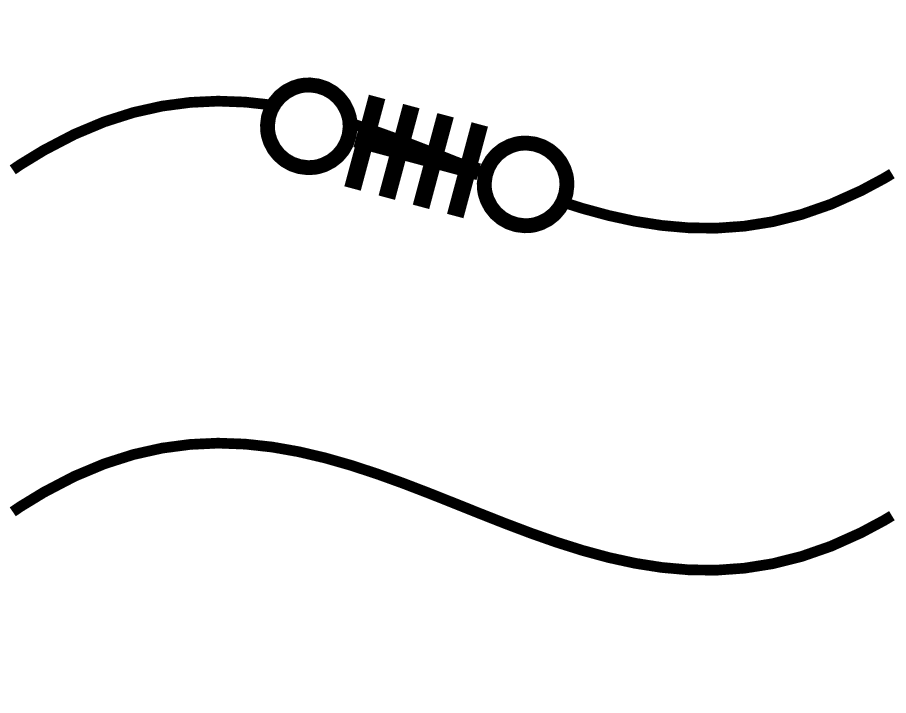} + \figu{d06} + \dots \, ,
\end{equation}
and
\begin{equation}
\label{wl_29}
K_{12} = \figu{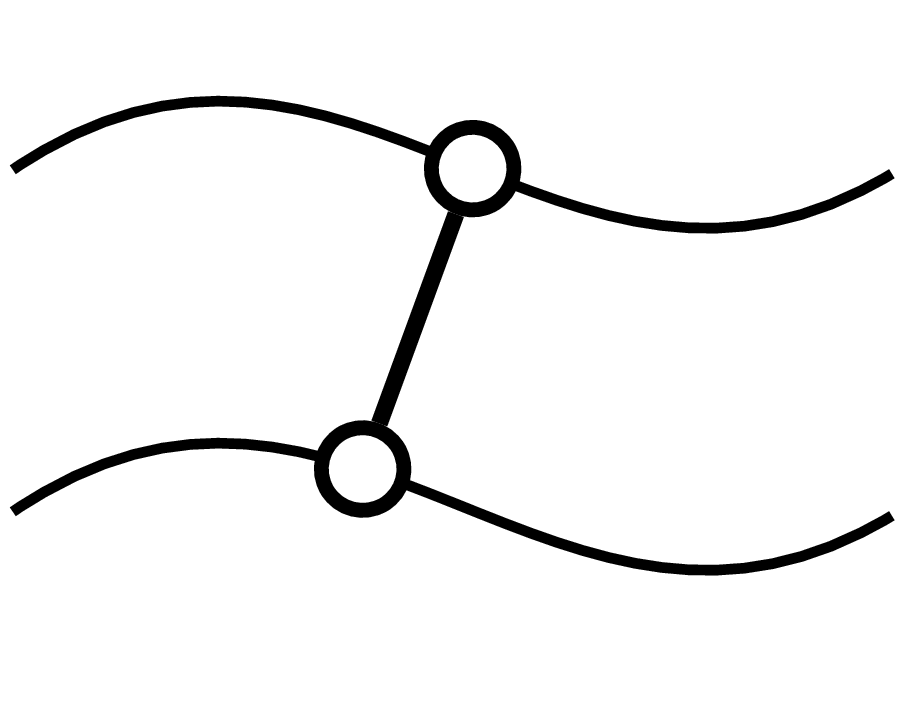} \, \qquad K_{21} = \figu{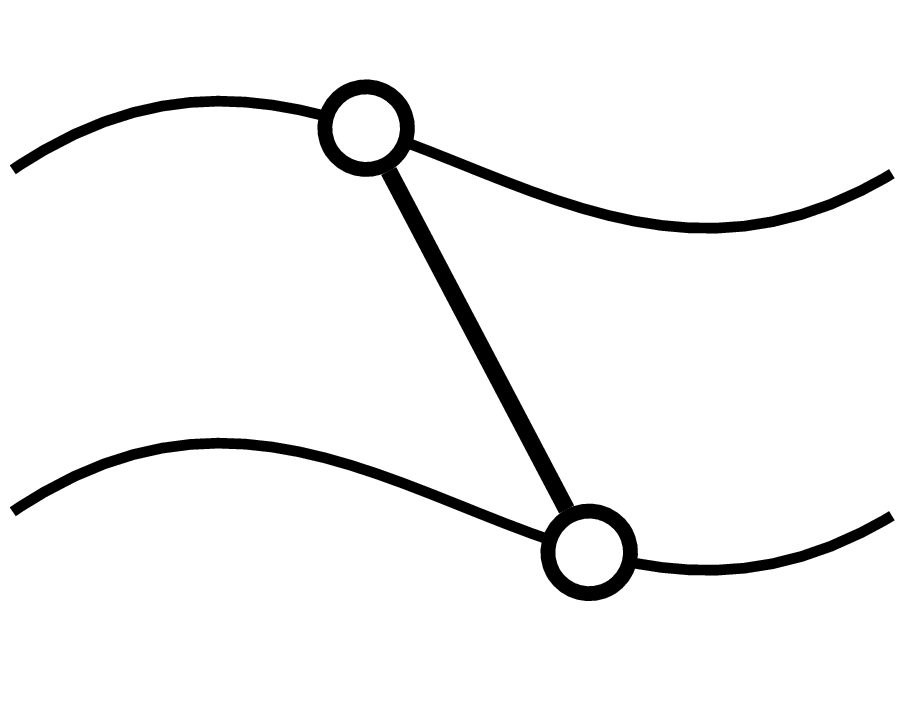} \, .
\end{equation}
with $K_{12} = K_{21}^{\dag}$. As a result, the kernel of the integral operator $\mathsf{N}$, $N(\mathbf{s},\mathbf{s}^\prime)$, admits the diagrammatic expansion
\begin{eqnarray} \nonumber
\label{wl_31}
&& N(\mathbf{s},\mathbf{s}^\prime) = \figu{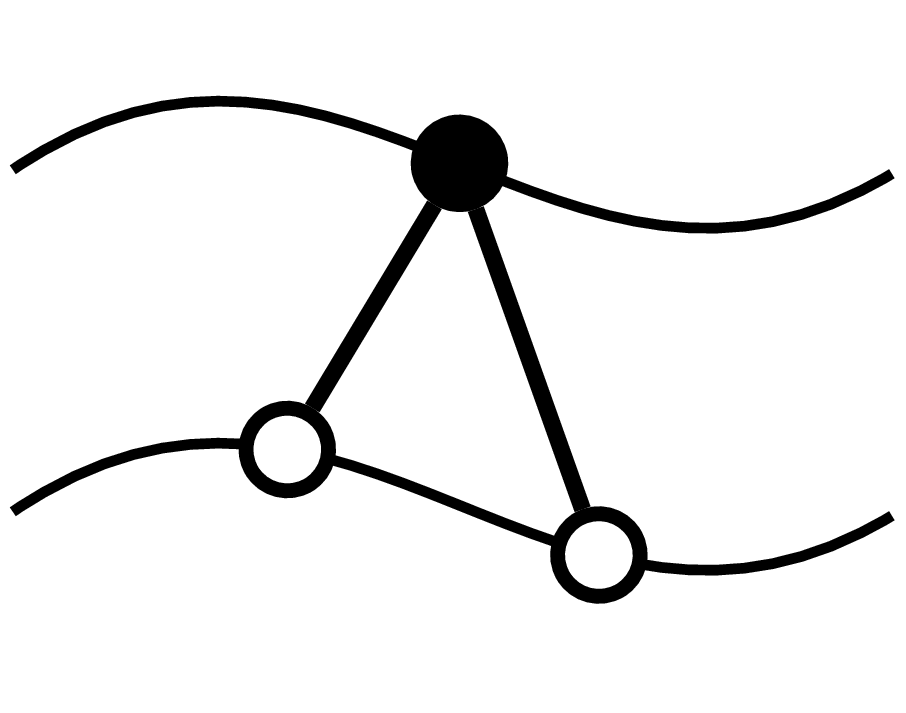} - \figu{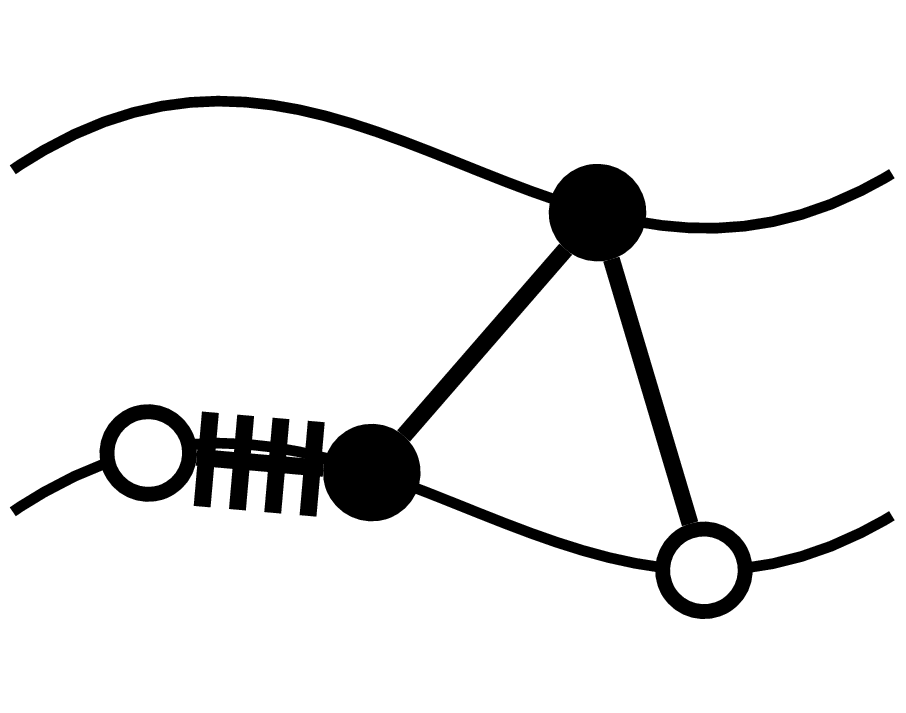} \\ \nonumber
&& - \fig{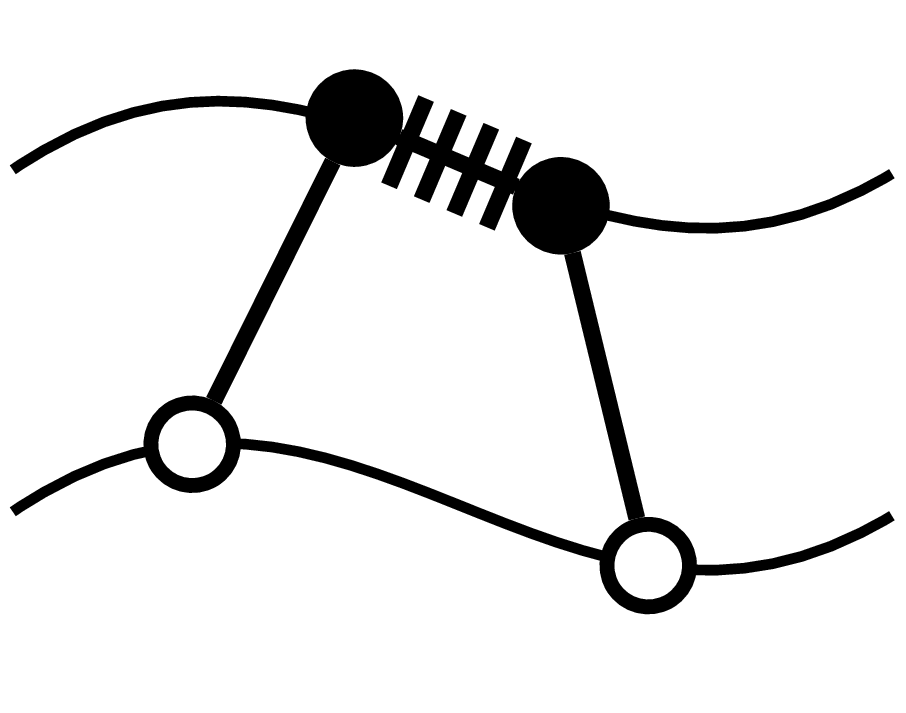} + \fig{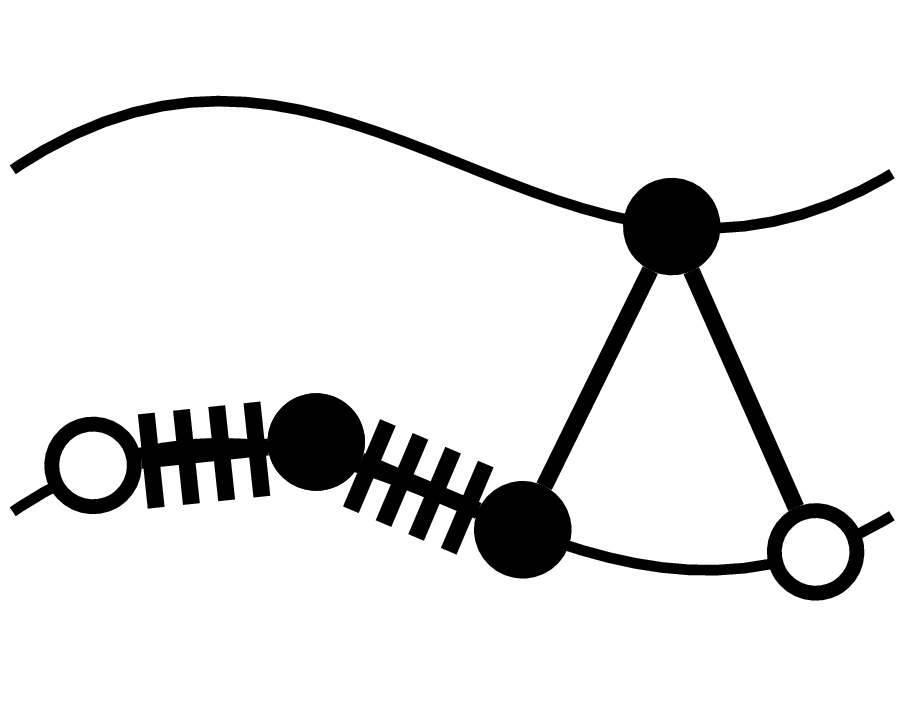} + \fig{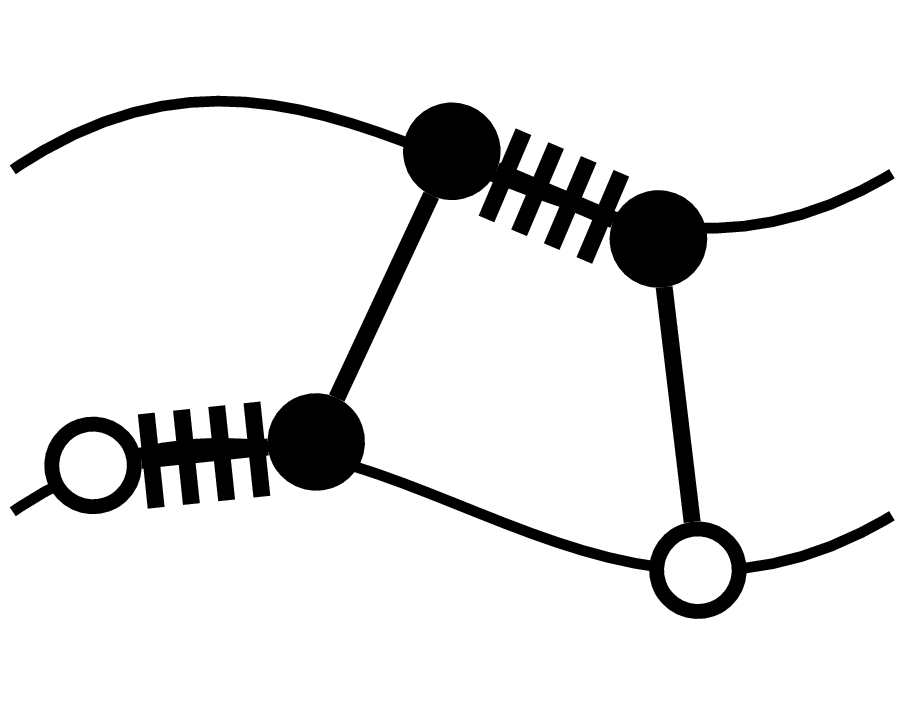} \\
&& + \fig{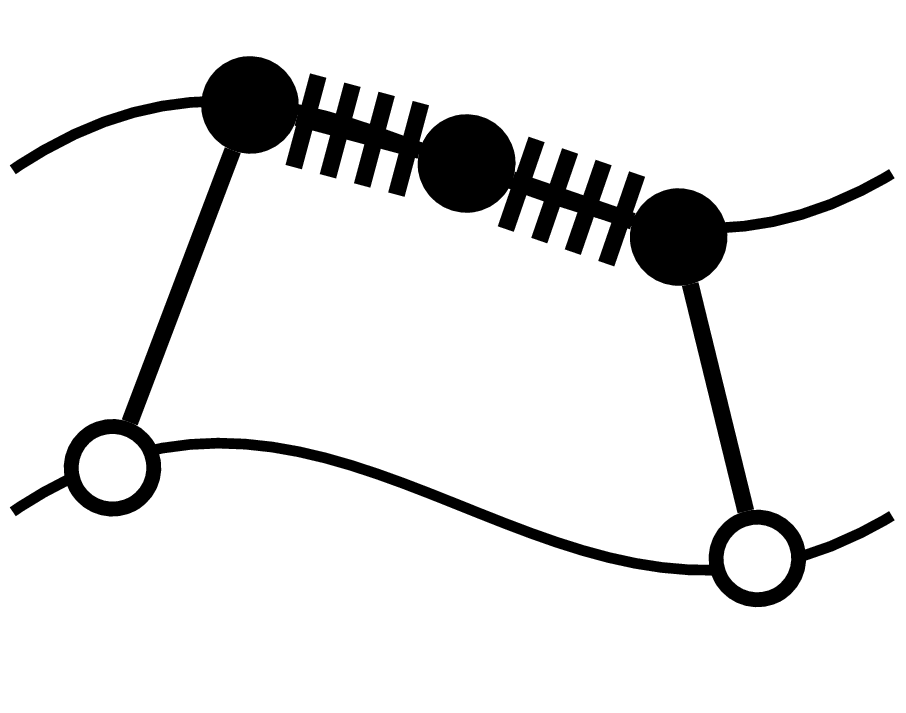} + \dots \, ,
\end{eqnarray}
where the open circles represent both $\mathbf{s}$ and $\mathbf{s}^\prime$, and the filled circles mean integration over the surface with the area element. The latter expansion allows us to express the Casimir interaction in a diagrammatic way by inserting (\ref{wl_31}) into (\ref{omega_n_n}) and recalling that performing the trace $\tr_{S_2}$ corresponds to glue together the end of the diagrams and integrating over them. Note that, for each $n$, there will be $2n$ legs connecting both interfaces. Thus, the 2-leg contribution $\left( \Omega_{\rm Cas}\right)_{2}^{2}$ is the functional
\begin{eqnarray} \nonumber
\label{wl_34}
\left( \Omega_C\right)_{1}^{1} & = & - \figu{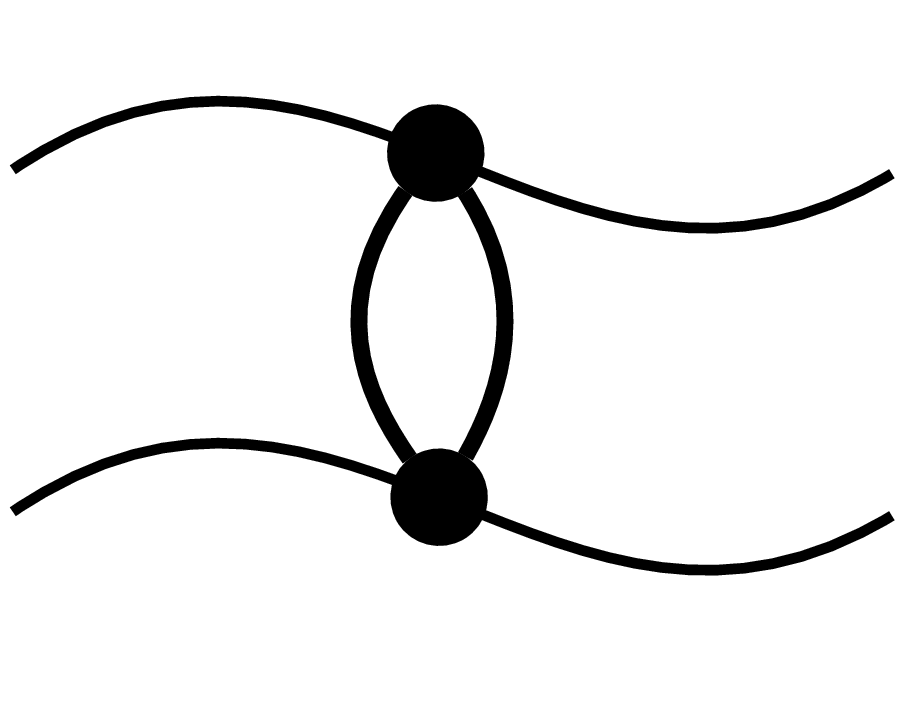} + \figu{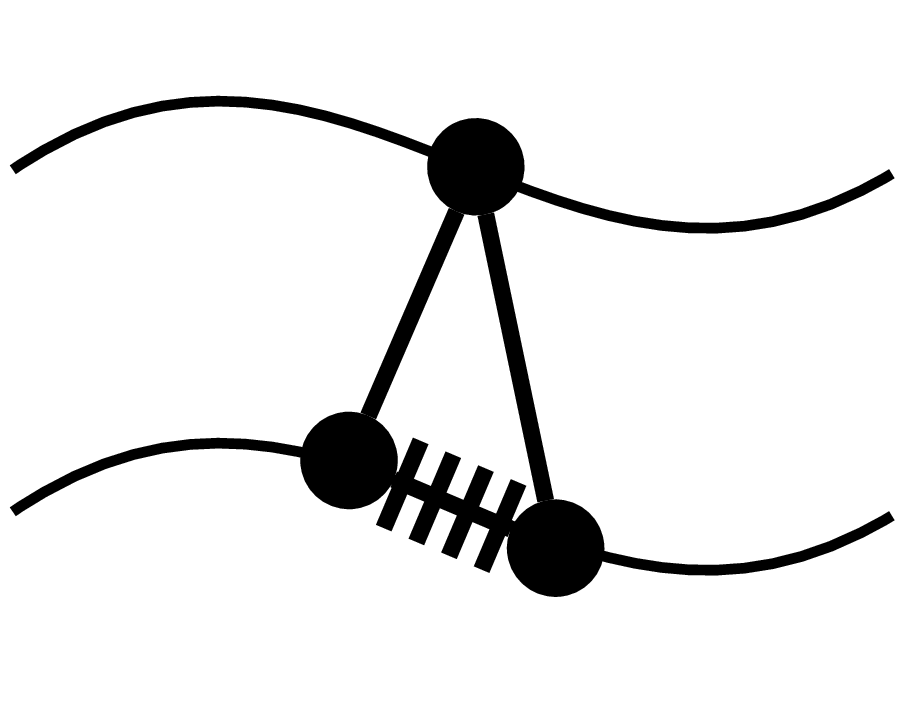} \\
& + & \figu{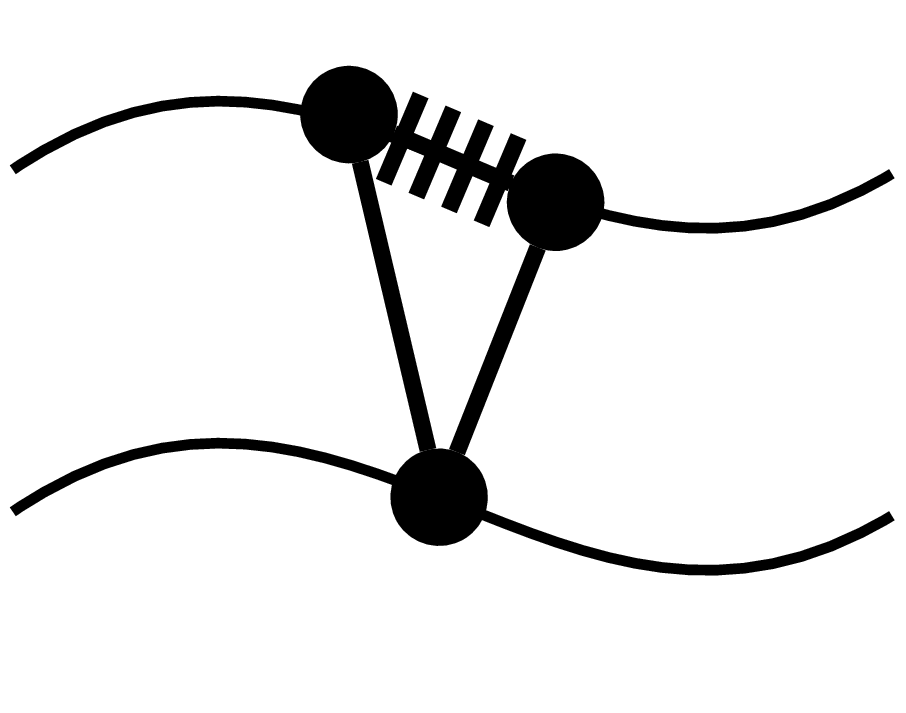} + \dots \, ,
\end{eqnarray}
while for the 4-leg contribution
\begin{equation}
\label{wl_35}
\left( \Omega_C\right)_{2}^{2}  = - \frac{1}{2} \figu{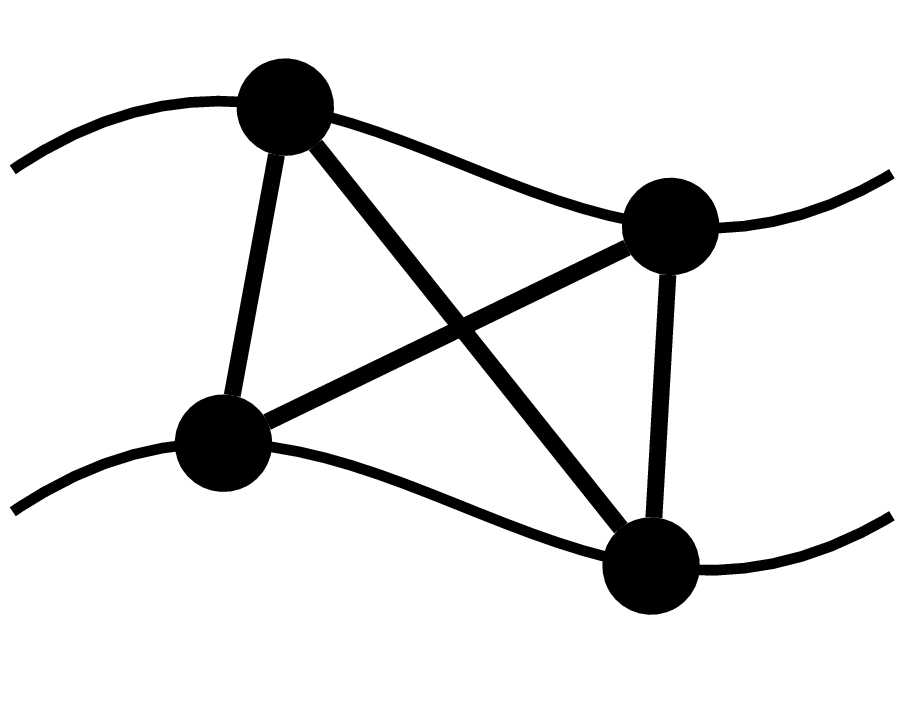} + \frac{1}{2} \figu{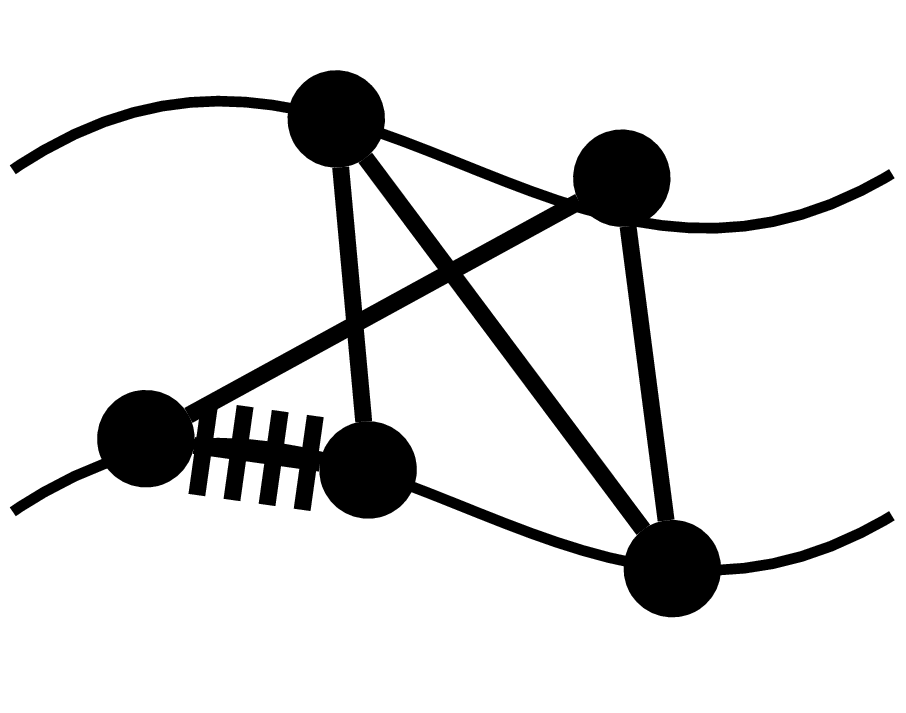} + \dots \, ,
\end{equation}
and similarly for larger values of $n$. The overall coefficient in front of each diagram is
\begin{equation}
\label{wl_36}
\frac{1}{(n_{\rm legs}/2)} (-1)^{n_{1}+n_{2}+1} \, ,
\end{equation}
where $n_{\rm legs}$ is the number of legs connecting both interfaces, and $n_{i}$ is the number of $U$-bonds on the $i$-th interface.

In general, the decorated $U$-bond diagrams lead to curvature corrections of the bare diagrams (without $U$-bonds), but only in a global way \cite{RESPG_2018}. Otherwise, $U-$bonds renormalize the $K-$kernels connecting the wall and the interface by using the identities
\begin{eqnarray}\nonumber
\label{wl_37}
&& \figu{d14} - \figu{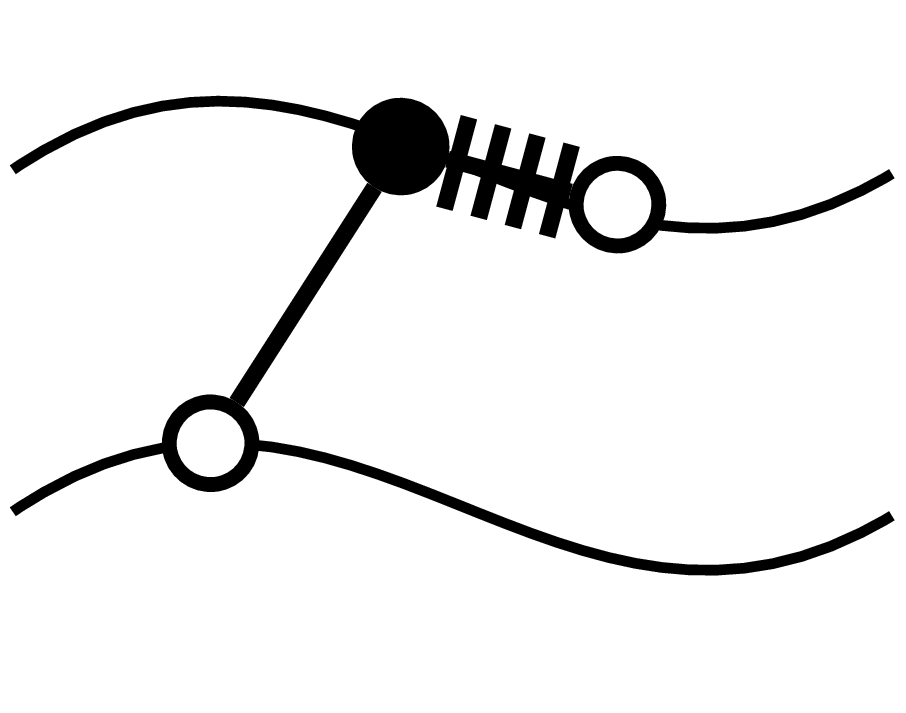} + \figu{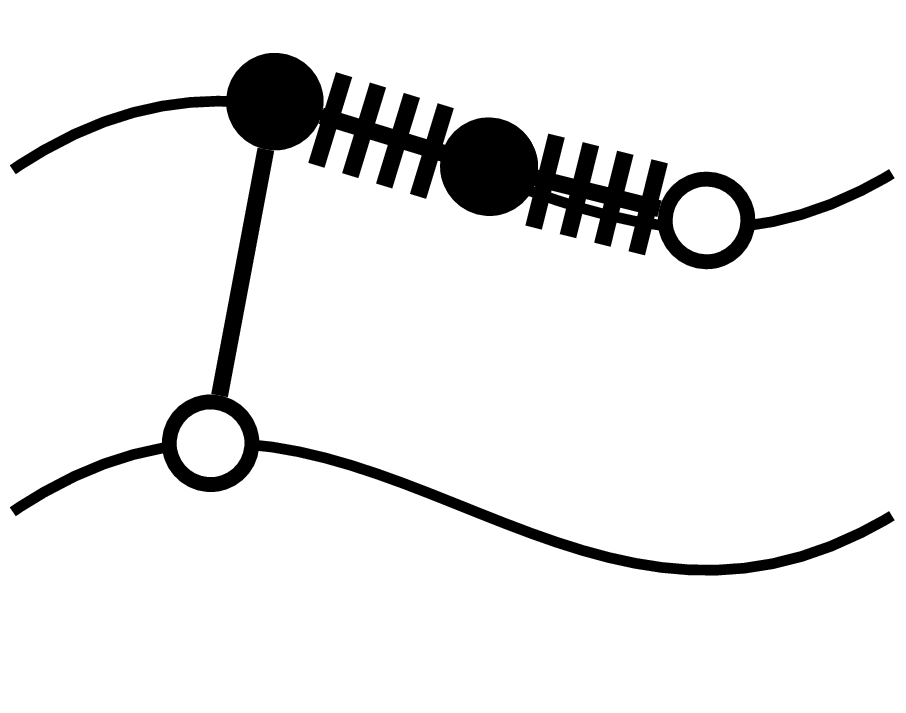} + \dots \\
&& = - \figu{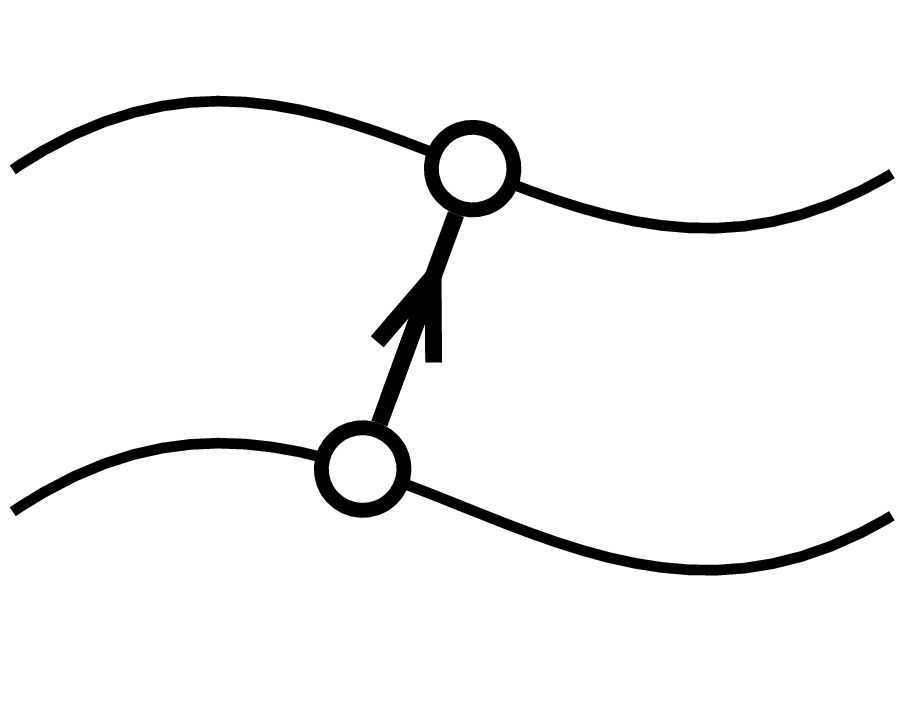} - \figu{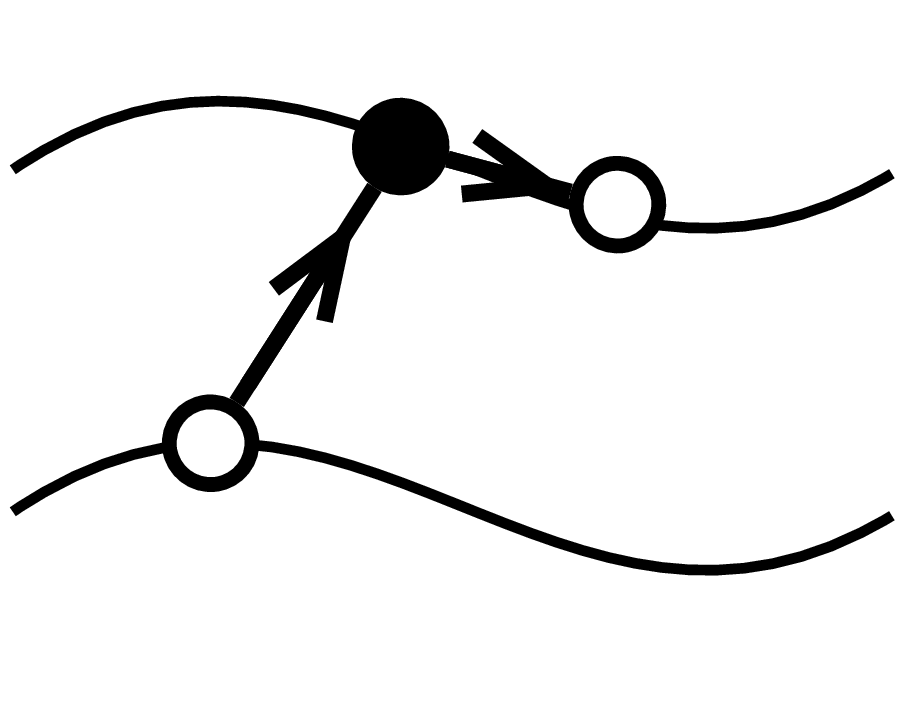} - \figu{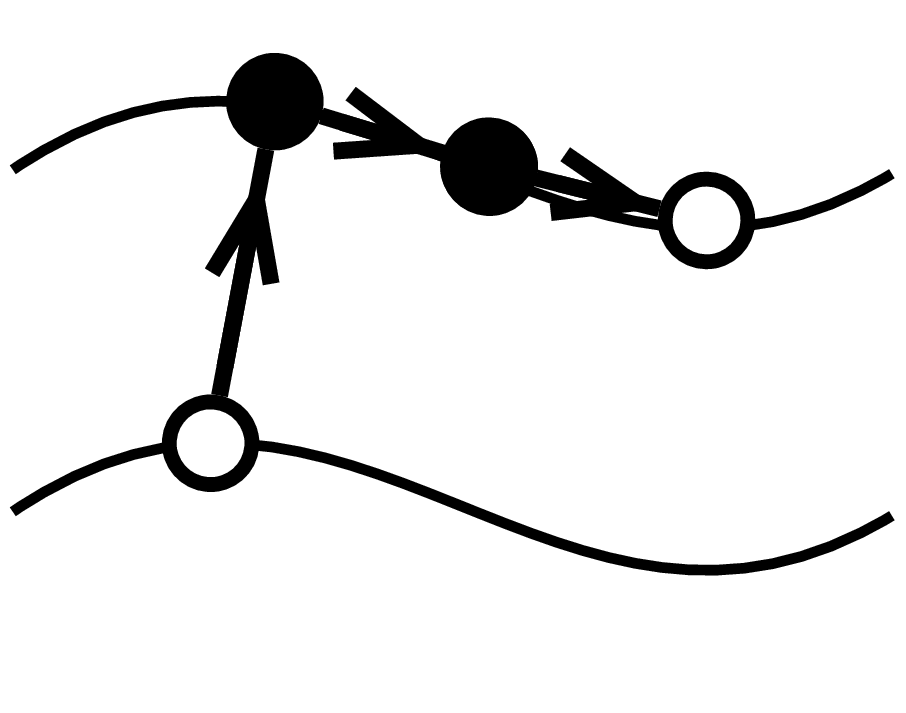} + \dots
\end{eqnarray}
and
\begin{eqnarray}\nonumber
\label{wl_38}
&& \figu{d14} - \figu{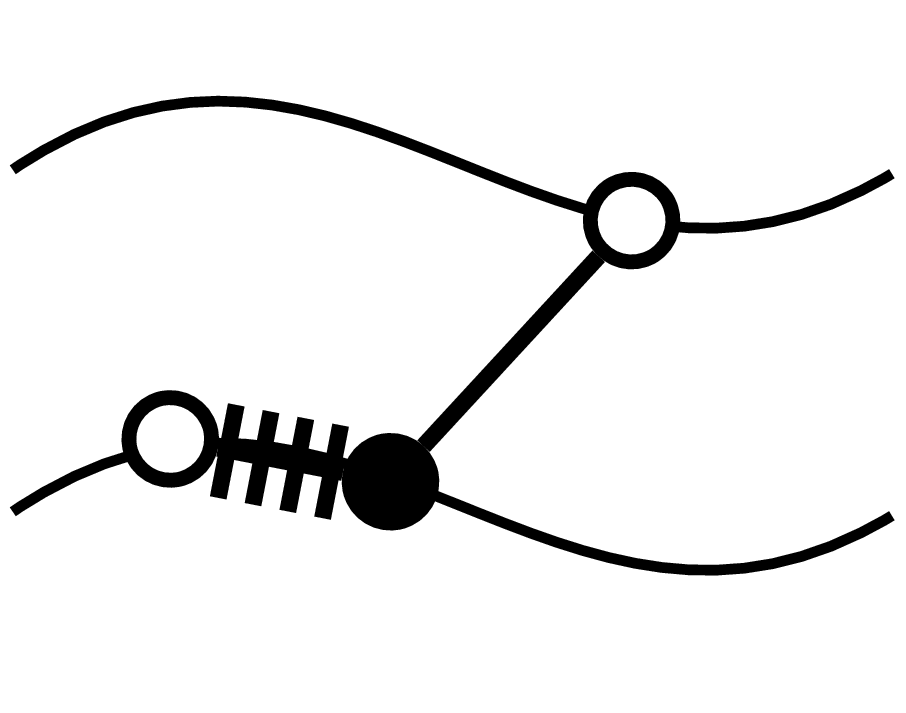} + \figu{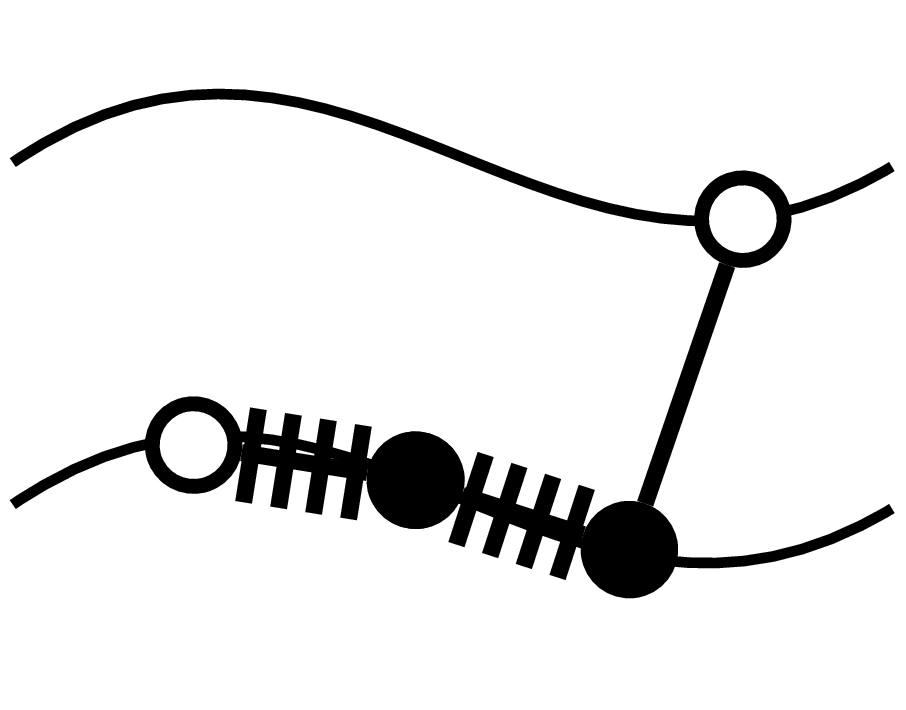} \\
&& = \figu{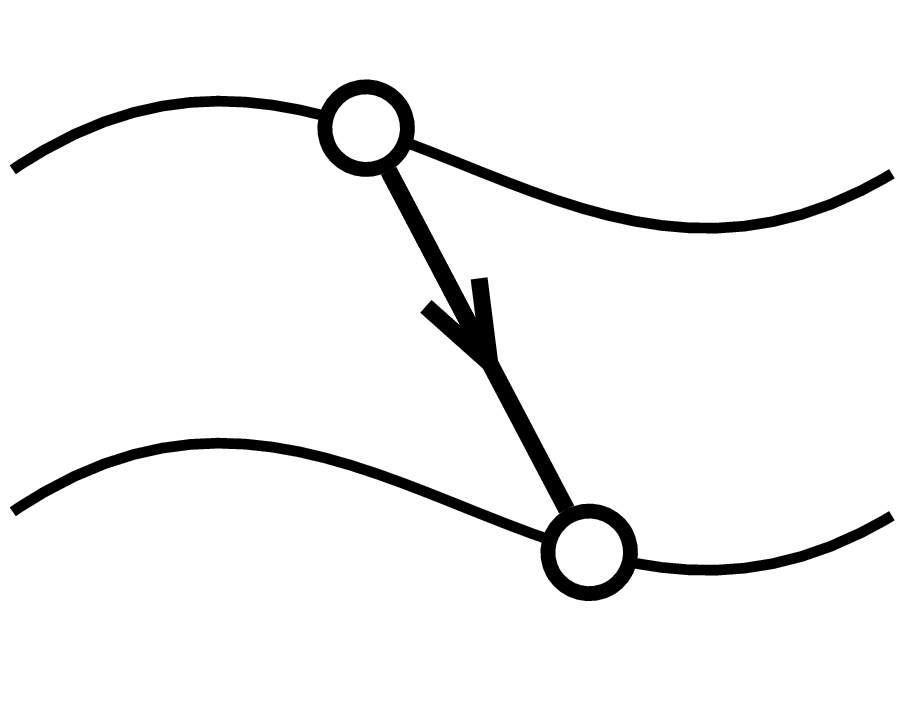} - \figu{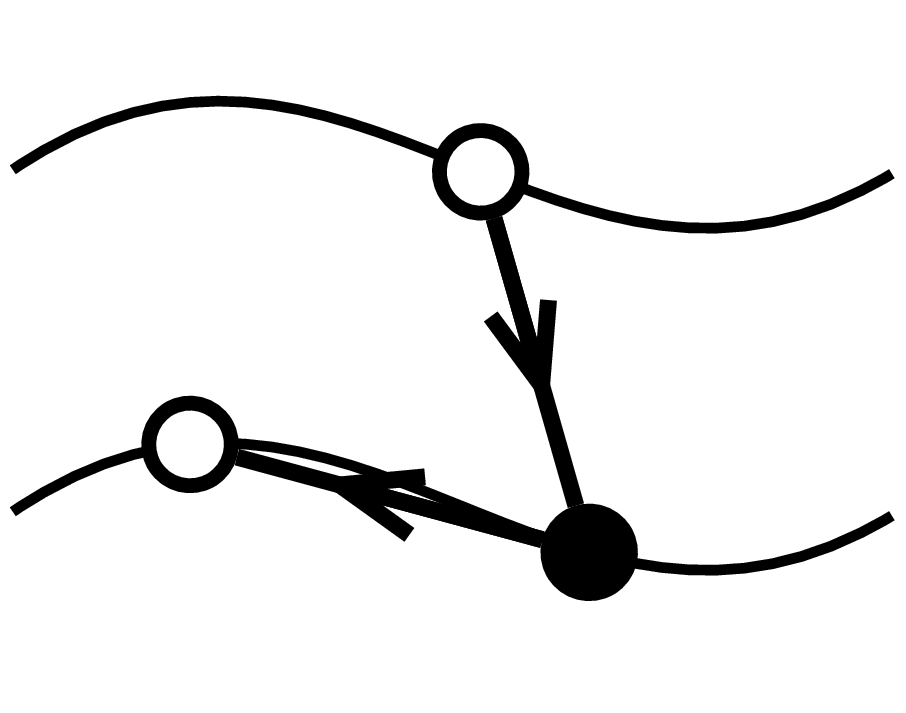} + \figu{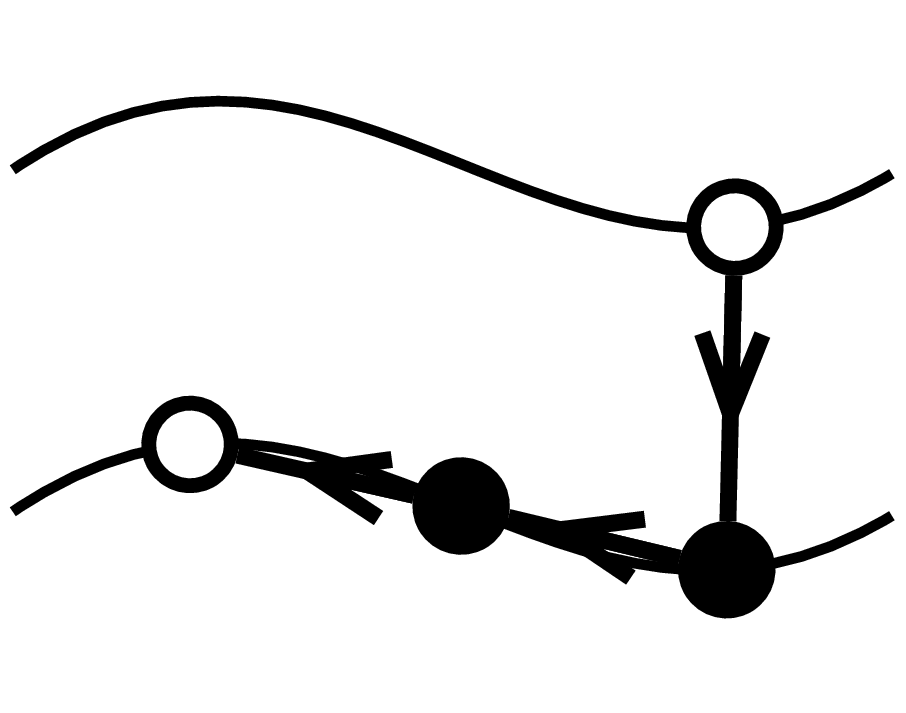} + \dots \, .
\end{eqnarray}
Note that $\kappa^{-1}\partial_{n}K$ bonds on both the wall and the interface are $\mathcal{O}(H/\kappa)$, where $H$ is the mean curvature of the boundary \cite{RESPG_2018} Then, Eq. (\ref{wl_31}) becomes
\begin{eqnarray}
\label{wl_39}\nonumber
&& N =  - \figu{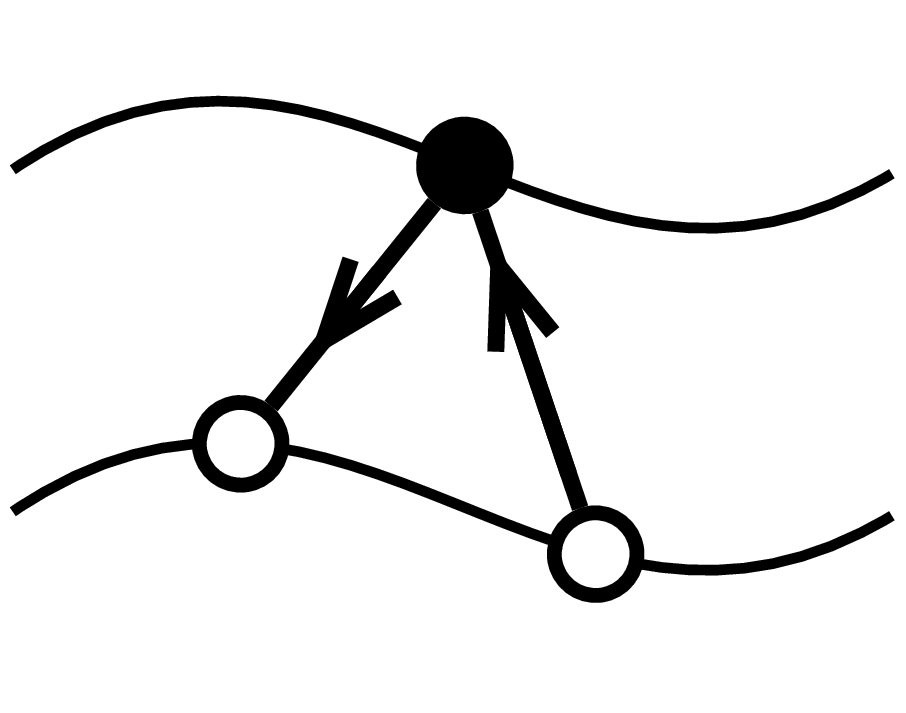} + \figu{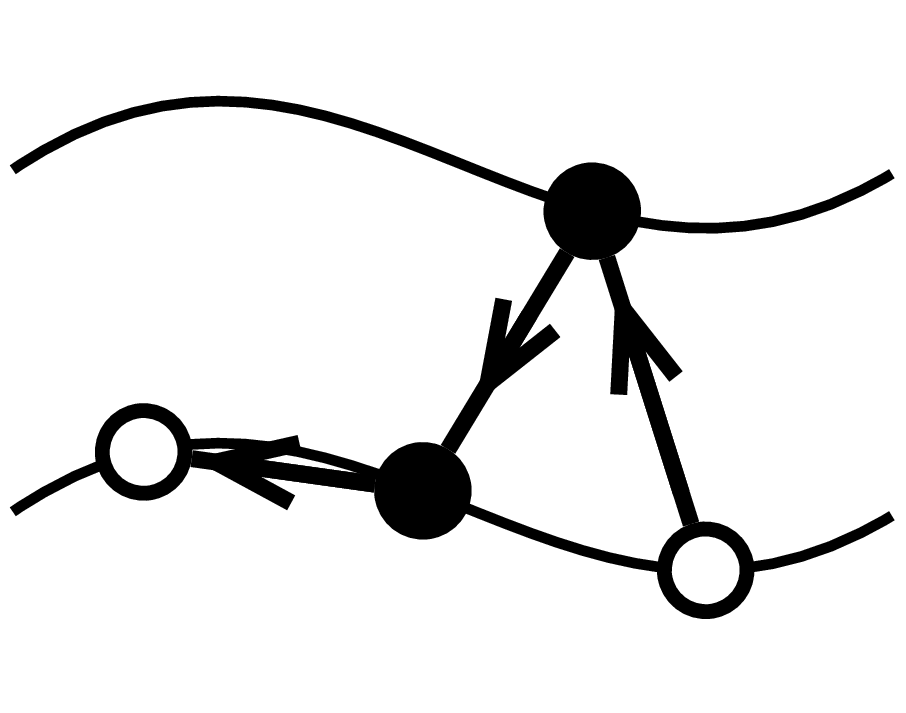} - \figu{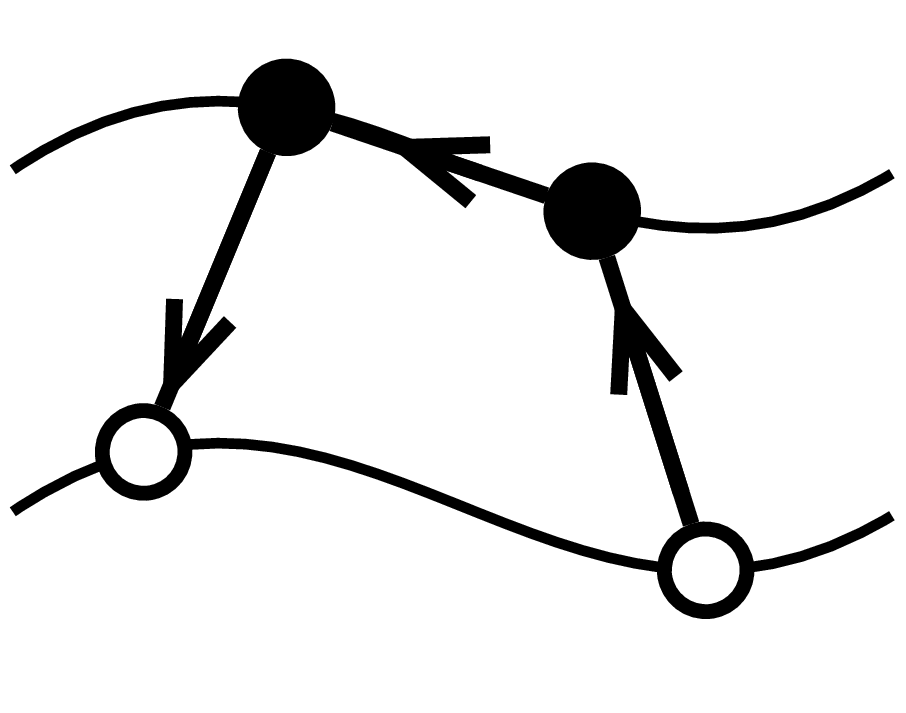} + \dots \, .\\ 
\end{eqnarray}
The functionals (\ref{wl_34}) and (\ref{wl_35}) become
\begin{eqnarray} \nonumber
\label{wl_40}
\left( \Omega_C\right)_{1}^{1} & = & \figu{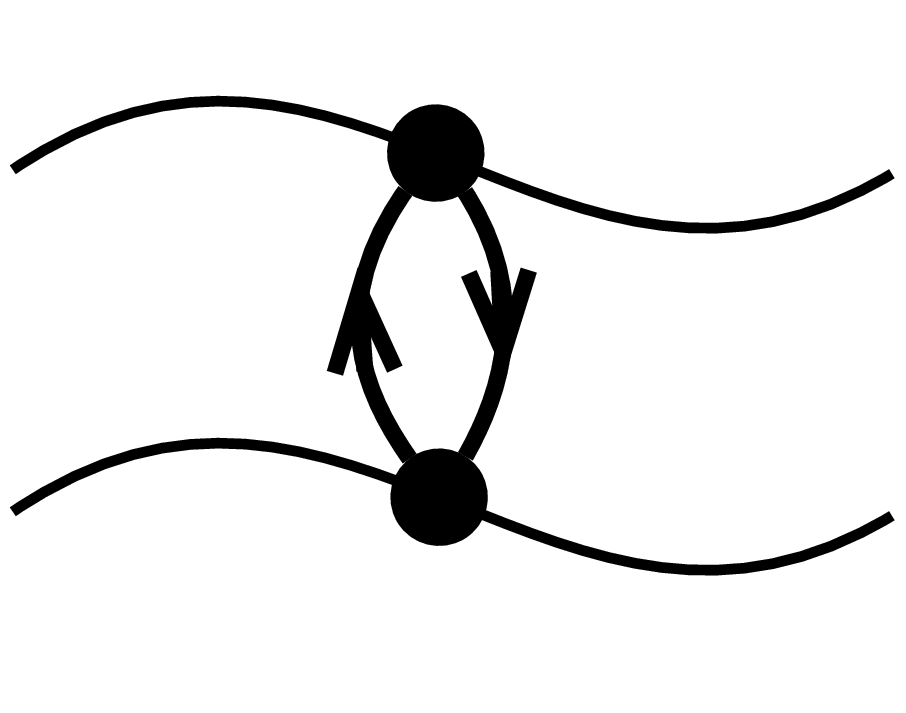} - \figu{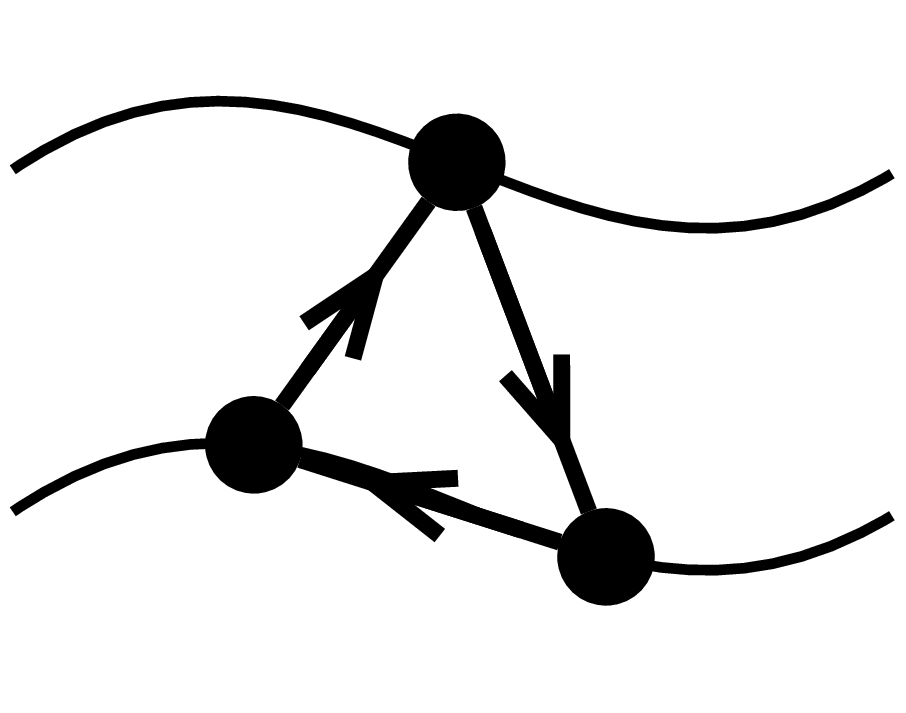} + \figu{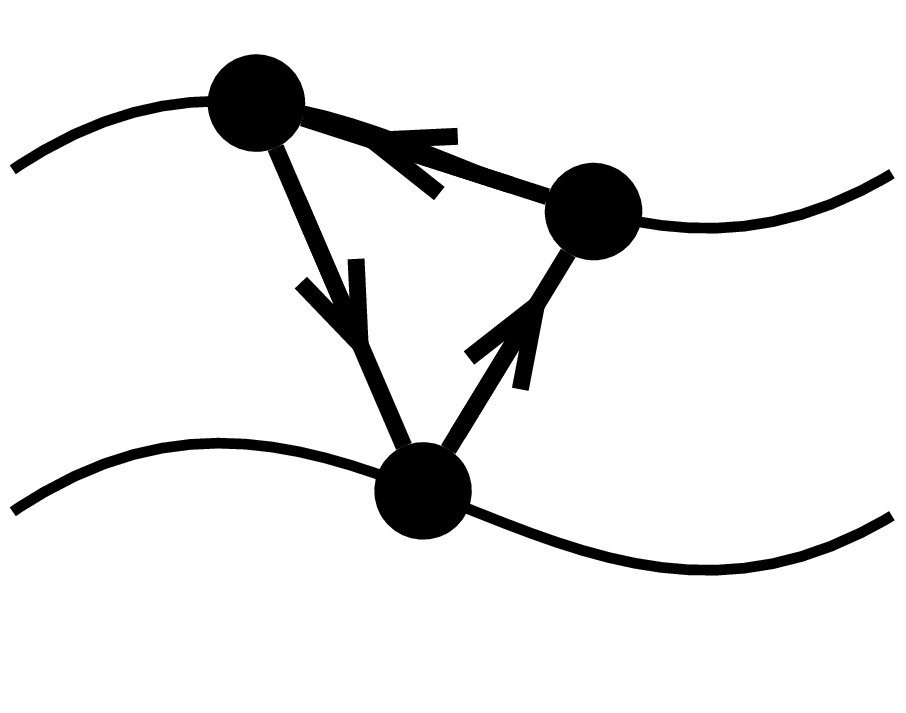} \\
& - & \figu{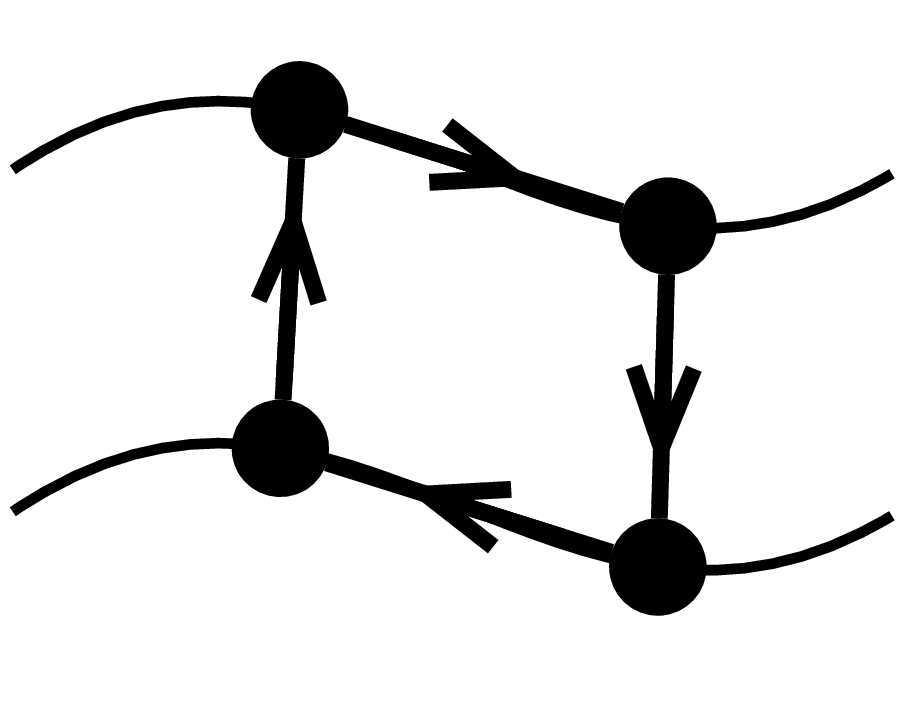} + \dots \, ,
\end{eqnarray}
and
\begin{eqnarray}
\label{wl_41}
\left( \Omega_{C}\right)_{2}^{2} & = & -\frac{1}{2} \figu{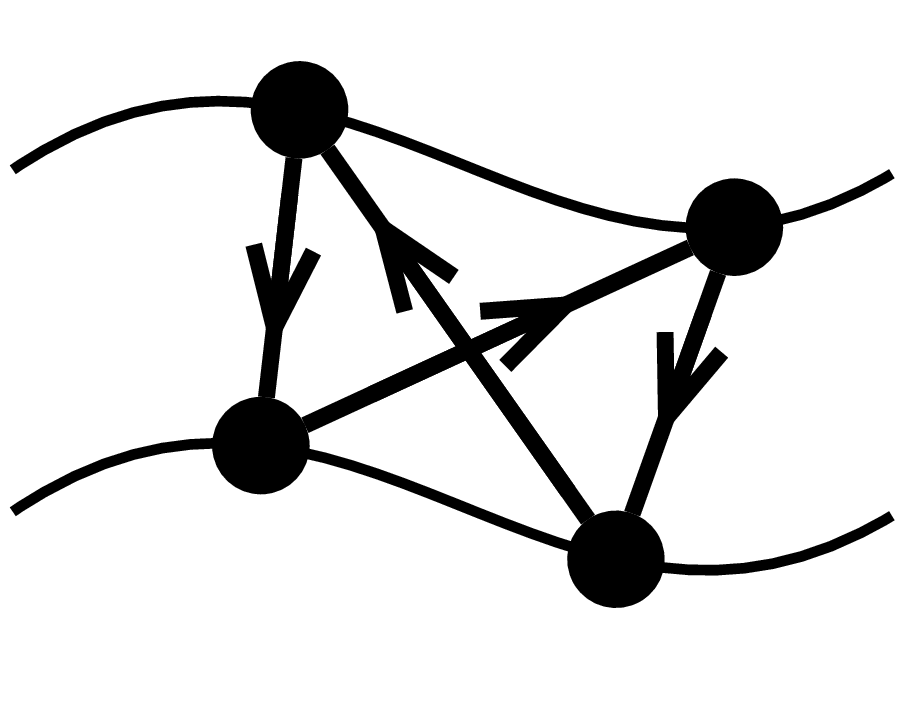} + \frac{1}{2} \figu{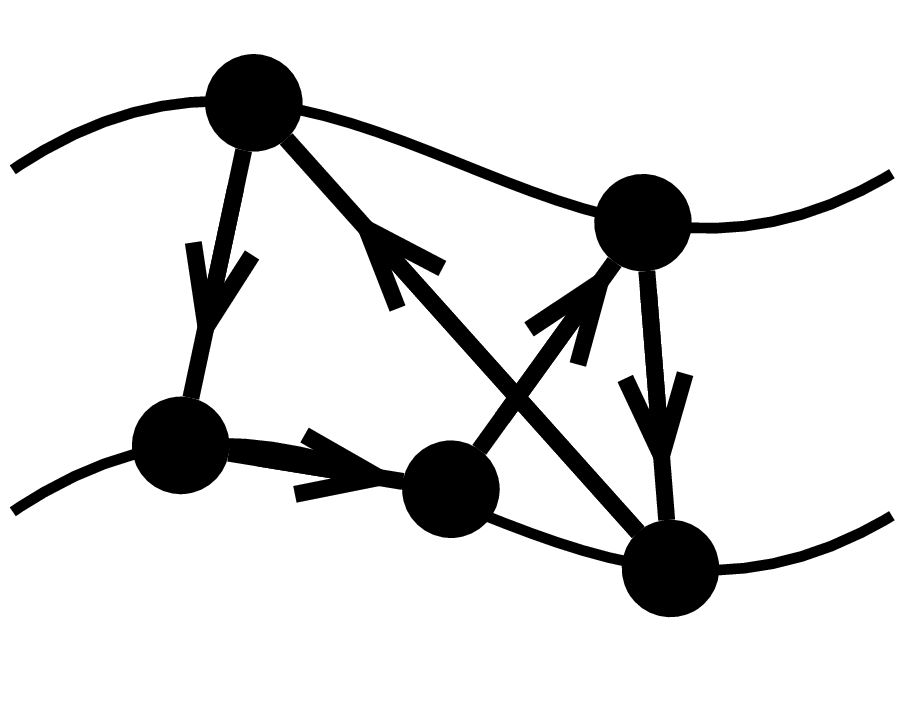} + \dots \, .
\end{eqnarray}
In this case, the coefficient will be
\begin{equation}
\label{wl_42}
\frac{1}{(n_{\rm legs}/2)} (-1)^{n_{1}+n_{\rm legs}/2+1} \, ,
\end{equation}
where $n_{1}$ is the number of $\kappa^{-1}\partial_{n}K$ bonds along the surface $S_{1}$ (the normal $\mathbf{n}$ points upward for both $S_{1}$ and $S_{2}$). This prediction is consistent with the prediction from the slab geometry \cite{SK_2017}. As mentioned above, the on-boundary arrow diagrams provide curvature corrections. The presence of $H/\kappa$ corrections are confirmed by the exact solutions for 3D cylindrical and spherical shells,  as it is shown in Appendix \ref{sec_cylinders}. If these are neglected, the Casimir interaction takes the form
\begin{equation}
\beta W_{\rm C}[\ell,\psi] \approx \frac{1}{2}\left(\fig{d34}-\frac{1}{2}\fig{d38}+\ldots\right).
\label{wcasimirdiagrammatic0}
\end{equation} 

Now we turn to the case of the surface field case. The kernel of $\mathsf{N}$, $N(\mathbf{s},\mathbf{s}^\prime)$, can also be written diagrammatically similarly to the Dirichlet case, but the expression is more involved. Instead, we will focus on the representation for large curvature radii. For this purpose, we define the bond
\begin{eqnarray}
\fig{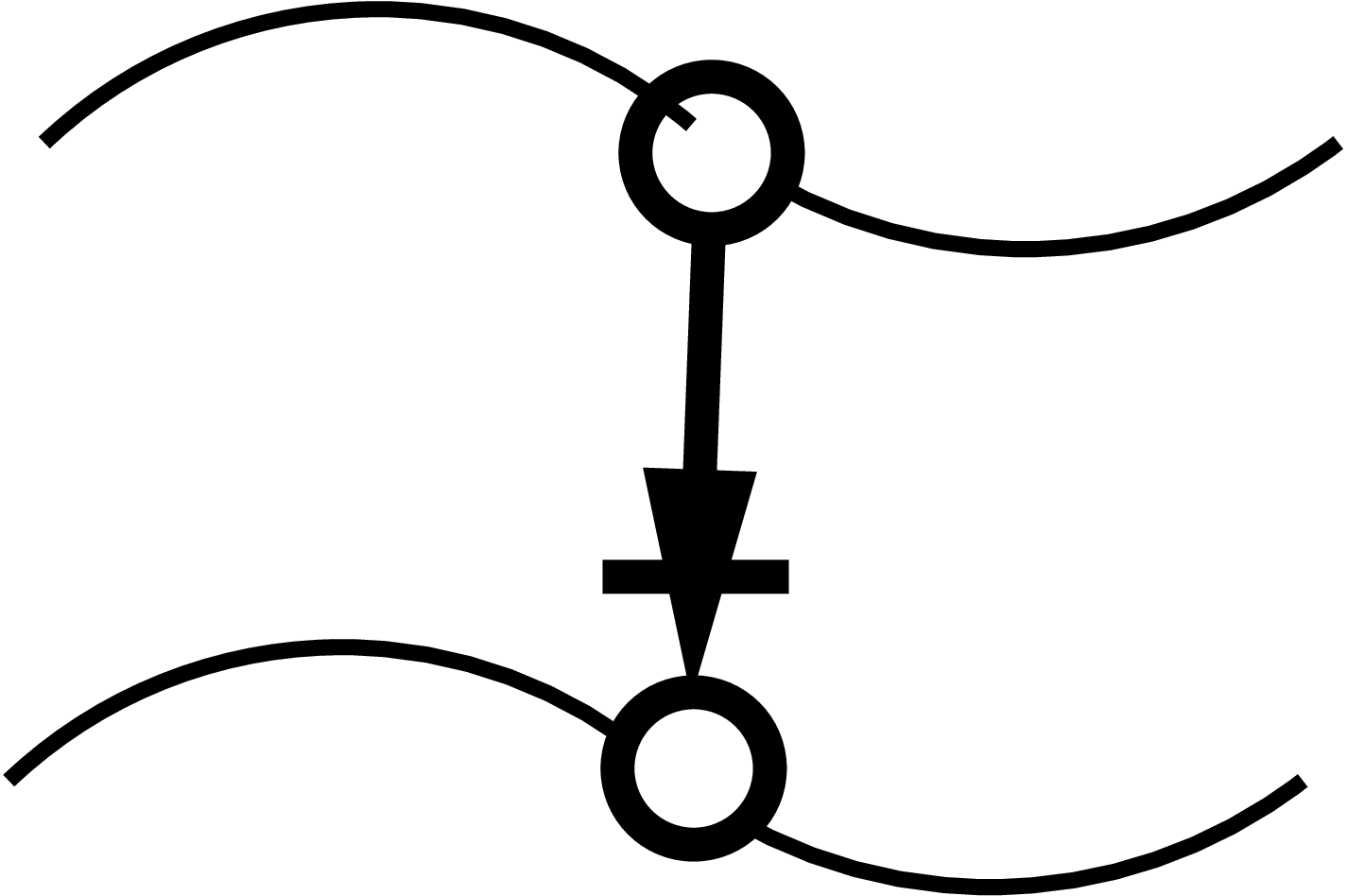}=\int_{\mathcal{S}_{1}} d\mathbf{s}_1\left(K+\frac{1}{g}\mathbf{n}\mathbf{\cdot}\boldsymbol\nabla K\right)X \, ,
 \label{defdiagrams2-0}
\end{eqnarray}
where $K=K(\textbf{s},\textbf{s}_{1})$ and $X=X(\textbf{s}^{\prime},\textbf{s}_{1})$. Here, as earlier, the upper wavy line represents the interface, the lower wavy lines represent the wall and $\textbf{s}$, $\textbf{s}^{\prime}$ are the fixed locations on them (open circles). Neglecting $\kappa^{-1}\partial_n K-$diagrams on $S_1$, which are ${\mathcal O}(H/\kappa)$, the new diagram can be decomposed as
\begin{eqnarray}
\fig{fig6} \approx\,  \alpha\mathcal{I}+(1-\alpha)\mathcal{J} \, ,
\label{defdiagrams2}
\end{eqnarray}
with
\begin{equation}
\alpha = g/(g-\kappa) \, .
\end{equation}
The expression (\ref{defdiagrams2-0}) for $N$ then determines that $\mathcal{I}$ and $\mathcal{J}$ can be written as expansions involving the original elementary diagrams
\begin{equation}
\mathcal{I}=\fig{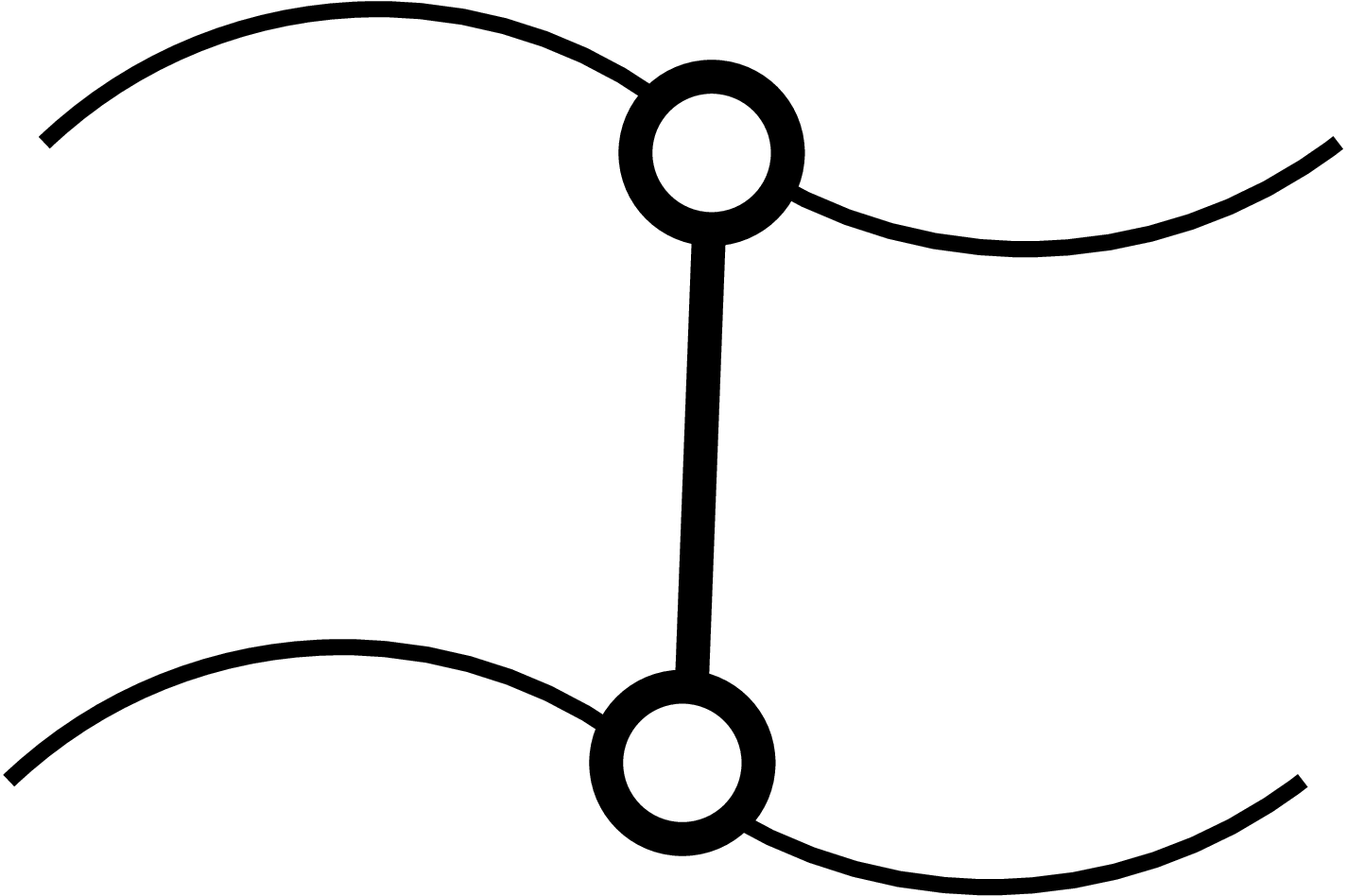}-\alpha \fig{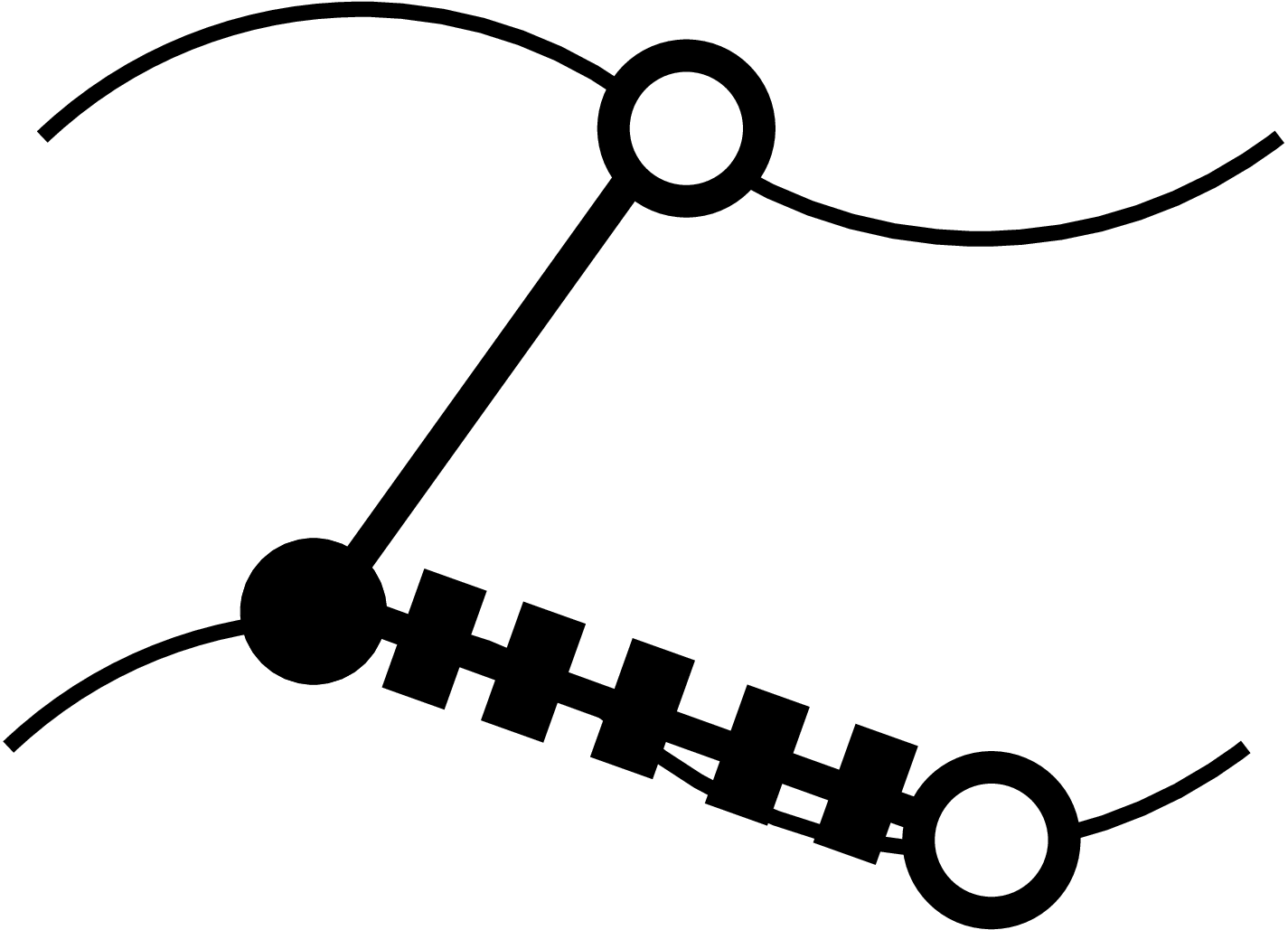} + \alpha^2 \fig{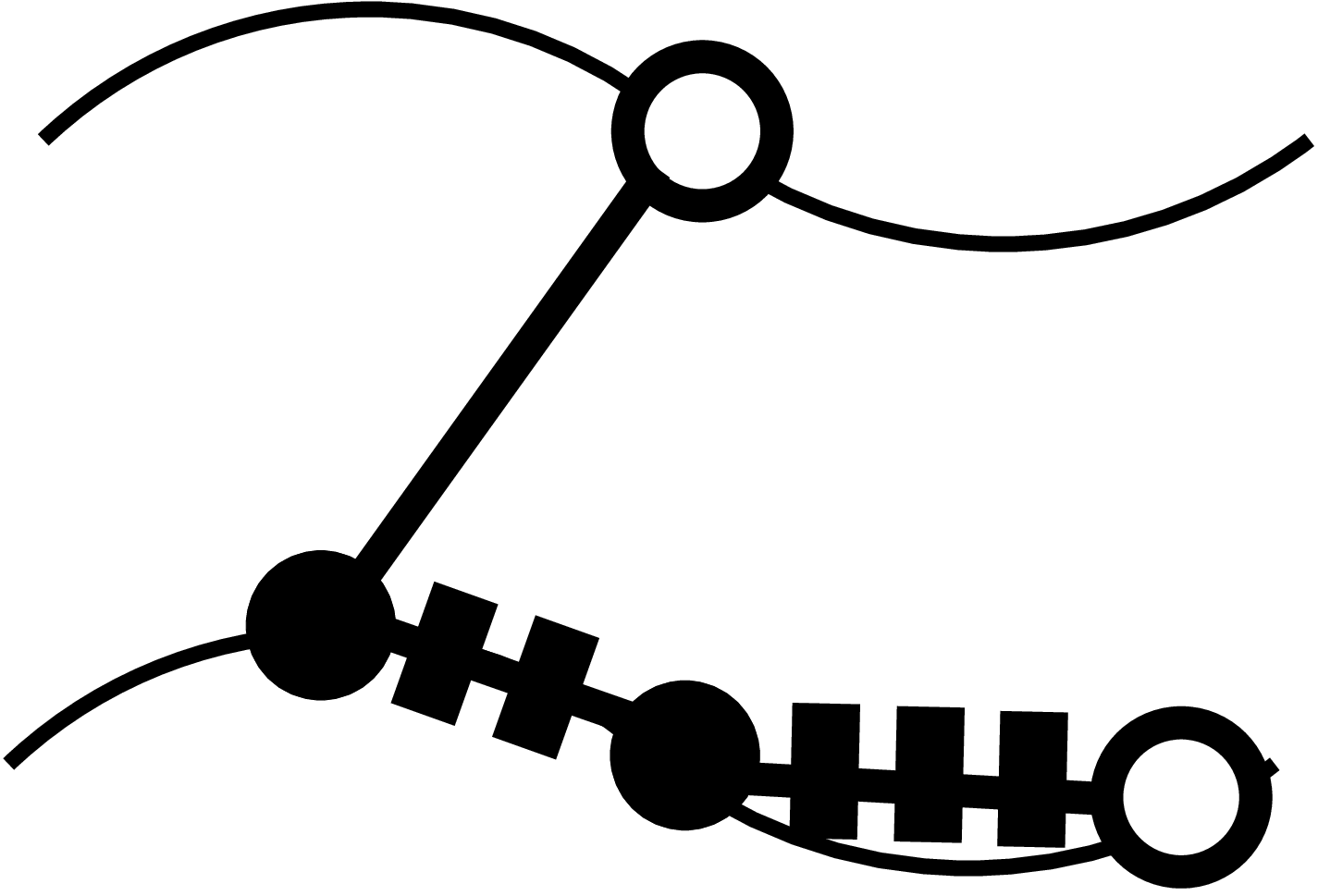}+\ldots 
\label{defdiagrams2-1}
\end{equation}
and
\begin{equation}
\mathcal{J}=\fig{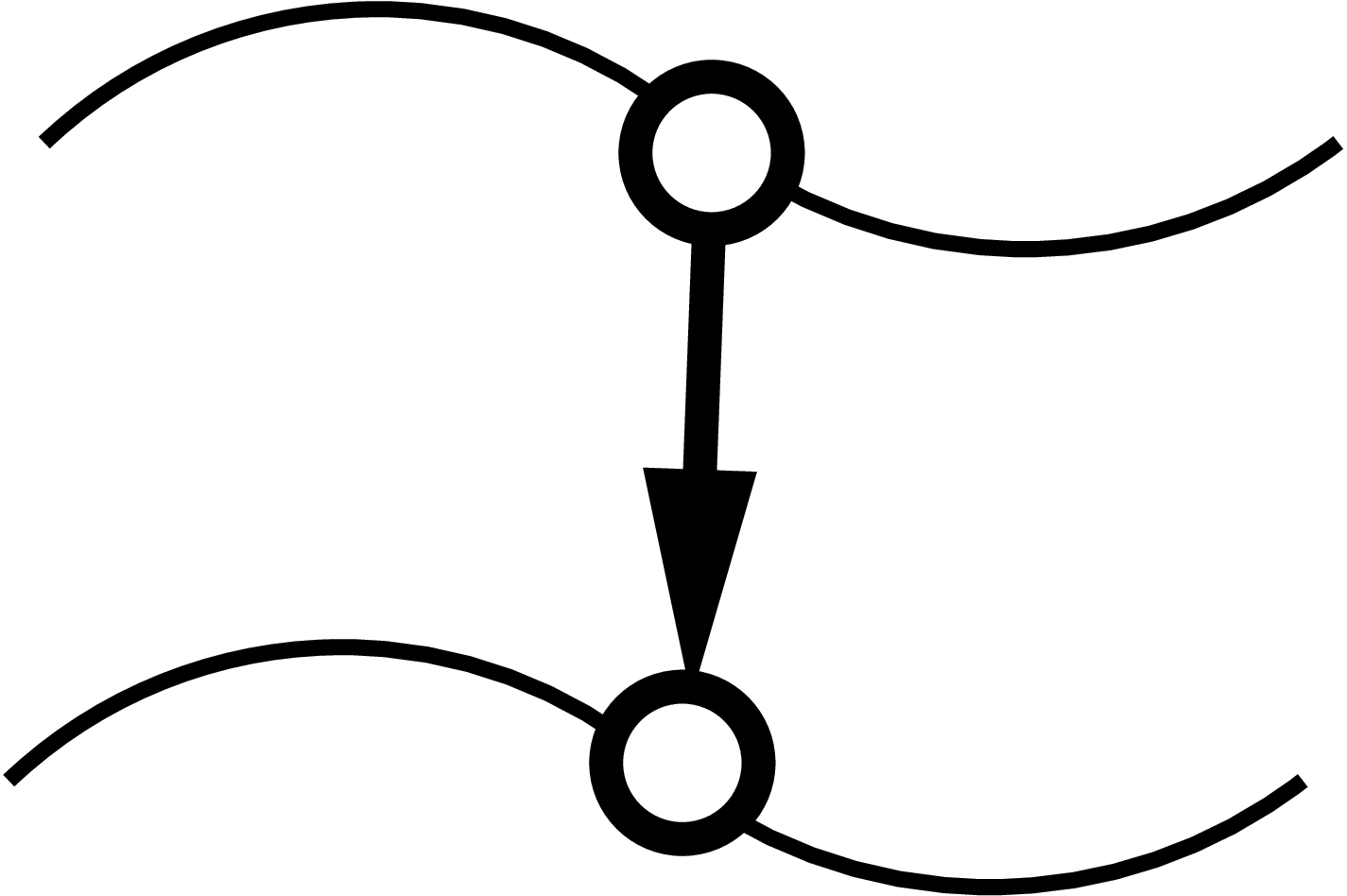}-\alpha \fig{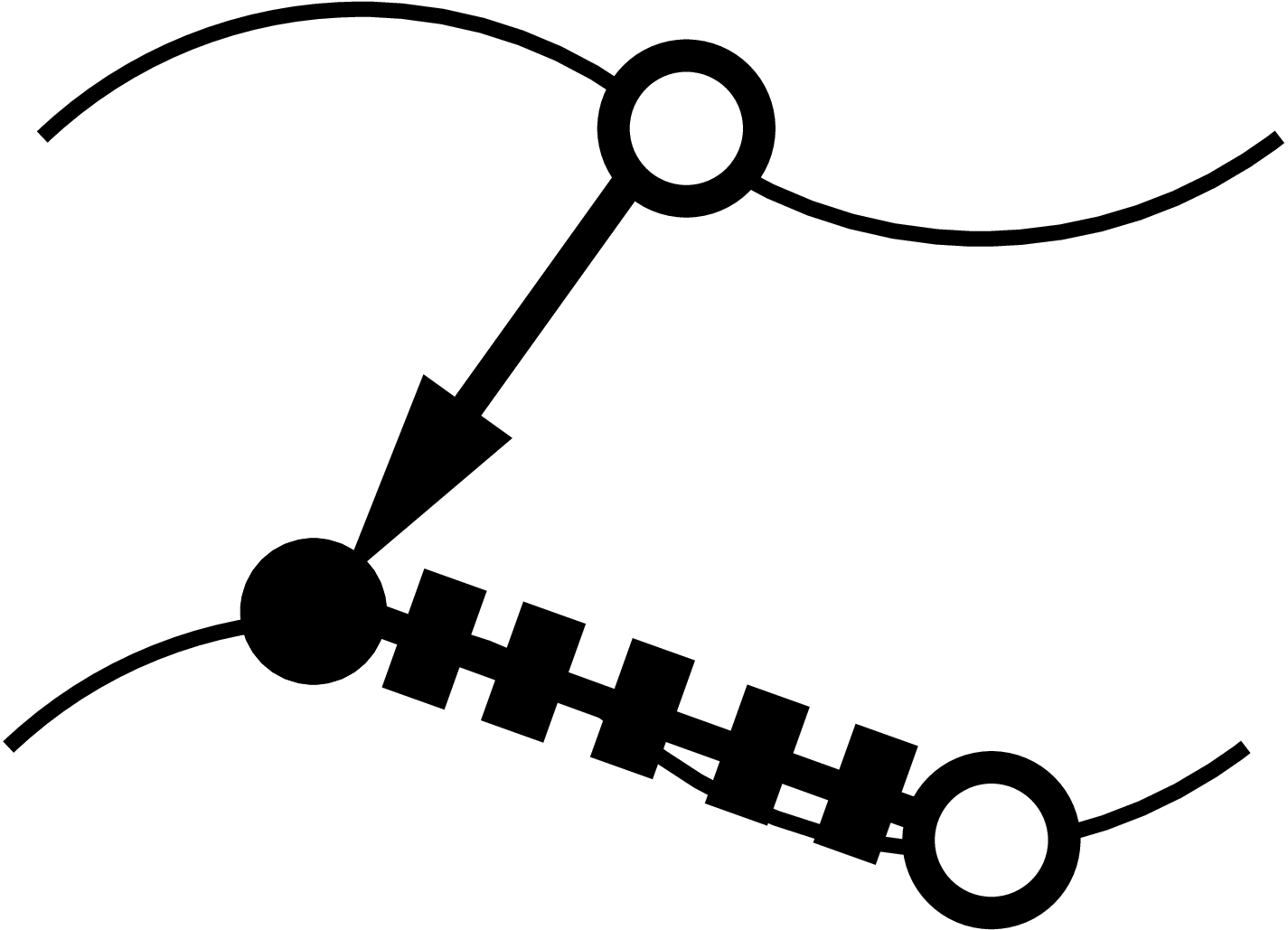} + \alpha^2 \fig{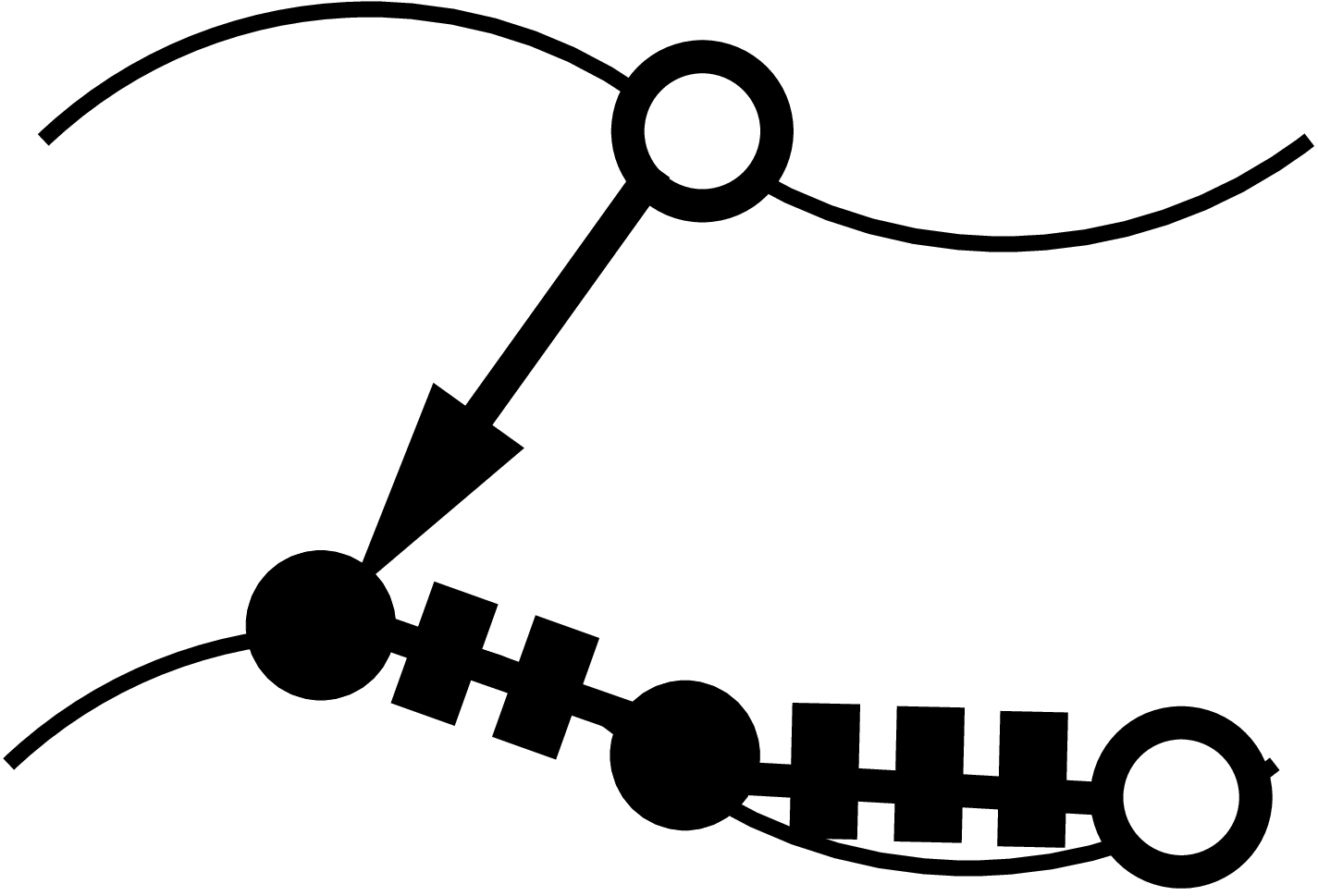}+\ldots\ , 
\label{defdiagrams2-2}
\end{equation}
where the black dot means integration over the surface with the area element. Using the explicit expressions for $K$ and $\nabla K$, it is shown in Appendix \ref{sec_planarwallcurvedinterface} that these resum to give
\begin{eqnarray}
\fig{fig6} \approx 
\int_{0}^{\infty}\rd q \left(\frac{q}{2\pi}\right)^{\frac{d-1}{2}} \, \frac{g+\kappa_{q}}{g-\kappa_{q}} \textrm{e}^{-\kappa_{q}\ell} \frac{J_{\frac{d-3}{2}}(q\rho)}{\rho^{\frac{d-3}{2}}} \, , \nonumber \\
\label{defdiagrams2bisbis}
\end{eqnarray}
where $\rho$ and $\ell$ are, respectively, the transverse and normal coordinates of $\mathbf{s}-\mathbf{s}'$, $J_\nu(z)$ is the  Bessel function of the first kind and order $\nu$ and  $\kappa_q\equiv \sqrt{\kappa^2+q^2}$. The kernel $N$ also contains the integral over $K$ and its inverse which resums, as we saw previously and up to $\mathcal{O}(H/\kappa)-$corrections, to the arrow diagram
\begin{equation}
\int_{\mathcal{S}_1} d\mathbf{s}_1 K(\mathbf{s}',\mathbf{s}_1)K^{-1}(\mathbf{s}_1,\mathbf{s})
 \approx -\fig{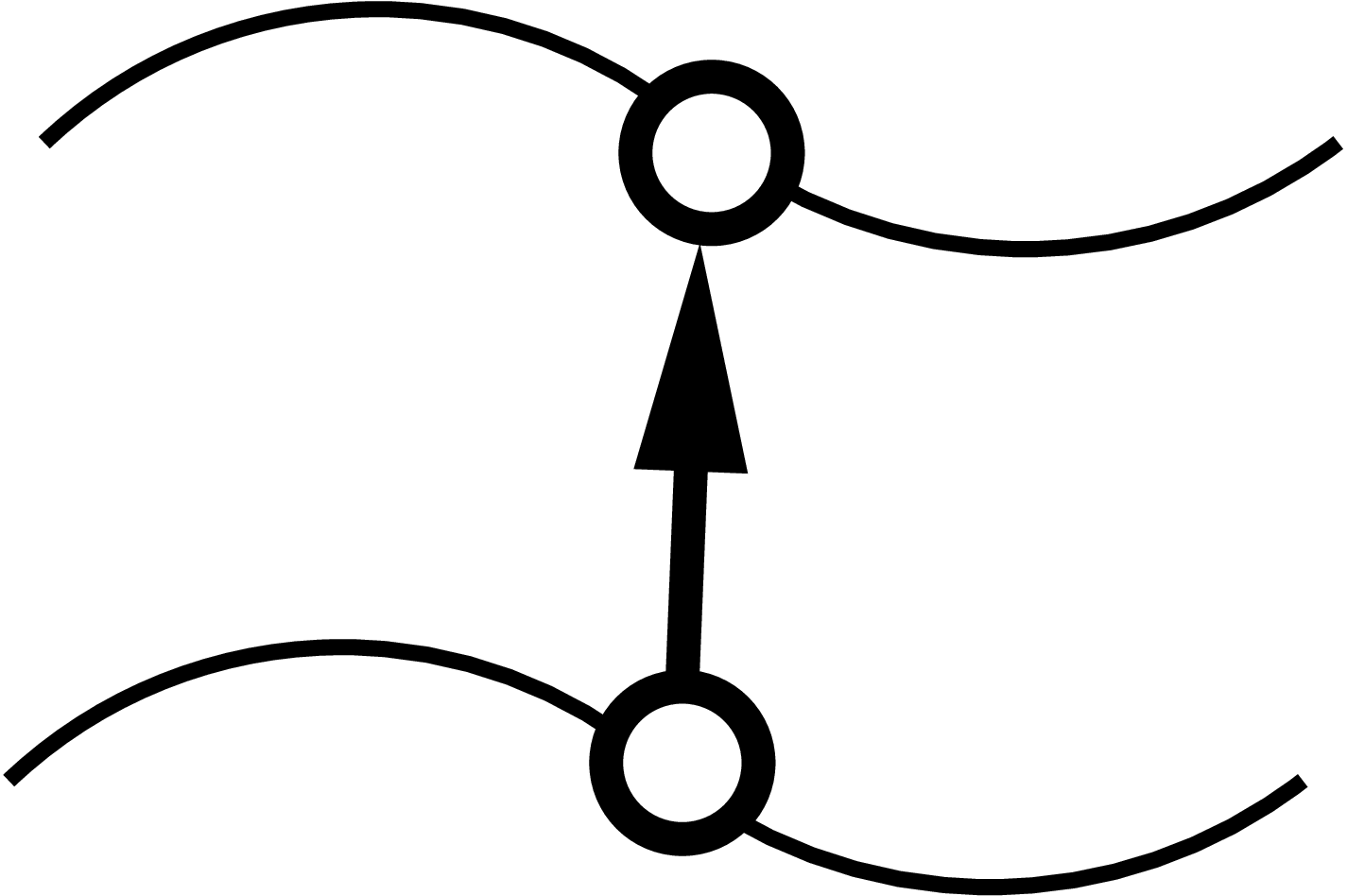} .\label{defdiagrams3}
\end{equation}
given explicitly by
\begin{equation}
\fig{fig13} =  2 \mathbf{n}(\mathbf{s}) \mathbf{\cdot}
\frac{\mathbf{s}-\mathbf{s}^{\prime}}{|\mathbf{s}-\mathbf{s}^{\prime}|^{\frac{d}{2}}} \left(\frac{\kappa}{2\pi}\right)^{\frac{d}{2}} K_{\frac{d}{2}}\left(\kappa|\textbf{s}-\textbf{s}^{\prime}|\right) \, .
\label{defdiagrams3bisbis}
\end{equation}
Together these identify the Casimir contribution to the binding potential functional, up to ${\mathcal O}(H/\kappa)$ corrections, as the diagrammatic expansion
\begin{equation}
\beta W_{\rm C}[\ell,\psi] \approx \frac{1}{2}\left(\fig{fig14}-\frac{1}{2}\fig{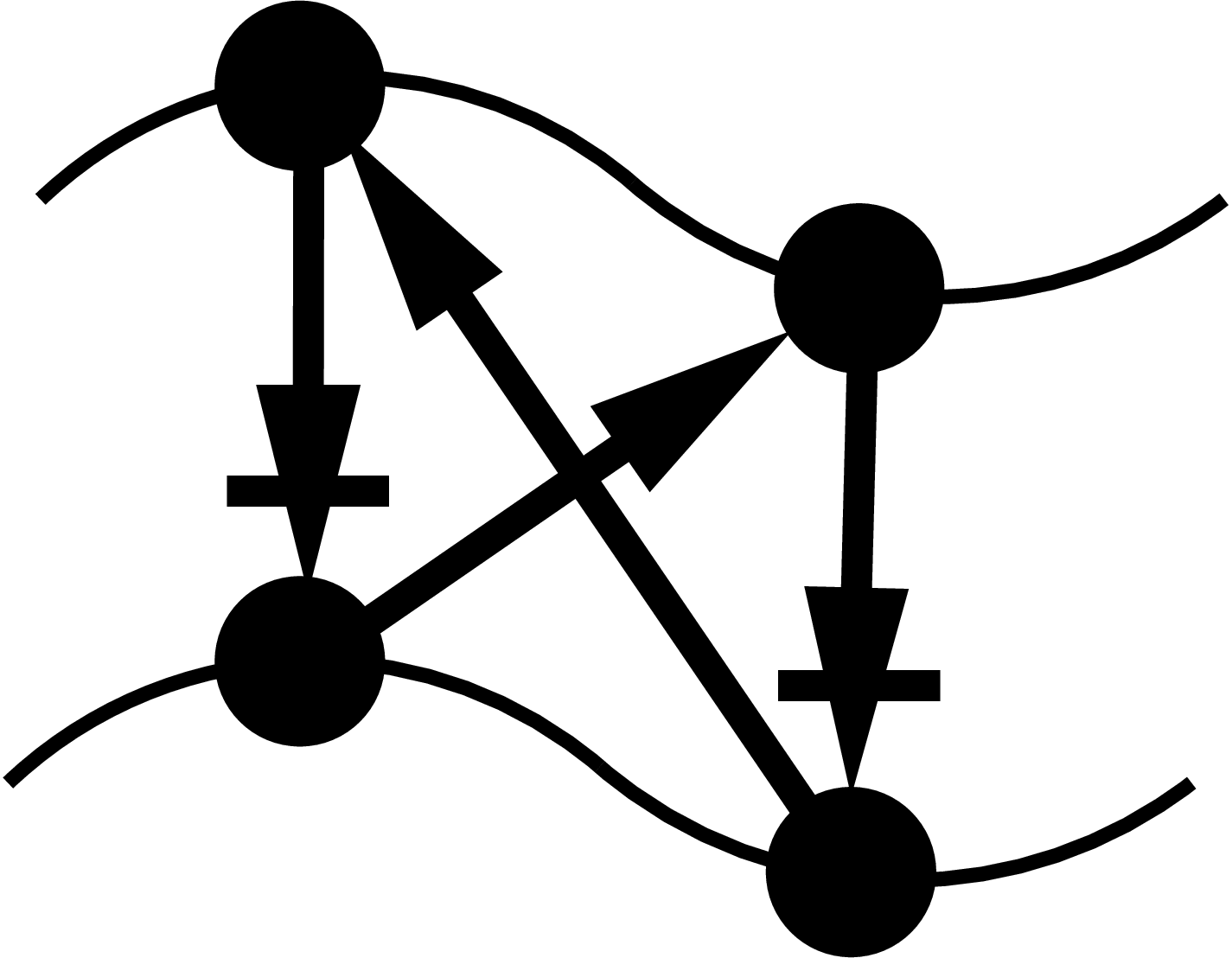}+\ldots\right),
\label{wcasimirdiagrammatic}
\end{equation} 
which is the central result of this section. For thick wetting films, only the first term
\begin{equation}
(\Omega_{\rm C})_{1}^{1} \approx \fig{fig14}
\label{wcasimirdiagrammatic2}
\end{equation}
is required and generates the leading order exponential decay.

\section{The Influence of the Casimir contribution on wetting transitions at mean-field level.}
In this Section, we discuss the influence of the entropic, or thermal Casimir contribution to the binding potential, on critical, tricritical and first-order wetting transitions of the fluid in presence of a flat wall. To do this systematically we first inquire how the mean-field predictions and Nakanishi-Fisher phase diagram are altered by the inclusion of the Casimir contribution. Strictly speaking, in order to do this, it is also needed to go beyond the DP model since the latter has singular behavior at MF tricritical wetting. However this is not a problem and can be done perturbatively to include quartic terms present in the LGW $m^4$ model \cite{PRBRE_2007}. In this sense, the DP Casimir term can be understood as the leading-order term of the one-loop expression for the full Casimir contribution. Having determined the influence of the Casimir term on MF singularities we can then understand its effect on the interfacial fluctuations and non-universal critical singularities, occurring at 3D critical and tricritical wetting transitions. The RG theory of this will be left for a forthcoming publication.

\subsection{The Casimir contribution to the binding potential}
The binding potential functional can be evaluated in closed form for a number of specific interfacial and wall configurations. By far the most important of these is a flat wall ($\psi=0$) with a uniform wetting layer of thickness $\ell$, i.e., both wall and interfacial are planar and parallel. In that case, the binding potential functional, per unit wall area, reduces to the binding potential function $w(\ell)$ appearing in the local interfacial Hamiltonian. The Casimir contribution can be determined analytically using a number of techniques without having to resum each diagrammatic contribution. For example, in Ref. \cite{SREP2023}, we showed how $w_C(\ell)$ may be determined using a path integral method which maps the problem onto the propagator for the quantum simple harmonic oscillator. This determines that in dimension $d$ the full expression for the Casimir binding potential function is given by Eq. (\ref{slitgeneral_21}), i.e.
 \begin{equation}
 \label{wcd}
\beta w_{C}(\ell) = \frac{1}{2} 
\int\frac{\textrm{d}^{d-1}\textbf{q}}{(2\pi)^{d-1}} 
\ln\left( 1 - \frac{g+\kappa_{q}}{g-\kappa_{q}}
e^{-2\kappa_{q}\ell} \right) \, 
\end{equation}
so that in three dimensions
\begin{equation}
\label{wc3}
\beta w_{C}(\ell) = \frac{1}{4\pi} 
\int dq 
q\ln\left( 1 - \frac{g+\kappa_{q}}{g-\kappa_{q}}
e^{-2\kappa_{q}\ell} \right) \, .
\end{equation}

The qualitative form of the Casimir contribution $w_{C}(\ell)$ is similar to the MF contribution $w_{MF}(\ell)$ but is controlled by the surface enhancement $g$ rather than the scaling field $t$ of which it is independent. We note that for $\kappa \ell  \ll 1$, the Casimir contribution diverges as $w_C(\ell) \propto 1/\ell^2$ (and that more generally $w_{C}(\ell) \propto \ell^{-(d-1)}$, in dimension $d$)) which is the familiar power-law for the critical Casimir effect. However, this limit is not of interest to us and, indeed, we stress that the DP potential is only a reliable physical model of adsorption for thick wetting films, with $\kappa \ell \gg 1$. In general, when $-g > \kappa$, corresponding to the MF regime of critical wetting, the potential is repulsive at short-distances and attractive at large distances, possessing a minimum, the location of which diverges continuously as $-g$ approaches $\kappa$. On the other hand for $-g=\kappa$, which recall is the condition for MF tricriticality, and $-g<\kappa$, corresponding to the first-order wetting regime, the potential is purely repulsive.

We provide some results about the asymptotic behavior for short separations. Regardless of the coupling strength, the Casimir force behaves as $\ell^{-2}$ in $d=3$. For infinite coupling the force turns out to be attractive and asymptotically it behaves as $\beta w_{C}(\ell) \sim - \mathfrak{a}_{3}/\ell^{2}$ with $\mathfrak{a}_{3}=\zeta(3)/16\pi$ where $\zeta(3)$ is Apéry's constant. For finite coupling instead the force is repulsive and it behaves as $\beta w_{C}(\ell) \sim \mathfrak{b}_{3}/\ell^{2}$ with amplitude $\mathfrak{b}_{3}=3 \zeta(3)/64\pi$. In general dimension $d$, the amplitude of the attractive force is
\begin{equation}
\mathfrak{a}_{d} = \frac{\Gamma(d/2) \zeta(d)}{2^{d}\pi^{d/2}} \, ,
\end{equation}
while instead in the repulsive case,
\begin{equation}
\mathfrak{b}_{d} = (1-2^{1-d}) \mathfrak{a}_{d} \, .
\end{equation}

Remarkably, the 3D Casimir binding potential $w_{C}(\ell)$  is described to near perfect accuracy, over the whole range of length scales $\kappa \ell$, and for all values of the surface enhancement $g$, by the leading order contribution 
coming from the $(\Omega_C)_1^1$ diagram. This is a huge simplification in the analysis of fluctuation effects at wetting transitions and means we can safely ignore terms in $w_C(\ell)$, and more generally the terms
$W_C[\ell,\psi]$, coming from higher-order diagrams that are $\mathcal{O} (e^{-4\kappa \ell})$, which will not be discussed further. 

\begin{figure}[htbp]
\includegraphics[width=\columnwidth]{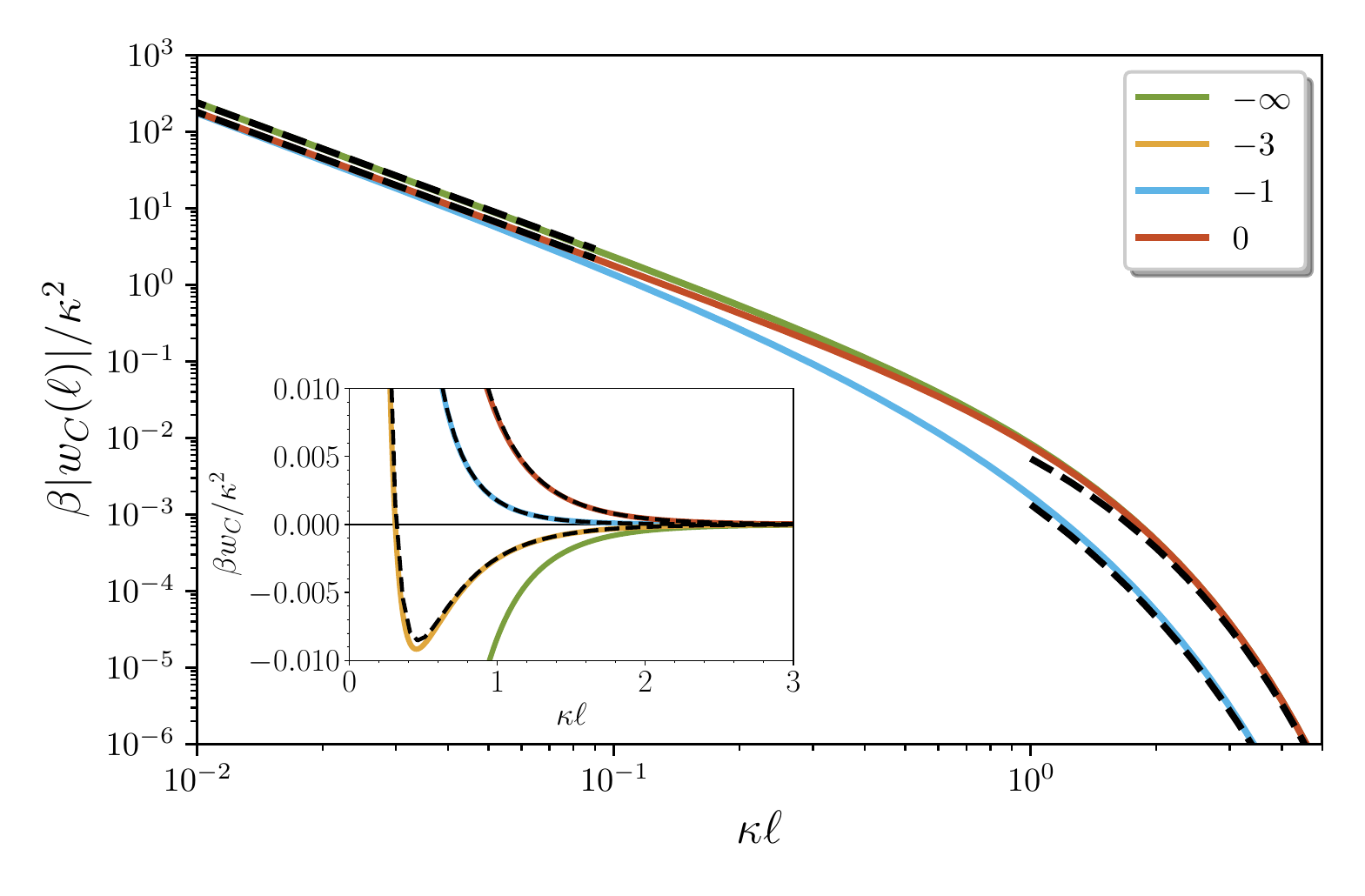}
\caption{Main plot: the magnitude of the Casimir potential in $d=3$ (solid curves) illustrating the crossover from short- to large-distance behavior for the indicated values of $g/\kappa$. The large-distance (\eqref{24072025_1132} and \eqref{17042025_1741}) and short-distance asymptotic behavior ($-\mathfrak{a}_{3}/\ell^{2}$ and $\mathfrak{b}_{3}/\ell^{2}$) is shown with dashed lines. In inset: the Casimir potential with its sign (solid lines) compared with the approximated expression \eqref{24072025_1129} (dashed); the solid green line corresponds to $-(1+2\kappa\ell)/(16\pi \kappa^{2}\ell^{2})$, corresponding to \eqref{24072025_1129} for $g=-\infty$.}
\label{fig_asymptotic}
\end{figure}

It is immediately apparent that the leading-order exponential term in the expansion of the logarithm controls the large-distance behavior. Therefore  when $\kappa\ell>>1$ we can expand
\begin{equation}
\label{wc3-1}
\beta w_{C}(\ell) \approx -\frac{1}{2} 
\int\frac{\textrm{d}^{d-1}\textbf{q}}{(2\pi)^{d-1}}
\frac{g+\kappa_{q}}{g-\kappa_{q}}
e^{-2\kappa_{q}\ell} \, ,
\end{equation}
which is the same as the contribution the diagram $(\Omega_C)_{1}^{1}$ when evaluated for a uniform wetting layer, i.e. Eq.~(\ref{wcasimirdiagrammatic2}), after substituting Eq.~(\ref{alt_08}) into Eq.~(\ref{omega_n_n}) for $n=1$. The integral can be evaluated using spherical coordinates.
At MF tricriticality, we find
\begin{equation}
\label{24072025_1132}
\beta w_{C}(\ell)/\kappa^{d-1} \sim \mathfrak{t}_{d} \frac{\textrm{e}^{-2\kappa\ell}}{(\kappa\ell)^{\frac{d+1}{2}}} \, ,
\end{equation}
with the amplitude
\begin{equation}
\label{17042025_1740}
\mathfrak{t}_{d} = \frac{\Gamma(\frac{d+1}{2})\Gamma(\frac{d-1}{2})}{2^{d+2}\pi^{(d-1)/2}} \, ,
\end{equation}
while instead for $g=0$ and $g=\infty$, we find, respectively,
\begin{equation}
\label{17042025_1741}
\beta w_{C}(\ell)/\kappa^{d-1} \sim \pm \mathfrak{c}_{d} \frac{\textrm{e}^{-2\kappa\ell}}{(\kappa\ell)^{\frac{d-1}{2}}} \, ,
\end{equation}
with
\begin{equation}
\mathfrak{c}_{d} = \frac{\Gamma^{2}(\frac{d-1}{2})}{2^{d}\pi^{(d-1)/2}} \, .
\end{equation}
These asymptotic behaviors for both short- and large-distances are plotted in Fig.~\ref{fig_asymptotic}.

In three dimensions we can proceed further, and evaluate exactly the leading order $(\Omega_C)_1^1$ contribution, for arbitrary $g$, in terms of special functions, yielding
\begin{eqnarray} 
\label{24072025_1129}
&& \beta w_{C}(\ell) = - \frac{1}{4\pi} \int_{0}^{\infty} dq \, q \frac{g+\kappa_{q}}{g-\kappa_{q}}e^{-2\kappa_{q}\ell} \\ \nonumber
&& = \frac{\kappa^{2}}{2\pi}  \biggl[ \frac{1+(2+4v)u}{ 8u^2 } e^{ - 2 u } + v^2 e^{-2 uv} \Gamma(0,2 (1-v)u) \biggr] 
\end{eqnarray}
where $u=\kappa\ell$, $v=g/\kappa$ and $\Gamma(0,z)$ is the upper incomplete gamma function. Further simplification is possible for specific values of the surface enhancement. For example, when $g=-\infty$, (corresponding to critical wetting with a fixed value of the surface magnetization) we obtain $\beta w_{C}(\ell) = - (1+2\kappa \ell) e^{-2\kappa \ell}/\pi\ell^2$
which is equivalent to the result obtained in \cite{KD_2010} for a Gaussian
 field theory, in a parallel plate geometry with Dirichlet boundary conditions. Parenthetically, we observe that this approximate result is correct within $3\%$ for $\kappa\ell>1$. Intriguingly, this result for the leading order decay is dual to the 
case $g=0$, corresponding to a first-order wetting transition, 
for which 
$\beta w_{C}(\ell) = (1+2\kappa \ell) e^{-2\kappa \ell}/\pi\ell^2$
 although the duality does not extend to the higher-order contributions.
 
In the vicinity 
of the MF tricritical point, $-g \approx \kappa$, where the qualitative 
change in the form of $w_{C}(\ell)$ occurs, the Casimir contribution
 behaves as
\begin{equation}
\beta w_{C}(\ell) \approx \frac{ e^{-2\kappa\ell} }{ 32\pi\ell^{2} } \bigl[ 2(\kappa+g)\ell + 1 \bigr] \, ,
\end{equation}
determining that the minimum in $w_{C}(\ell)$ occurs at $\ell_{\min} 
\approx -1/2(g+\kappa)$. This simple expression, for the approximate form of the
 Casimir contribution, valid for large $\kappa\ell$ and $-g\approx\kappa$, 
is one of the central results of our paper.

 The same qualitative change in $w_C(\ell)$, from 
an attractive to repulsive decay, occurs at $-g=\kappa$ in general 
dimension $d$. In this case,
provided that
$-g\ne\kappa$, the Casimir contributions decays as
$w_{C}(\ell) \sim e^{-2\kappa\ell}/\ell^{(d-1)/2}$ and is attractive for
 critical wetting ($-g>\kappa$) and repulsive for first-order wetting
 $-g<\kappa$. However, exactly $-g=\kappa$, where the potential is
 first purely repulsive, it decays, provided $d>1$ as 
$\beta w_{C}(\ell) \sim e^{-2\kappa\ell}/ \ell^{(d+1)/2}$
 recovering our previous expressions when $d=3$. These results are of interest when
 discussing the properties of first-order and tricritical wetting in
 dimension $d>3$ and, in particular, the revaluation of the
 meaning of the upper critical dimension for tricritical wetting.

\subsection{Mean-field theory and Casimir effect}
If we ignore the role played by interfacial fluctuation effects, the equilibrium film wetting film thickness, $\langle \ell\rangle$, follows from simple minimization of the full binding-potential
\begin{equation}
w(\ell)=w_{MF}(\ell)+w_C(\ell)
\end{equation}
which in three dimensions, and in the vicinity of the MF tricritical point, reads
\begin{equation}
w(\ell)= ae^{-\kappa\ell}+be^{-2\kappa\ell}+\frac{e^{-2\kappa\ell}}{32\beta\pi\ell^2}(2(\kappa+g)\ell+1)
\end{equation}
where recall, from (33) that, $a\propto -t$ and $b\propto -(g+\kappa)$ so that $b$ is positive for critical wetting, negative for first-order wetting and vanishes at tricriticality. Using this potential it is straightforward to see what aspects of wetting unchanged, and what aspects are altered, from predictions of MF theory.

{\it{The Surface Phase Diagram}}. The inclusion of the Casimir contribution does not change the qualitative form of the Nakanishi-Fisher surface diagram. That is, critical wetting still occurs for $-g>\kappa$ at $t=0$ and tricritical wetting still occurs at $t=0$ when $-g=\kappa$. First-order wetting transitions occur when $-g<\kappa$ and at a non-vanishing (negative) value of the scaling field $t_w$ determined by $g$. The line of first-order transitions, in the surface diagram, still meets tangentially the line of critical wetting transitions at the tricritical point, as shown earlier in Fig. \ref{NF_phase_diagram}. All these qualitative aspects of the wetting phase diagram are unchanged from the predictions of standard MF theory and are completely consistent with the findings of Ising model simulations.

However, there is a significant quantitative change to the value of the field $t_w$ at which first-order wetting occurs. Recall that, at MF level and in the vicinity of the tricritical point, the value of the scaling field $t_w$ behaves as $t_w\propto -(g+\kappa)^2$ on approaching the tricritical point. When the Casimir potential is included however this is altered to (ignoring unimportant constants)
\begin{equation}
t_w\propto -(g+\kappa)^{3/2} e^{-\sqrt{\omega/4}/\sqrt{1+g/\kappa}}
\end{equation}
(where $\omega$ is the dimensionless wetting parameter) which, close to tricriticality, lies extremely close to the analytic extension of the line of critical wetting $t=0$. This is a first indication of the importance of including the Casimir contribution to the binding potential.

{\it{Critical wetting}}. The Casimir effect has no influence on the critical exponents for critical wetting transitions, since when $-g<\kappa$, the Casimir contribution is higher-order than the repulsive contribution $\propto e^{-2\kappa\ell}$ appearing in $w_{MF}(\ell)$. Thus, in compliance with the expectations of MF theory, the wetting film thickness still diverges as $\kappa\ell\approx \ln t^{-1}$, while, from the standard identification that $\xi_\parallel^2=\gamma/w''(\langle\ell\rangle)$, we still have $\xi_\parallel\propto t^{-1}$ so that $\nu_\parallel=1$. The upper critical dimension for critical wetting remains $d^*=3$ and MF is valid for $d>3$.

{\it{Tricritical Wetting}}. The inclusion of the Casimir contribution has a dramatic effect on the characteristic singularities of tricritical wetting since at the condition for tricriticality, $-g=\kappa$, Casimir term is non-vanishing (provided $d>1$) , and is larger than the MF repulsion which recall decays as $\mathcal{O}(e^{-3\kappa\ell})$. Thus, instead, of the usual MF predictions, $\kappa\ell\approx 2\ln t^{-1}$ and $\xi_\parallel\propto t^{-3/4}$, the Casimir term determines that these are altered to
\begin{equation}
\kappa\langle\ell\rangle \approx \ln t^{-1} - 2
\ln\ln t^{-1} + \cdots
\label{triMFC1}
\end{equation}
and
\begin{equation}
\xi_\parallel\sim \frac{1}{t|\ln t|}
\label{triMFC2}
\end{equation}
so that $\nu_\parallel=1$ (with a log correction) compared to the MF prediction $\nu_\parallel=3/4$. The singularities are therefore much more similar to the MF predictions for critical wetting.

These results for tricritical wetting generalize to higher dimensions, $d>3$, so that on combining the MF and Casimir contributions to the binding potential, we find $\kappa\langle\ell\rangle\approx \ln t^{-1}- [(d+1)/2]\ln\ln t^{-1}$ and $\xi_\parallel\sim 1/t |\ln t|^{(d+1)/4}$. These results are remarkable, since they show that the predictions of MF theory are not valid for $d>3$, as has always been though previously. Even in these higher dimensions the predictions of MF theory are altered by this second source of thermal fluctuations associated with the Casimir effect and equivalently the entropy of the many microscopic states that correspond to a given interfacial configuration. The predictions of MF theory are only obtained on artificially setting $k_B T=0$ or in the limit $d\to \infty$. This means that there is a subtle mistake in previous interpretations of the Ginzburg criterion for short-ranged tricritical wetting. These studies used a one-loop calculation to determine that the predictions of MF theory are invalidated by interfacial fluctuations when $d=3$. However, the present study shows that, for tricritical wetting, the predictions of MF theory are also invalidated by the small bulk-like fluctuations about the MF profile, which give rise to the entropic/Casimir interaction. Three-dimensions remains an upper critical dimension for tricritical wetting as regards the role played by interfacial fluctuations in the sense that the predictions (\ref{triMFC1}) and (\ref{triMFC2}) will be altered by interfacial fluctuations in $d=3$. However, the predictions of MF theory are not valid above this upper critical dimension since the Casimir contribution still has to be included for these higher dimensions. 

{\it{First-order Wetting}}. Similar, quantitative corrections to MF theory apply for first-order wetting transitions. At a first-order wetting transition, the equilibrium film thickness and parallel correlation length, jumps from finite, microscopic values, at $T=T_w^-$ to infinity. These microscopic values, characterizing the properties of the thin wetting layer, diverge smoothly as we follow the line of first-order wetting transitions towards the tricritical point at $-g=\kappa$. We are not aware that these singularities have been discussed before, even at MF level.  It is straightforward to show that these singularities are the same as those obtained on setting $t=0$ and then letting the (negative) coefficient $b\propto -(g+\kappa)$ vanish as $-g\to \kappa^-$. According to MF theory, for which, recall $w_{MF}(\ell)\approx b e^{-2\kappa\ell} + e^{-3\kappa\ell}$, we obtain, in $d=3$
\begin{equation}
\kappa\langle\ell\rangle\approx -\ln\frac{g+\kappa}{\kappa}, \hspace{1cm}\xi_\parallel\sim \frac{1}{(g+\kappa)^{3/2}}
\end{equation}
However, these MF predictions are altered substantially when we include the Casimir contribution, because in that case the total binding potential, for $t=0$ is $w(\ell)\approx (b+c/\ell^2)e^{-2\kappa\ell}$ where $c=1/32\pi\beta$ is the coefficient of the Casimir repulsion at tricriticaliy determined earlier. The strict MF predictions for the film thickness and parallel correlation length are then modified to
\begin{equation}
\kappa\langle \ell\rangle\approx\frac{1}{\sqrt{\frac{4}{\omega}(1+\frac{g}{\kappa})}}
\label{MFCell}
\end{equation}
and
\begin{equation}
\xi_\parallel\sim \frac{\exp\left(1/{\sqrt{\frac{4}{\omega}(1+\frac{g}{\kappa})}}\right)}{(1+\frac{g}{\kappa})^{3/4}}
\label{MFCxi}
\end{equation}
where $\omega=k_BT\kappa^2/4\pi\gamma$ is the dimensionless wetting parameter. This, parameter, which is indicative of the presence of thermal fluctuation effects, is therefore already present via the Casimir contribution, even before we consider the role of interfacial fluctuations. The Casimir contribution therefore dramatically increases the wetting film thickness for weakly first-order wetting transitions. We remark, in passing, that the exponential divergence of $\xi_\parallel$ on approaching the tricritical point is strikingly similar to that characteristic of the Kosterlitz-Thouless phase transition as pertinent to 3D roughening and melting.
 These, Casimir induced, modifications to MF theory also occur for $d>3$. For example, the (thin) film thickness along the line of first-order wetting (or equivalently at $t=0$) scales as $\langle\ell\rangle \sim (g+\kappa)^{-2/(d+1)}$, implying, again that the MF predictions are only obtained in the limit $d\to \infty$.

\section{Summary and Discussion}

In this paper we have provided comprehensive details of how a Casimir or entropic contribution to the binding potential for short-ranged wetting phenomena, arising from the many microscopic configurations that correspond to a given interfacial one, may be determined exactly for the LGW Hamiltonian in the reliable double parabola approximation. The Casimir contribution is present even at low temperatures, far from the bulk critical point and decays on the scale of the bulk correlation length, similar to the usual MF contribution, competing with it. To determine $W_C$ and it's functional dependence on the interface and wall shape, we have extended the rigorous boundary-integral method, previously used to determine the MF contribution to the binding potential, which allows us to express the MF and Casimir contributions as different diagrammatic expressions representing successively higher-order exponentially decaying contributions to $W$. This diagrammatic method works for non-planar interface and wall configurations allowing us to determine whether they correspond to local or non-local contributions to the  interfacial Hamiltonian.
The decay of the Casimir contribution, which is determined almost exactly by its leading-order diagrammatic contribution $(\Omega_C)_1^1$, depends on the surface enhancement $g$ and is qualitatively different for critical wetting and first-order wetting transitions, changing form precisely at the MF tricritical point.

The presence of the Casimir contribution changes the interpretation of thermal fluctuation effects at wetting transitions which arise both from it and from the capillary-wave-like fluctuations of the unbinding fluid interface. The Casimir contribution leads to substantial changes to predictions for critical singularities at tricritical wetting and also for the film thickness at first-order wetting for all dimensions $d\ge 3$. Our predictions for first-order wetting in 3D can be tested in simulation studies of the Ising model and also molecular liquids and hopefully are experimentally accessible – for example, in studies of wetting in colloid-polymer mixtures or ionic fluids where the forces are effectively short-ranged.

The presence of two fluctuation regimes for short-ranged wetting, depending on the dimension $d$, is not dissimilar to the fluctuation theory of critical wetting in systems with long-ranged forces, where we must distinguish between the weak (WFL) and strong (SFL) fluctuation regimes, which have separate marginal dimensions {\color{blue}{\cite{LF_1987}}}. For systems with short-ranged forces the MF behavior is also modified in the regime $d>3$, by Casimir contributions (analogous to the WFL regime) and then in dimension $d\le 3$, by interfacial fluctuations, which remains the SFL regime. Thus $d=3$ is still an upper critical dimension, but distinguishes the SFL regime from one where the Casimir effect modifies MF behavior, which is only obtained as $d\to\infty$. Having determined the Casimir contribution to the binding potential  we may now return to the original controversy surrounding the predicted nonuniversality for critical wetting in $d=3$. The renormalization group analysis of this is discussed in a further paper.

\section*{Acknowledgments}

JMRE acknowledges financial support by Ministerio de Ciencia e Innovación (Spain) through Grant No PID2021-126348NB-I00.


\appendix

\section{Representation of the 3D kernel $K(\mathbf{r},\mathbf{r'})$ in Cartesian, cylindrical and spherical coordinates}
\label{repKcoordinates}
In this Appendix we will represent the kernel $K(\mathbf{r},\mathbf{r'})$ in 3D
\begin{equation}
    \label{appendixa1}
    K(\mathbf{r},\mathbf{r'})=\frac{\kappa}{2\pi}\frac{\exp(|\mathbf{r}-\mathbf{r}'|)}{|\mathbf{r}-\mathbf{r}'|} \, ,
\end{equation}
which is solution of the following equation
\begin{equation}
  \label{appendixa2} 
   (-\nabla_{\mathbf{r}}^2+\kappa^2)K(\mathbf{r},\mathbf{r'})=2\kappa \delta(\mathbf{r}-\mathbf{r}') \, .
\end{equation}
For Cartesian coordinates, a partial Fourier series representation in the transversal coordinates (i.e. $x$ and $y$) can be obtained as follows
\begin{equation}
\label{appendixa3}
K(\mathbf{r},\mathbf{r'})=\sum_{\mathbf{q}} K_{\mathbf{q}}(z-z')\frac{e^{\textrm{i}\mathbf{q}\cdot (\mathbf{x}-\mathbf{x}')}}{L_\parallel^2} \, , 
\end{equation}
where $\mathbf{q}=(2\pi/L_\parallel)(n_x,n_y)$ with $n_x,n_y=0,\pm 1, \pm 2, \ldots$, $\mathbf{x}=(x,y)$ and $L_\parallel$ is the transversal dimension of the system. Substituting (\ref{appendixa3}) into (\ref{appendixa2}), and taking into account the Fourier representation of Dirac delta
\begin{equation}
    \label{appendixa4}
    \delta(x-x_0)=\frac{1}{L}\sum_{n=-\infty}^\infty e^{\frac{2\pi\textrm{i}n(x-x_0)}{L}} \, ,
\end{equation}
we get the following equation for $K_{\mathbf{q}}$
\begin{equation}
  \label{appendixa5} 
   \left(-\frac{d^2}{dz^2}+\kappa^2_q\right)K_{\mathbf{q}}(z-z')=2\kappa \delta(z-z') \, ,
\end{equation}
where $\kappa_q=\sqrt{\kappa^2+q^2}$, which has solution
\begin{equation}
    \label{appendixa6}
    K_\mathbf{q}=\frac{\kappa}{\kappa_q}e^{-\kappa_q|z-z'|} \, .
\end{equation}
Thus, Eq. (\ref{appendixa3}) can be recast as
\begin{equation}
  \label{appendixa7}
  K(\mathbf{r},\mathbf{r'})=\sum_{\mathbf{q}}\frac{\kappa}{\kappa_q}e^{-\kappa_q|z-z'|} \phi_{\mathbf{q}}(\mathbf{x})(\phi_{\mathbf{q}}(\mathbf{x'}))^*\ , 
\end{equation}
where $\{\phi_{\mathbf{q}}(\mathbf{x})\equiv \exp(\textrm{i}\mathbf{q}\cdot \mathbf{x})/\sqrt{L_\parallel^2}\}$ is an orthonormal complete set for functions on any plane perpendicular to the $z$ direction. We note that Eq. (\ref{appendixa7}) is valid for any dimension $d$, just substituting $L_\parallel^2$ by $L_\parallel^{(d-1)}$.

Now we turn to the case of cylindrical coordinates. As $|\mathbf{r}-\mathbf{r}'|=\sqrt{\rho^2 +\rho'^2 - 2\rho\rho'\cos(\varphi-\varphi')+(z-z')^2}$, we can again obtain a partial Fourier series representation in the angular variable $\varphi$ and the coordinate $z$ along the axis of the cylinder (which is supposed to have a length $L_z$)
\begin{equation}
\label{appendixa8}
K(\mathbf{r},\mathbf{r'})=\sum_{n=-\infty}^\infty\sum_{q_z} K_{n,q_z}(\rho,\rho')\frac{e^{\textrm{i}n(\varphi-\varphi')}}{2\pi}\frac{e^{\textrm{i}q_z(z-z')}}{L_z} \, , 
\end{equation}
where $q_z=2\pi m/L_z$ with $m=0,\pm 1,\pm 2,\ldots$. Substitution of (\ref{appendixa8}) into (\ref{appendixa2}) leads to the following equation
\begin{eqnarray}
\frac{\partial}{\partial \rho}\left(\rho \frac{\partial K_{n,q_z}(\rho,\rho')}{\partial \rho}\right)&-&\left(\kappa_{q_z}^2\rho+\frac{n^2}{\rho}\right)K_{n,q_z}(\rho,\rho')\nonumber\\&=&-2\kappa \delta (\rho-\rho') \, ,
\label{appendixa9}
\end{eqnarray}
where $\kappa_{q_z}=\sqrt{\kappa^2+q_z^2}$, which has as solution
\begin{equation}
\label{appendixa9}
K_{n,q_z}(\rho,\rho')=2\kappa K_{|n|}(\kappa_{q_z}\rho_>)I_{|n|}(\kappa_{q_z}\rho_<)\, ,
\end{equation}
where $\rho_<=\min(\rho,\rho')$, $\rho_>=\max(\rho,\rho')$, and $I_\nu$ and $K_\nu$ are the modified Bessel functions of first and second type, respectively. Now Eq. (\ref{appendixa8}) can be expressed as
\begin{eqnarray}
K(\mathbf{r},\mathbf{r'})&=&\sum_{n=-\infty}^\infty\sum_{q_z} 2\kappa\sqrt{\rho \rho'} K_{|n|}(\kappa_{q_z}\rho_>) I_{|n|}(\kappa_{q_z}\rho_<)
\nonumber\\
&\times&\phi_{n,q_z}(\rho,\varphi,z) (\phi_{n,q_z}(\rho',\varphi',z'))^*\, , 
\label{appendixa10}
\end{eqnarray}
where $\{\phi_{n,q_z}(\rho,\varphi,z)\equiv \exp(\textrm{i}(n\varphi+q_z z))/\sqrt{2\pi \rho L_z}\}$ is an orthonormal complete set for functions with support on the cylindrical surface of radius $\rho$.

Finally we turn to the case of spherical coordinates. Now $|\mathbf{r}-\mathbf{r}'|=\sqrt{r^2 +r'^2 - 2rr'\cos(\theta-\theta')}$, so we expand $K(\mathbf{r},\mathbf{r}')$ as a series on the Legendre polynomials $P_l(\cos(\theta-\theta'))$, $l=0,1,\ldots$
\begin{equation}
K(\mathbf{r},\mathbf{r'})=\sum_{l=0}^\infty \frac{2l+1}{4\pi}K_l(r,r')P_l(\cos(\theta-\theta')) \, .
\label{appendixa11}
\end{equation}
Taking into account the addition theorem
\begin{equation}
\label{appendixa12}
P_l(\cos(\theta-\theta'))=\frac{4\pi}{2l+1}\sum_{m=-l}^l Y_l^m(\theta,\varphi)
(Y_l^m(\theta',\varphi'))^*\, ,
\end{equation}
where $Y_l^m$ are the spherical harmonics, which have the following closure condition
\begin{align}
& \delta(\cos(\theta)-\cos(\theta'))\delta(\varphi-\varphi') \nonumber \\
& = \sum_{l=0}^\infty\sum_{m=-l}^l Y_l^m(\theta,\varphi)
(Y_l^m(\theta',\varphi'))^*\, .
\label{appendixa13}
\end{align}
Thus, Eq. (\ref{appendixa11}) reads
\begin{equation}
K(\mathbf{r},\mathbf{r'})=\sum_{l=0}^\infty \sum_{m=-l}^l K_l(r,r')Y_l^m(\theta,\varphi)
(Y_l^m(\theta',\varphi'))^*  \, ,
\label{appendixa14}
\end{equation}
Substitution of (\ref{appendixa14}) into (\ref{appendixa2}) leads to
\begin{eqnarray}
\frac{\partial}{\partial r}\left(r^2 \frac{\partial K_{l}(r,r')}{\partial r}\right)&-&\left(\kappa^2r^2+l(l+1)\right)K_{l}(r,r')\nonumber\\&=&-2\kappa \delta (r-r') \, ,
\label{appendixa15}
\end{eqnarray}
which has as solution
\begin{eqnarray}
K_{l}(r,r')=2\kappa \frac{K_{l+1/2}(\kappa r_>)I_{l+1/2}(\kappa r_<)}{\sqrt{r r'}}\, ,
\label{appendixa16}
\end{eqnarray}
with $r_<=\min(r,r')$ and $r_>=\max(r,r')$. Thus, Eq. (\ref{appendixa14}) can be recast as
\begin{eqnarray}
K(\mathbf{r},\mathbf{r'})&=&\sum_{l=0}^\infty \sum_{m=-l}^l 2\kappa\sqrt{rr'}K_{l+1/2}(\kappa r_>)I_{l+1/2}(\kappa r_<)\nonumber\\
&\times&\phi_{l,m}(r,\theta,\varphi)(\phi_{l,m}(r',\theta',\varphi'))^*  \, ,
\label{appendixa14}
\end{eqnarray}
where $\{\phi_{l,m}(r,\theta,\varphi)\equiv Y_l^m(\theta,\varphi)/r\}$ is an orthonormal complete set for functions with support on the spherical surface of radius $r$.

\section{Casimir interaction in cylindrical and spherical shells with Dirichlet boundary conditions}
\label{sec_cylinders}
A natural question that arises concerns the role played by surface curvature in the Casimir effect. In order to address this issue, we consider the simplest geometries that allow  a clear identification of curvature-induced effects on the Casimir force. To this end we will examine the cases of two concentric cylinders and two concentric spherical shells. These types of geometries have been extensively studied in the context of the quantum Casimir effect; we refer to, e.g., \cite{Teo} and references therein.

The analysis of the Casimir effect (either quantum or thermal) requires the analysis of a fluctuation operator describing the quantum/classical fluctuations about a classical/mean-field solution. We will restrict ourselves to the case of Dirichlet boundary conditions. This type of investigations passes through the study and calculation of functional determinants, which are themselves an interesting field in mathematical physics that finds several applications in various areas of theoretical physics \cite{Dunne, Kirsten}.

Several techniques exist for the calculation of functional determinants, we refer to \cite{Dunne} for an introductory review of them; the technique we will employ in this paper is the zeta function regularization. The starting point of our analysis is the eigenvalue problem
\begin{equation}
\mathcal{M} \Psi(\textbf{r}) = \lambda \Psi(\textbf{r}) \, .
\end{equation}
For the problem at hand the operator $\mathcal{M}$ is the Helmholtz operator $\mathcal{M} = -\nabla^{2} + \kappa^2$, where  $\nabla^2$ is the $d$-dimensional Laplace operator.
The $\zeta$-regularization proceeds by introducing the corresponding $\zeta$-function
\begin{equation}
\label{11072024_1247}
\zeta(s) = \textrm{Tr}\Bigl\{\frac{1}{\mathcal{M}^s}\Bigr\} = \sum_{n} \frac{1}{\lambda_{n}^s} \, ,
\end{equation}
in such a way that
\begin{equation}
\label{}
\zeta^{\prime}(s) = - \sum_{n} \frac{\ln\lambda_{n}}{\lambda_{n}^s} \, ,
\end{equation}
and
\begin{equation}
\label{}
\zeta^{\prime}(0) = - \ln \prod_{n} \lambda_{n} \, ;
\end{equation}
the formal definition of the functional determinant of the operator $\mathcal{M}$ is therefore
\begin{equation}
\label{}
\det \mathcal{M} = \exp(-\zeta^{\prime}(0)) \, .
\end{equation}
These formal definitions are not free of possible ambiguities and subtleties, such as the convergence of sums and products as well as the analytic continuation to $s=0$, these issues have been carefully addressed in the review by G. Dunne \cite{Dunne}; see also \cite{Kirsten}.

At this stage it is meaningful to reconsider the $d$-dimensional slit geometry. For the discretized case, the fluctuation operator determinant can be obtained explicitly \cite{Ossipov,SREP2023}. A slit of dimensions $L_\parallel^{d-1}\times \ell$ with periodic boundary conditions along the transversal directions is considered. Let $\epsilon$ is the lattice spacing, i.e., $\epsilon=\ell/N$ and $M \epsilon = L_{\parallel} \gg \epsilon$. The transverse discrete $(d-1)$-Laplacian operator has as eigenvalues \cite{Ossipov}:
\begin{equation}
\label{}
\lambda_{\bm l} = -\frac{2}{\epsilon^2} 
\sum_{i=1}^{d-1} \left( 1 - \cos\frac{\pi l_{i}}{M}\right) \, ,
\end{equation}
where $\bm{l} = (l_{1},\dots,l_{d-1})$ is a vector of integers with $l_{i} \in \{1,\dots,M-1\}$. The determinant $\Delta$ of the discretized version of $-\nabla^2+\kappa^2$ is \cite{Ossipov,SREP2023}
\begin{equation}
\label{}
\Delta = \prod_{\bm l}\frac{\sinh \gamma_{\bm l} N}{\sinh \gamma_{\bm l}} \, ,
\end{equation}
where 
\begin{equation}
\label{e027}
\gamma_{\bm l}=\textrm{arccosh}\left(1+\frac{\kappa_q ^2 \epsilon^2}{2}\right)\, ,
\end{equation}
and $\kappa_q^2=\kappa^2-\lambda_{\bm l}^2$.
As it was discussed in Ref. \cite{SREP2023}, $\Delta$ can be split in two parts
\begin{equation}
\Delta=\left(\prod_{\bm l} \frac{e^{\gamma_{\bm l}N}}{2\sinh \gamma_{\bm l}}\right)\times
\left(\prod_{\bm l} \left(1-\textrm{e}^{-2N \gamma_{\bm l} }\right) \right)  \, ,
\end{equation}
where the first factor arises from fluctuation effects on the contributions to the free energy from bulk and isolated surfaces, while the second factor leads to the Casimir contribution. For the latter, only small values of $| \bm l |$ will contribute for large $N$ and $M$, i.e. small $\epsilon$. With this in mind, we are allowed to expand to lowest order in $l_{i}$ and retain the first contribution in such an expansion. Therefore we write
\begin{eqnarray}
\label{e028}
\gamma_{\bm l}\approx \kappa_q \epsilon \approx \sqrt{ \kappa^{2} + q^2}\epsilon.
\end{eqnarray}
where the wavenumber $q$ is defined as
\begin{equation}
\label{e029}
q^{2} \equiv \frac{\pi^{2}}{L_{\parallel}^{2}} \sum_{i=1}^{d-1}l_{i}^{2} \, .
\end{equation}
Therefore 
\begin{eqnarray}
\label{e030}
\textrm{e}^{-2N \gamma_{\bm \ell}} & \approx & \textrm{e}^{-2 \sqrt{ \kappa^{2} + q^{2} } \ell} \approx \textrm{e}^{-2\kappa_{q}\ell} \, ,
\end{eqnarray}
and we arrive at the following expression the Casimir contribution to $\ln \Delta$
\begin{eqnarray} \nonumber
&& \lim_{M \rightarrow \infty} \sum_{\bm \ell} \ln\left( 1 - \textrm{e}^{-2\gamma_{\bm l} N} \right) \\ \nonumber
&& = \frac{L_{\parallel}^{d-1}}{\pi^{d-1}} \int_{q_{i}>0} \rd \bm{q} \ln\left( 1 - \textrm{e}^{-2\kappa_{q}\ell } \right) \\ \label{e031}
&& = \frac{L_{\parallel}^{d-1}}{(2\pi)^{d-1}} \int_{\mathbb{R}^{d-1}} \rd \bm{q} \ln\left( 1 - \textrm{e}^{-2\kappa_{q}\ell } \right) \, ,
\end{eqnarray}
as we proved in \cite{SREP2023}.

Let us obtain the corresponding result by means of zeta-regularization. We consider the continuum limit for the $z$ direction but keep the discretization on the transversal directions. We introduce
\begin{equation}
\label{e032}
\zeta(s) \equiv \sum_{\bm l} \zeta_{\bm l}(s) \, ,
\end{equation}
where $\zeta_{\bm l}$ is the $\zeta-$function (\ref{11072024_1247}) associated to the operator $\mathcal{M}=-d^2/dz^2+\kappa_q^2$. Thus,
\begin{equation}
\label{e033}
\exp(-\zeta_{\bm l}^{\prime}(0)) = \det \left( - \frac{\rd^{2}}{\rd z^{2}} + \kappa_q^{2} \right) \, ,
\end{equation}
The eigenvalue problem for this 1D operator takes the form
\begin{equation}
\label{e034}
\left( - \frac{\rd^{2}}{\rd z^{2}} + \kappa_{q}^{2} \right) u_{\lambda}(z) = \lambda u_{\lambda}(z) \, ,
\end{equation}
with $u_{\lambda}(0)=0$ and $u_{\lambda}(\ell)=0$. Up to a multiplicative factor, the solution is given by $u_{\lambda}(z) = \sin(\sqrt{\lambda-\kappa_{q}^{2}}z)$. Notice that $u_\lambda(\ell)$ vanishes at the eigenvalues of Eq. (\ref{e034}) and for $\lambda=\kappa_q^2$. So, in order to eliminate this spurious root of the eigenvalue equation, we choose $F_{\bm l}(\lambda)=\sin(\sqrt{\lambda-\kappa_{q}^{2}}z)/\sqrt{\lambda-\kappa_{q}^{2}}$ as the function which vanishes exactly at the eigenvalues of the eigenvalue problem (\ref{e034}). By using the residue theorem, Eq. (\ref{11072024_1247}) can be recast as an integral in the $\lambda-$complex plane \cite{Dunne}
\begin{equation}
\label{e035}
\zeta_{\bm l}(s) = \frac{1}{2\pi \im} \int_{\gamma} \rd \lambda \, \lambda^{-s} \frac{\rd}{\rd\lambda} \ln F_{\bm l}(\lambda) \, ,
\end{equation}
where $\gamma$ is a counterclockwise contour which encloses all the poles of the integrand, i.e. the real positive eigenvalues of (\ref{e034}). Note that there is a branch cut due to the factor $\lambda^{-s}$ in the integrand, which is chosen on the negative real $\lambda-$axis. We now deform $\gamma$ into a contour that encloses the negative $\lambda-$real axis, so Eq. (\ref{e035}) yields
\begin{eqnarray}
\label{e036} \nonumber
\zeta_{\bm l}(s) & = & \frac{\sin \pi s}{\pi} \int_{0}^{\infty} \rd t \, t^{-s} \frac{\rd}{\rd t} \ln F_{\bm l}(-t) \, , \\ \nonumber
& = & \frac{\sin \pi s}{\pi} \int_{0}^{\infty} \rd t \, t^{-s} \frac{\rd}{\rd t} \ln \frac{ \sinh\sqrt{t+\kappa_{q}^{2}}\ell }{ \sqrt{t+\kappa_{q}^{2}} }  \, , \\
& = & \zeta_{\bm l}^{(1)}(s) + \zeta_{\bm l}^{(2)}(s) \, ,
\end{eqnarray}
with
\begin{equation}
\label{e037}
\zeta_{\bm l}^{(1)}(s) = \frac{\sin\pi s}{\pi} \int_{0}^{\infty}\rd t \, t^{-s} \frac{\rd}{\rd t} \ln \frac{ \textrm{e}^{\sqrt{t+\kappa_{q}^{2}}\ell} }{ 2 \sqrt{t+\kappa_{q}^{2}} } \, ,
\end{equation}
and
\begin{equation}
\label{e038}
\zeta_{\bm l}^{(2)}(s) = \frac{\sin\pi s}{\pi} \int_{0}^{\infty}\rd t \, t^{-s} \frac{\rd}{\rd t} \ln\biggl[ 1 - \textrm{e}^{-2\sqrt{t+\kappa_{q}^{2}}\ell} \biggr] \, .
\end{equation}
The first term needs to be regularized. Note that, as $t \rightarrow \infty$, 
\begin{equation}
\label{e039}
\frac{ \textrm{e}^{\sqrt{t+\kappa_{q}^{2}}\ell} }{ 2 \sqrt{t+\kappa_{q}^{2}} } \approx \frac{ \textrm{e}^{\sqrt{t}\ell} }{ 2\sqrt{t} } \, ,
\end{equation}
therefore, following \cite{Kirsten}, we have
\begin{equation}
\label{e040}
\zeta_{\bm l}^{(1)} = \zeta_{\bm l, \rho}^{(1)} + \zeta_{\bm l, \rm as}^{(1)} \, ,
\end{equation}
where
\begin{eqnarray}
\label{e041} \nonumber
\zeta_{\bm l, \rho}^{(1)} & = & \frac{\sin\pi s}{\pi} \int_{0}^{\ell^{-2}} \rd t \, t^{-s} \frac{\rd}{\rd t} \ln \frac{ \textrm{e}^{\sqrt{t+\kappa_{q}^{2}}\ell} }{ 2\sqrt{t+\kappa_{q}^{2}}} \\ \nonumber
& + & \frac{\sin\pi s}{\pi} \int_{\ell^{-2}}^{\infty} \rd t \, t^{-s} \frac{\rd}{\rd t} \ln \biggl[ \sqrt{\frac{t}{t+\kappa_{q}^{2}}} \textrm{e}^{\sqrt{t+\kappa_{q}^{2}}\ell-\sqrt{t}\ell} \biggr]
\end{eqnarray}
and
\begin{eqnarray}
\label{e042}
\zeta_{\bm l, \rm as}^{(1)} & = & \frac{\sin\pi s}{\pi} \int_{\ell^{-2}}^{\infty} \rd t \, t^{-s} \frac{\rd}{\rd t} \ln \frac{ \textrm{e}^{\sqrt{t}\ell} }{ \sqrt{t} }
\end{eqnarray}
is obtained from the asymptotic result (\ref{e039}). The function $\zeta_{\bm l, \rho}^{(1)}$ is analytic in $s=0$, therefore
\begin{eqnarray}
\label{e043}
\left( \zeta_{\bm l, \rho}^{(1)}\right)^{\prime}(0) & = & 1+ \ln \left( \frac{\kappa_{q}\ell}{ \textrm{e}^{\kappa_{q}\ell} } \right) \, .
\end{eqnarray}
On the other hand, $\zeta_{\bm l, \rm as}^{(1)}$ can be analytic continued to
\begin{eqnarray}
\label{e044}
\zeta_{\bm l, \rm as}^{(1)}(s) & = & \ell^{2s} \frac{\sin \pi s}{2\pi} \left( \frac{1}{s-1/2} - \frac{1}{s} \right),
\end{eqnarray}
so
\begin{eqnarray}
\label{e045}
\left( \zeta_{\bm l, \rm as}^{(1)}\right)^{\prime}(0) & = & -1 -\ln \ell\, .
\end{eqnarray}
Summing up the two terms, we find
\begin{eqnarray}
\label{e046}
\left( \zeta_{\bm l}^{(1)}\right)^{\prime}(0) & = & -\kappa_{q}\ell + \ln \kappa_{q} \, .
\end{eqnarray}
The term $-\kappa_{q}\ell$ gives a fluctuation-induced bulk contribution in between the plates and $\ln\kappa_q$ a contribution to the isolated boundaries, so they do not contribute to the Casimir term. On the other hand,
\begin{eqnarray}
\label{e047}
- \left( \zeta_{\bm l}^{(2)}\right)^{\prime}(0) & = & \ln\left( 1 - \textrm{e}^{-2\kappa_{q}\ell} \right) \, ,
\end{eqnarray}
which means that the Casimir term is given by
\begin{equation}
\label{e048}
\sum_{ \bm l }  \ln\left( 1 - \textrm{e}^{-2\kappa_{q}\ell} \right) \stackrel{ L_{\parallel} \rightarrow \infty }{ = } \frac{ L_{\parallel}^{d-1} }{ (2\pi)^{d-1} } \int\rd \bm{q} \ln\left( 1 - \textrm{e}^{-2\kappa_{q}\ell} \right) \, ,
\end{equation}
in agreement with Eq. (\ref{e031}).

Let us turn to the eigenvalue problem (\ref{e015}) for the Helmholtz operator in $d-$dimensional spherical coordinates.
This operator falls in those of the form $\mathcal{M} = -\nabla^{2} + V(r)$,  and $V(r)$ is a radially symmetric potential. By using the separation of angular and radial variables, we can write $\Psi(\textbf{r}) = r^{-(d-1)/2} \phi_{m}(r)Y_{m}(\bm{\theta})$, where $Y_{m}(\bm{\theta})$ are hyperspherical harmonics. Then, the radial eigenfunctions satisfy
\begin{equation}
\label{e015}
\left( - \frac{ \rd^{2} }{ \rd r^{2} } + \frac{U(m,d)}{r^{2}} + \kappa^{2} \right) \phi_{m}(r) = \lambda \phi_{m}(r) \, ,
\end{equation}
where $\phi_{m}(r)$ is subjected to the boundary conditions $\phi_{m}(R)=\phi_{m}(R+\ell)=0$,
\begin{equation}
\label{}
U(m,d) = \left( m+ \frac{d-3}{2} \right) \left( m+ \frac{d-1}{2} \right) \, .
\end{equation}

Strictly speaking, the summation in (\ref{11072024_1247}) has to be adapted by taking into account that for $d \geqslant 2$ the eigenvalue $\lambda_{m}$ carries the degeneracy factor
\begin{equation}
\label{}
g(m,d) = \frac{(2m+d-2)(m+d-3)!}{m! (d-2)!} \, ,
\end{equation}
for $m=0,1,2, \dots$. 
However, as shown by Dunne \cite{Dunne}, the naive extension from ordinary ($d=1$) to partial differential equations ($d \geqslant 2$) might fail, as happens when the 1D Gel’fand–Yaglom expression for the functional determinant of the eigenvalue problem operator Eq. (\ref{e015}) is used for the separable problem we are discussing. Indeed, by taking into account the degeneracy factor for all values of the angular momentum $m$, the product of radial determinants diverges, so a renormalization procedure is needed to obtain the $d-$dimensional functional determinant \cite{Dunne}. For this reason we will make use of the zeta function regularization in the rest of this appendix.

By adapting the above notations to (\ref{e015}), the functional determinant we need to compute is formally defined by
\begin{equation}
\label{e013}
\det\left( - \frac{\rd^{2}}{\rd r^{2}} + \frac{U(m,d)}{r^{2}} + \kappa^{2} \right) = \exp(-\zeta_{m}^{\prime}(0)) \, .
\end{equation}
The zeta function in (\ref{e013}) is defined by
\begin{equation}
\label{e014}
\zeta_{m}(s) = \frac{1}{2\pi \im} \int_{\gamma} \rd \lambda \, \lambda^{-s} \frac{\rd}{\rd \lambda} \ln F_{m}(\lambda) \, ,
\end{equation}
where $\gamma$ is a contour in the complex $\lambda$-plane that surrounds the zeros of the eigenvalues of the operator appearing in (\ref{e013}). The function $F_{m}(\lambda)$ can be identified from the eigenvalue problem (\ref{e015}). Its eigenfunctions have the form
\begin{eqnarray} \nonumber
\label{e016}
\phi_{m}(r) & = & \alpha_{m} \sqrt{r} \biggl[ J_{m+\frac{d}{2}-1}(\bar{\lambda}R) Y_{m+\frac{d}{2}-1}(\bar{\lambda} r) \\
& - & J_{m+\frac{d}{2}-1}(\bar{\lambda} r) Y_{m+\frac{d}{2}-1}(\bar{\lambda} R) \biggr] \, ,
\end{eqnarray}
where $\bar{\lambda} = \sqrt{\lambda-\kappa^{2}}$. The boundary condition $\phi_{m}(R)=0$ is evidently satisfied, while  the fulfillment of the condition $\phi_{m}(R+\ell)=0$ allows us to identify the function $F_{m}(\lambda)$,
\begin{eqnarray} \nonumber
\label{e017}
F_{m}(\lambda) & = & J_{m+\frac{d}{2}-1}(\bar{\lambda}R) Y_{m+\frac{d}{2}-1}(\bar{\lambda}(R+\ell)) \\
& - & J_{m+\frac{d}{2}-1}(\bar{\lambda}(R+\ell)) Y_{m+\frac{d}{2}-1}(\bar{\lambda}R) \, .
\end{eqnarray}
We can now plug this expression for $F_{m}(\lambda)$ into (\ref{e014}) and make progress with the calculation. To this end, we perform the change of variable in the integrand $\lambda=\mu^{2}+\kappa^{2}$. Successively, we shift the contour integral along the imaginary axis. This operation can be performed without impunity by preliminary noting the existence of branch cuts at $\mu = \pm \im \kappa$. Therefore, we can write the $\zeta$-function as the following integral along the imaginary axis
\begin{equation} \nonumber
\label{e018}
\zeta_{m}(s) = \frac{\sin \pi s}{\pi} \int_{\kappa}^{\infty} \rd t \left( t^{2}-\kappa^{2} \right)^{-s} \frac{\rd}{\rd t} \ln \chi_{m+\frac{d}{2}-1}(t; R, \ell)
\end{equation}
where
\begin{eqnarray} \nonumber
\label{e019}
\chi_{\nu}(t; R, \ell) & = & J_{\nu}(\im t R) Y_{\nu}(\im t R(1+\ell/R)) \\
& - & J_{\nu}(\im t R(1+\ell/R)) Y_{\nu}(\im t R) \, .
\end{eqnarray}
Since the Bessel functions $J_{\nu}(z)$ and $Y_{\nu}(z)$ are evaluated at imaginary arguments, we can convert them into modified Bessel functions of the second type, $I_{\nu}(z)$ and $K_{\nu}(z)$. Therefore, (\ref{e019}) admits the equivalent rewriting
\begin{eqnarray} \nonumber
\label{e020}
\chi_{\nu}(t; R, \ell) & = & \frac{2}{\pi} \biggl[ K_{\nu}(t R) I_{\nu}(t R(1+\ell/R)) \\
& - & I_{\nu}(t R) K_{\nu}(t R(1+\ell/R)) \biggr] \, .
\end{eqnarray}
At this point we notice that by plugging (\ref{e020}) into (\ref{e018}) we obtain the same expression obtained in \cite{Teo} in the context of the electromagnetic Casimir effect.

The zeta function can be decomposed into three contributions, one of them stemming from the region inside the cavity where $r<R$, one from the space outside the larger cavity, $r>R+\ell$, and a third contribution originating from the interaction of the cavities. This decomposition entails
\begin{equation}
\label{e021}
\zeta_{m}(s) = \zeta_{m}^{\rm out}(s;R) + \zeta_{m}^{\rm in}(s;R+\ell) + \zeta_{m}^{\rm int}(s;R,R+\ell) \, .
\end{equation}
The various terms entering (\ref{e021}) are explicitly given by
\begin{eqnarray} \nonumber
\label{e022}
\zeta_{m}^{\rm out}(s;R) & = & \frac{\sin \pi s}{\pi} \int_{\kappa}^{\infty} \rd t \left( t^{2}-\kappa^{2} \right)^{-s}\\ &\times& \frac{\rd}{\rd t} \ln (t^{\nu}K_{\nu}(Rt)) \, , \\ \nonumber
\zeta_{m}^{\rm in}(s;R+\ell) & = & \frac{\sin \pi s}{\pi} \int_{\kappa}^{\infty} \rd t \left( t^{2}-\kappa^{2} \right)^{-s} \\
& \times &\frac{\rd}{\rd t} \ln (t^{-\nu}I_{\nu}((R+\ell)t) \, ,
\end{eqnarray}
and
\begin{eqnarray}
\label{e023}
\zeta_{m}^{\rm int}(s;R,R+\ell) & = & \frac{\sin \pi s}{\pi} \int_{\kappa}^{\infty} \rd t \left( t^{2}-\kappa^{2} \right)^{-s} \\ \nonumber
& \times & \frac{\rd}{\rd t} \ln \biggr[ 1 - \frac{ K_{\nu}(t(R+\ell)) I_{\nu}(tR) }{ I_{\nu}(t(R+\ell)) K_{\nu}(tR) } \biggr] \, .
\end{eqnarray}

Note that $\zeta_{m}^{\rm out}(s;R)$ and $\zeta_{m}^{\rm in}(s;R+\ell)$ are the fluctuation contributions outside an \emph{isolated} sphere of radius $R$ and inside an isolated sphere of radius $R+\ell$. Hence, these terms do not contribute to the Casimir interaction. Furthermore, the existence of these terms is the cause of the divergence in $\zeta_{m}(s)$; we refer the interested reader to \cite{Kirsten} for a detailed discussion of these terms.

The analytic behavior of the interaction term, $\zeta_{m}^{\rm int}(s;R,R+\ell)$, has been analyzed in Ref. \cite{Teo} and it is analytic in $s$. We are allowed to perform the derivative with respect to $s$, an operation that produces two terms: one upon acting on $\sin \pi s$, another one comes from the derivative of $(t^{2}-\kappa^{2})^{-s}$. The latter term originates a finite contribution provided $s<0$ because the corresponding integral is convergent. The derivative with respect to $s$ is performed as the following limiting operation
\begin{equation}
\label{e024}
\left( \zeta_{m}^{\rm int}(0) \right)^{\prime} \equiv \lim_{s \rightarrow 0^{-}} \left( \zeta_{m}^{\rm int}(s;R,R+\ell) \right)^{\prime} \, ,
\end{equation}
therefore we are left with
\begin{eqnarray}
\label{e025}
- \left( \zeta_{m}^{\rm int}(0) \right)^{\prime} & = & \ln \biggr[ 1 - \frac{ I_{\nu}(\kappa R) K_{\nu}(\kappa(R+\ell)) }{ I_{\nu}(\kappa(R+\ell)) K_{\nu}(\kappa R) } \biggr] \, .
\end{eqnarray}

With this result in mind, let us analyze the 3$D$ spherical and cylindrical geometries. For the spherical shell the Casimir term takes the form
\begin{equation}
\label{e049}
\sum_{m=0}^{\infty} (2m+1) \ln\biggl[ 1 - \frac{ I_{m+1/2}(\kappa R) K_{m+1/2}(\kappa (R+\ell)) }{ K_{m+1/2}(\kappa R) I_{m+1/2}(\kappa (R+\ell)) } \biggr]
\end{equation}
while for the cylindrical shell the Casimir term is
\begin{eqnarray} \nonumber
\label{e050}
&& \frac{L_{z}}{2\pi} \int_{-\infty}^{\infty} \rd q_{z} \biggr\{ \ln\biggl[ 1 - \frac{ I_{0}(\kappa_{q_{z}} R) K_{0}(\kappa_{q_{z}} (R+\ell)) }{ K_{0}(\kappa_{q_{z}} R) I_{0}(\kappa_{q_{z}} (R+\ell)) } \biggr] \\
&& + 2 \sum_{m=1}^{\infty} \ln\biggl[ 1 - \frac{ I_{m}(\kappa_{q_{z}} R) K_{m}(\kappa_{q_{z}} (R+\ell)) }{ K_{m}(\kappa_{q_{z}} R) I_{m}(\kappa_{q_{z}} (R+\ell)) } \biggr] \biggr \} \, ,
\end{eqnarray}
where $\kappa_{q_{z}} \equiv \sqrt{ \kappa^{2} + q_{z}^{2} }$. The cylindrical shell differs from the 2$D$ case (unlike the mean-field expression) because of the fluctuations along the cylinder axis ($z$). The effect is: i) to sum over the fluctuations along $z$ (i.e., plane waves) and ii) to modify the transversal fluctuations by modifying $\kappa^{2}$ by $\kappa^{2}+q_{z}^{2}=\kappa_{q_{z}}^{2}$ (in a similar way as sketched in the slit geometry).

Let us consider now the regime $R \gg \ell$, $R \gg 1/\kappa$. For this purpose, we will consider Debye's expansion of the modified Bessel functions \cite{Abramowitz_Stegun}. For large index $\nu$, the expansions read:
\begin{eqnarray}\nonumber
\label{e051}
I_{\nu}(\nu z) & \sim & \frac{1}{\sqrt{2\pi \nu}} \frac{ \textrm{e}^{\nu/2} }{ (1+z^{2})^{1/4} } \biggl[ 1 + \sum_{k=1}^{\infty} \frac{u_{k}(t)}{\nu^{k}} \biggr] \\ \nonumber
K_{\nu}(\nu z) & \sim & \sqrt{\frac{\pi}{2\nu}} \frac{ \textrm{e}^{-\nu/2} }{ (1+z^{2})^{1/4} } \biggl[ 1 + \sum_{k=1}^{\infty} (-1)^{k} \frac{u_{k}(t)}{\nu^{k}} \biggr] \, ,
\end{eqnarray}
where $t=1/\sqrt{1+z^{2}}$, $\eta=\sqrt{1+z^{2}} + \ln(z/[1+\sqrt{1+z^{2}}])$ and $u_{k}(t)$ for $k=0,1,2,\dots$ are polynomials in $t$, which satisfy
\begin{equation}\nonumber
\label{e052}
u_{k+1}(t) = \frac{1}{2} t^{2} (1-t^{2}) u_{k}^{\prime}(t) + \frac{1}{8} \int_{0}^{t} \rd \tau \, (1 - 5 \tau^{2} ) u_{k}(\tau) \, ,
\end{equation}
with $u_{0}(t)=1$. This representation is valid for large $\nu$ but fixed $z$, which corresponds to large $R$ and large $m$. Thus, we define
\begin{equation}
\label{e053}
\nu = m + \frac{d}{2} - 1 = 
\begin{cases}
l      & d=2 \, , \\
l+\frac{1}{2}      & d=3 \, ,
\end{cases}
\end{equation}
where $l$ is an integer. Then, we introduce the shorthand notation
\begin{equation}
\label{e054}
z =  
\begin{cases}
\kappa_{q_{z}}R/l \equiv \kappa_{q_{z}}/q   & d=2 \, , \\
\kappa R/(l+1/2) \equiv \kappa/q     & d=3 \, ,
\end{cases}
\end{equation}
with $q \equiv \nu/R$; we therefore adopt $\kappa$ for $\kappa_{q_{z}}$ when $d=2$. Leaving the lengthy calculations, the ratio of Bessel functions ($z=\kappa/\nu$) is
\begin{eqnarray}
\label{e055} \nonumber
\frac{ I_{\nu}(\kappa R) K_{\nu}(\kappa (R+\ell)) }{ K_{\nu}(\kappa R) I_{\nu}(\kappa (R+\ell)) } & = & \textrm{e}^{-2\kappa_q\ell} \biggl[ 1 +  \frac{q^{2}\ell^{2}}{\kappa_q R} + O\left( R^{-2} \right) \biggr] \, ,
\end{eqnarray}
where $\kappa_q=\kappa^2+q^2$ ($=\kappa^2+q^2+q_z^2$ in the cylindrical case). Therefore, coming back to (\ref{e049}), for the spherical shell we have
\begin{eqnarray}
\label{e056} \nonumber
&& 2\beta W_{C} = \\ \nonumber
&& \sum_{m=0}^{\infty} (2m+1) \ln\biggl[ 1 - \frac{ I_{m+1/2}(\kappa R) K_{m+1/2}(\kappa (R+\ell)) }{ K_{m+1/2}(\kappa R) I_{m+1/2}(\kappa (R+\ell)) } \biggr] \, .
\end{eqnarray}
We introduce the discrete wave number $q_{m}=(m+1/2)/R$ and $ \Delta q_{m}=1/R$, so the right hand side of the above becomes
\begin{equation}\nonumber
\label{e057}
2R^{2} \sum_{m=0}^{\infty} \Delta q_{m} q_{m} \ln\biggl[ 1 - \frac{ I_{q_{m}R}(\kappa R) K_{q_{m}R}(\kappa (R+\ell)) }{ K_{q_{m}R}(\kappa R) I_{q_{m}R}(\kappa (R+\ell)) } \biggr] \, .
\end{equation}
Now we apply Euler-Maclaurin formula in order to convert the sum into an integral; hence, the above becomes
\begin{eqnarray}
\label{e058} \nonumber
&& 2R^{2} \int_{0}^{\infty}\rd q \, q \ln\biggl[ 1 - \frac{ I_{qR}(\kappa R) K_{qR}(\kappa (R+\ell)) }{ K_{qR}(\kappa R) I_{qR}(\kappa (R+\ell)) } \biggr] \\ \nonumber
&& + \frac{1}{2} \ln\biggl[ 1 - \frac{ I_{1/2}(\kappa R) K_{1/2}(\kappa (R+\ell)) }{ K_{1/2}(\kappa R) I_{1/2}(\kappa (R+\ell)) } \biggr] + O(R^{-1}) \, ,
\end{eqnarray}
where the second term can be simplified as
\begin{equation}\nonumber
\label{e059}
\frac{ I_{1/2}(\kappa R) K_{1/2}(\kappa (R+\ell)) }{ K_{1/2}(\kappa R) I_{1/2}(\kappa (R+\ell)) } = \textrm{e}^{-2\kappa\ell} \frac{1 - \textrm{e}^{-2\kappa(R+\ell)}}{1 - \textrm{e}^{-2\kappa R}} \, .
\end{equation}
Provided $R \gg \ell$ and $R \gg 1/\kappa$, a simple rearrangement allows us to write
\begin{eqnarray}\nonumber
\label{e060}
2\beta W_{C} & \approx & \frac{A}{(2\pi)^{2}} \int_{\mathbb{R}^{2}} \rd \bm{q} \ln \left( 1 - \textrm{e}^{-2\sqrt{\kappa^{2}+\bm{q}^{2}}} \right) \, ,
\end{eqnarray}
where $\rd \bm{q} = \rd q_{x} \rd q_{y}$, $\bm{q}^{2} = q_{x}^{2} + q_{y}^{2}$, and $A=4\pi R^{2}$ is the area of the sphere. A completely analogous reasoning applies to the cylinder, in this case the only difference is the expression of the area, $A=2\pi R L_{z}$. Consequently, for both concentric spherical and cylindrical shell geometries, the result of the slit geometry is obtained in the limit $R \rightarrow \infty$ as expected.

However, the results for the spherical and cylindrical shells exhibit differences when $\ell$ is finite and in this case the two geometries differ by subleading corrections at order $R^{-1}$. These correction, which vanish in the limit $R \rightarrow \infty$, allow to retrieve the slit geometry. These corrections can be investigated as explained below. For large $\kappa \ell$, we write, 
\begin{eqnarray}\nonumber
\label{e061}
&&\ln\biggl[ 1 - \frac{ I_{\nu}(\kappa R) K_{\nu}(\kappa (R+\ell)) }{ K_{\nu}(\kappa R) I_{\nu}(\kappa (R+\ell)) }\biggr ] \\ 
\nonumber & \approx &\ln \biggl[ 1 - \textrm{e}^{-2\kappa_{q}\ell} \left( 1 + \frac{q^{2}\ell^{2}}{\kappa_{q}R} \right)\biggr] \, , \\ 
& \approx & - \textrm{e}^{-2\kappa_{q}\ell} \left( 1 + \frac{q^{2}\ell^{2}}{\kappa_{q}R} \right) \, .
\end{eqnarray}
We recall that for the cylinder geometry we have to replace $\kappa$ with $(\kappa^{2}+q_{z}^{2})^{1/2}$. With this in mind, for the spherical shell, we obtain
\begin{eqnarray}
\label{e062} \nonumber
\frac{2\beta W_{C}^{\rm sph-s}}{A} & = & \underbrace{- \frac{1}{(2\pi)^{2}} \int_{\mathbb{R}^{2}}\rd\bm{q} \, \textrm{e}^{-2\kappa_{q}\ell} }_{\rm planar} \\
& - & \frac{\ell^{2}}{2\pi R} \int_{0}^{\infty} \rd q \, q^{3} \frac{ \textrm{e}^{-2\kappa_{q}\ell} }{\kappa_{q}} \, ,
\end{eqnarray}
the last integral can be conveniently evaluated by using the change of variables $s=\sqrt{\kappa^{2}+q^{2}}/\kappa$, which brings the integral in the form
\begin{eqnarray}
\label{e063} \nonumber
\frac{2\beta W_{C}^{\rm sph-s}}{A} & = & \frac{2\beta W_{C}^{\rm planar}}{A} - \frac{ (\kappa\ell)^{3} }{ 2\pi R } \int_{1}^{\infty}\rd s \, (s^{2}-1) \textrm{e}^{-2\kappa \ell s} \, ,
\end{eqnarray}
and therefore
\begin{eqnarray}
\label{e064} \nonumber
\frac{2\beta W_{C}^{\rm sph-s}}{A} & = & \frac{2\beta W_{C}^{\rm planar}}{A} - \frac{ 1+2\kappa\ell }{ 8\pi R\ell } \textrm{e}^{-2\kappa\ell} \, .
\end{eqnarray}
Let us consider now the cylindrical shell. Now, (\ref{e062}) turns into
\begin{eqnarray}
\label{e065} \nonumber
\frac{2\beta W_{C}^{\rm cyl-s}}{A} & = & \frac{2\beta W_{C}^{\rm planar}}{A} - \frac{ \ell^{2} }{ (2\pi)^{2}R } \int_{0}^{2\pi}\rd\vartheta \\ \nonumber
& \times & \int_{0}^{\infty}\rd q \, q \frac{ q^{2}\cos^{2}\vartheta }{ \kappa_{q}} \textrm{e}^{-2\kappa_{q}\ell} \, .
\end{eqnarray}
After having performed the integration over the angle $\vartheta$ the resulting integral over the momentum $q$ can be evaluated as already done for the spherical shell, and the result is
\begin{eqnarray}
\label{e066} \nonumber
\frac{2\beta W_{C}^{\rm cyl-s}}{A} & = & \frac{2\beta W_{C}^{\rm planar}}{A} - \frac{ 1+2\kappa\ell }{ 16\pi R\ell } \textrm{e}^{-2\kappa\ell} \, .
\end{eqnarray}
In the cylinder, the correction is exactly $1/2$ that for the sphere. It is natural to speculate that such a factor is  related to the mean curvature of the surface.

\section{Evaluation of the downward barred arrow kernel}
\label{sec_planarwallcurvedinterface}
In this Appendix we will evaluate the kernel associated to the downward barred diagram defined by Eq. (\ref{defdiagrams2-0}). 
We will denote with $\textbf{s}_{2}$ the point on the interface and $\textbf{s}_{1}$ the point on the wall. This kernel is itself a convolution between the kernel $J$ between the wall and the surface
\begin{eqnarray}
J(\mathbf{s}_1,\mathbf{s}_2) & = & K(\mathbf{s}_1,\mathbf{s}_1) + g^{-1} \mathbf{n}(\mathbf{s}_1) \cdot \nabla_{\mathbf{s}_1} K(\mathbf{s}_1,\mathbf{s}_2) \,
\end{eqnarray}
where $\mathbf{n}(\mathbf{s}_1)$ is the upwards normal to the wall at $\mathbf{s}_1$, and the kernel $X$ of the inverse of the integral operator on $S_1$ with kernel $Y_{\mathbf{s}}K\equiv K+ g^{-1} \mathbf{n} \cdot \nabla K -(\kappa/g) \delta$. We want to obtain the leading-order expression of this kernel by neglecting wall curvature corrections. For this purpose, we can substitute the wall by the tangent plane on $\mathbf{s}_1$ and assuming that the kernel on the wall $Y_{\mathbf{s}_1}K(\mathbf{s}_1,\mathbf{s}_1^\prime)\approx K(\mathbf{s}_1,\mathbf{s}_1^\prime)-(\kappa/g)\delta(\mathbf{s}_1-\mathbf{s}_1^\prime)$. We denote $\ell=(\mathbf{s}_2-\mathbf{s}_1)\cdot \mathbf{n}(\mathbf{s}_1)$ and $\boldsymbol{\rho}=\mathbf{s}_2-\mathbf{s}_1-\ell \mathbf{n}(\mathbf{s}_1)$ is the projection of the vector $\textbf{s}_{1}-\textbf{s}_{2}$ on the horizontal surface $S_{1}$. By using the Fourier representation of the kernels $J$ and $X$ in the transversal components $\boldsymbol{\rho}$, we obtain
\begin{eqnarray} \nonumber
\label{star01}
\figu{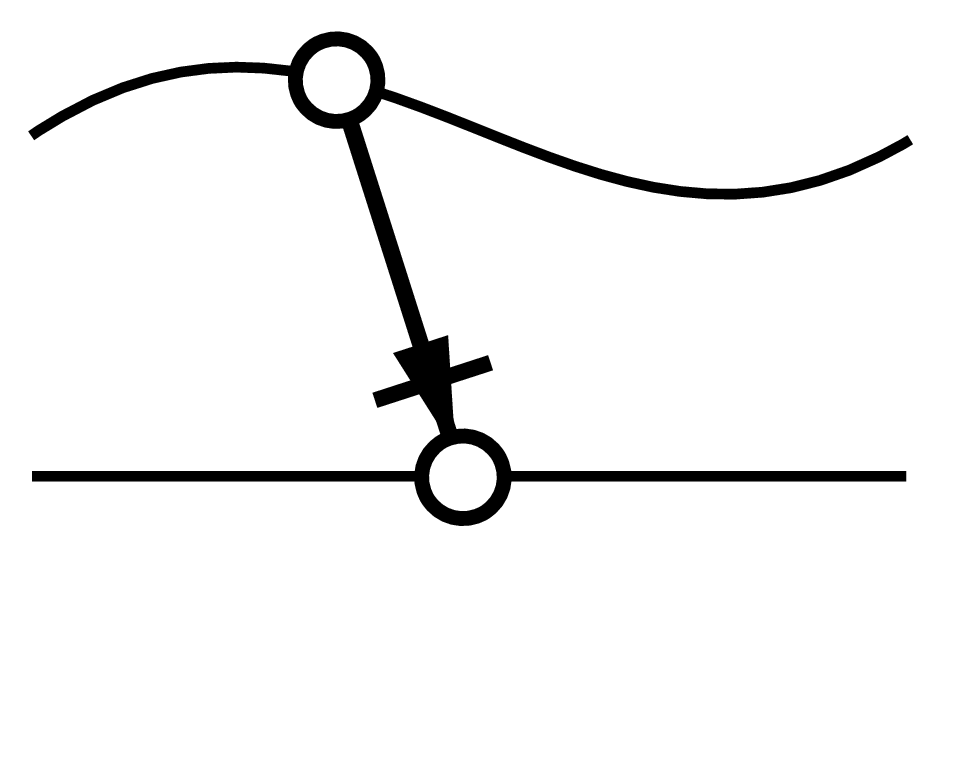} & = & \int\frac{\rd\textbf{q}}{(2\pi)^{d-1}} \, \textrm{e}^{\im\textbf{q}\cdot \boldsymbol{\rho}} \frac{\kappa}{\kappa_{q}} \left( 1 + \frac{\kappa_{q}}{g} \right)\\ &\times& \textrm{e}^{-\kappa_{q}\ell} \biggl[ \frac{\kappa}{\kappa_{q}} \left( 1 - \frac{\kappa_{q}}{g} \right) \biggr]^{-1}
\end{eqnarray}
and therefore
\begin{eqnarray}
\label{star02}
\figu{fig22} & = & \int\frac{\rd\textbf{q}}{(2\pi)^{d-1}} \, \frac{g+\kappa_{q}}{g-\kappa_{q}} \textrm{e}^{\im\textbf{q}\cdot \boldsymbol{\rho} -\kappa_{q}\ell} \, .
\end{eqnarray}
In order to make further progress we integrate over the orientations of $\mathbf{q}$. For $d=2$, Eq. (\ref{star02}) reduces to
\begin{align} \nonumber
\figu{fig22} & = \int_0^\infty\frac{\rd q}{2\pi} \, \frac{g+\kappa_{q}}{g-\kappa_{q}} \textrm{e}^{-\kappa_{q}\ell} 2\cos (q\rho)\\
& = \int_0^\infty \frac{\rd q \, q^{1/2}}{(2\pi)^{1/2}}\frac{g+\kappa_{q}}{g-\kappa_{q}} \textrm{e}^{-\kappa_{q}\ell} \frac{J_{-1/2}(q \rho)}{\rho^{-1/2}} \, ,
\label{star03}
\end{align}
where $\rho=|\boldsymbol{\rho}|$. For $d=3$, we pass to polar coordinates in momentum space and perform the integration with respect to the azimuthal angle $\theta$ to obtain
\begin{eqnarray} \nonumber
\figu{fig22} & = & \int_{0}^{\infty}\frac{\rd q \, q}{(2\pi)^{2}} \, \frac{g+\kappa_{q}}{g-\kappa_{q}} \textrm{e}^{-\kappa_{q}\ell} \int_{0}^{2\pi} \rd\theta \, \textrm{e}^{ \im q \rho \cos\theta} \\
& = & \int_{0}^{\infty}\frac{\rd q \, q}{2\pi} \, \frac{g+\kappa_{q}}{g-\kappa_{q}} \textrm{e}^{-\kappa_{q}\ell} J_{0}(q \rho) \, .
\label{star04}
\end{eqnarray}
In the last line, we have carried out the integral with respect to the angle $\theta$ thanks to the identity
\begin{equation}
\label{star05}
\int_{0}^{2\pi} \rd\theta \, \textrm{e}^{ \im  x\cos\theta} = 2\pi J_{0}(x) \, .
\end{equation}
For $d>3$, we use (hyper)spherical coordinates and integrate over the angular variables to obtain
\begin{eqnarray} \nonumber
\figu{fig22} & = & \frac{2\pi^{d/2-1}}{\Gamma(d/2-1)}\int_{0}^{\infty}\frac{\rd q \, q^{d-2}}{(2\pi)^{d-1}} \, \frac{g+\kappa_{q}}{g-\kappa_{q}} \textrm{e}^{-\kappa_{q}\ell} \nonumber\\
&\times&\int_{0}^{\pi} \rd\theta \, (\sin\theta)^{d-3}\textrm{e}^{ \im q \rho \cos\theta} \label{star06} \\
& = & \int_{0}^{\infty}\rd q \left(\frac{q}{2\pi}\right)^{\frac{d-1}{2}} \, \frac{g+\kappa_{q}}{g-\kappa_{q}} \textrm{e}^{-\kappa_{q}\ell} \frac{J_{\frac{d-3}{2}}(q\rho)}{\rho^{\frac{d-3}{2}}} \, ,
\nonumber
\end{eqnarray}
where we used the Poisson's integral representation of the Bessel function
\begin{equation}
\label{star07}
J_\nu(x)=\frac{(x/2)^\nu}{\sqrt{\pi}\Gamma(\nu+1/2)}\int_{0}^{\pi} \rd\theta \, \textrm{e}^{ \im  x\cos\theta} (\sin \theta)^{2\nu} \, .
\end{equation}
Note that the expressions for $d=2$ and $d=3$, Eqs. (\ref{star03}) and (\ref{star04}), also satisfy (\ref{star06}).

\section{On the bubble diagram in $d=2$}
A general feature characterizing the Casimir effect is its exponential decay at large distances. This behavior is encoded in the bubble diagram that represents the leading form of $\left( \Omega_C\right)_{1}^{1}$, as shown in (\ref{wl_34}) and (\ref{wcasimirdiagrammatic2}).

This result is reminiscent of closely related results for the Casimir effect for exactly solvable models. The partition function of certain integrable models on a strip with homogeneous finite width is expressed as bubble-like contributions like the one in \eqref{wl_34} \cite{LeClair_1995, DS2015}. Intriguingly, these diagrams also arise in the study of phase separation via intermediate phases nested in such bubbles \cite{DS_bubbles}. These interfacial structures also play a crucial role in the bubble model for correlation functions in the Ising model, as explored by Abraham and coworkers \cite{Abraham_1983}. Analogous diagrams have been discussed in relation to the Casimir effect for the Ising model on the square lattice wrapped on a cylinder \cite{AM2010}. In particular, for the Ising model, the free energy per unit of circumference can be expressed as a linked cluster expansion. As shown in \cite{Abraham_CMP_60_181_1978}, the latter can be represented diagrammatically as a weighted sum of disjoint loops analogous to bubble diagrams scattered between walls; in this case, the bubble represents the propagation of a pair of two fermions \cite{AM2010}.

\bibliographystyle{apsrev}
\bibliography{bibliography}

\end{document}